\documentclass[acmtog]{acmart}
\usepackage{booktabs}
\usepackage{soul} 
\usepackage{xcolor}
\usepackage[normalem]{ulem}   
\usepackage{multirow}
\usepackage{multicol}
\usepackage{amsmath}
\usepackage{bm}
\usepackage{enumitem}
\usepackage[labelfont=bf,textfont=it]{subcaption}
\usepackage{hyperref}
\usepackage[ruled]{algorithm2e} % For algorithms

\SetAlFnt{\small}
\SetAlCapFnt{\small}
\SetAlCapNameFnt{\small}
\SetAlCapHSkip{0pt}

\setcopyright{none}
\renewcommand\footnotetextcopyrightpermission[1]{}

\begin{document}

% Title portion
\title{Skinned Motion Retargeting with Spatially Adaptive Interaction Guidance}

\author{Soojin Choi}
\orcid{0000-0002-0832-6545}
\affiliation{%
 \institution{Visual Media Lab, KAIST}
 \country{Republic of Korea}
}
\email{97choisj@kaist.ac.kr}

\author{Seokhyeon Hong}
\orcid{0000-0002-8490-5338}
\affiliation{%
 \institution{Visual Media Lab, KAIST}
 \country{Republic of Korea}
}
\email{ghd3079@kaist.ac.kr}

\author{Chaelin Kim}
\orcid{0000-0001-8355-7522}
\affiliation{%
 \institution{Visual Media Lab, KAIST}
 \country{Republic of Korea}
}
\email{chaelin.kim@kaist.ac.kr}

\author{Junghyun Nam}
\orcid{0009-0003-6427-2013}
\affiliation{%
 \institution{Visual Media Lab, KAIST}
 \country{Republic of Korea}
}
\email{ys4990@kaist.ac.kr}

\author{Junhyuk Jeon}
\orcid{0009-0003-4152-8696}
\affiliation{%
 \institution{Visual Media Lab, KAIST}
 \country{Republic of Korea}
}
\email{jeonjh@kaist.ac.kr}

\author{Junyong Noh}
\orcid{0000-0003-1925-3326}
\affiliation{%
 \institution{Visual Media Lab, KAIST}
 \country{Republic of Korea}
}
\email{junyongnoh@kaist.ac.kr}

\renewcommand\shortauthors{S. Choi et al}

\begin{abstract}
Retargeting motion across characters with varying body shapes while preserving interaction semantics, such as self-contact and near-body proximity, remains a challenging problem. While recent geometry-aware approaches address this by maintaining spatial relationships between predefined corresponding regions, their reliance on static correspondences often struggles when the target character exhibits exaggerated body proportions. In this paper, we present a geometry-aware motion retargeting framework that preserves interaction semantics by performing proximity matching over spatially adaptive anchors. Unlike prior methods with static anchor definitions, the proposed method dynamically repositions anchors to reachable regions on the target character. This is achieved via a Transformer-based anchor refinement strategy that predicts anchor displacements and constrains the translated anchors to remain on the target character geometry through differentiable soft projection. By incorporating pose-dependent spatial structures from the source character, the adapted anchors provide structurally coherent guidance for interaction-aware retargeting. Conditioned on these anchors, a graph-based autoencoder predicts target skeletal motion that preserves the spatial configuration of the source. To encourage task-aligned optimization between anchor adaptation and motion retargeting, we adopt an alternating training scheme in which each module is optimized in turn. Through extensive evaluations, we demonstrate that our method outperforms state-of-the-art approaches in preserving interaction fidelity across diverse character geometries. Code is available at \href{https://suzyn.github.io/space_page/}{\textit{\textbf{Project Page}}}.
\end{abstract}


\begin{teaserfigure}
  \includegraphics[width=\textwidth]{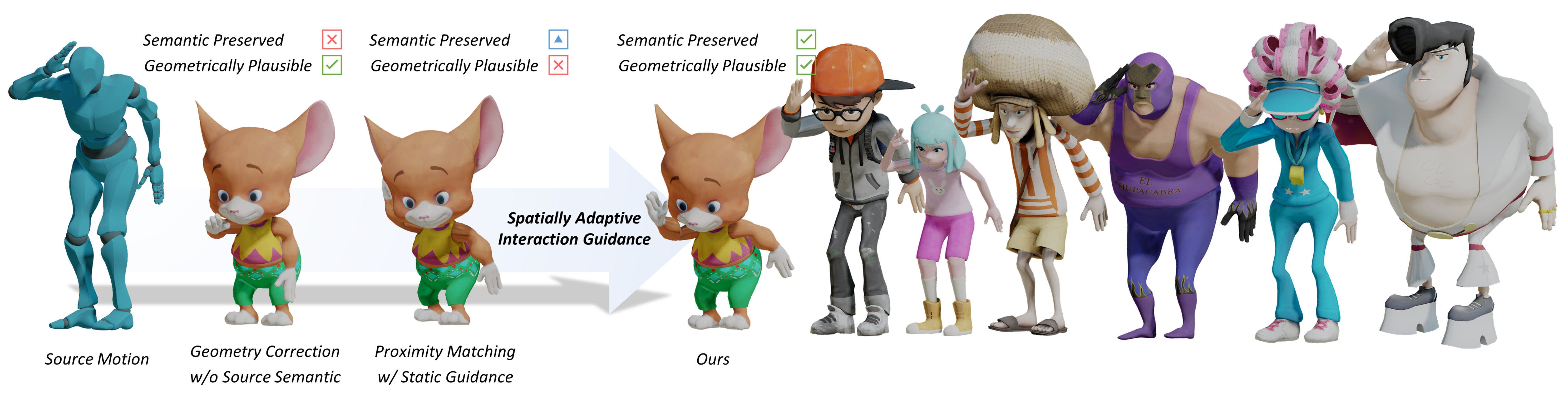} 
  \caption{We propose a geometry-aware motion retargeting framework that transfers motion between skinned characters by leveraging proximity matching over spatially adaptive anchors. Our method dynamically repositions anchors on the target mesh to semantically plausible and reachable locations, enabling the transfer of self-contact and near-body interactions across characters with exaggerated body proportions.}
  \label{fig:tesasr}
\end{teaserfigure}
\maketitle

% [arXiv] Remove only the ACM footer text while keeping the running header
\makeatletter
% \fancyfoot{}
% \thispagestyle{fancy}%
% \makeatother
\fancypagestyle{firstpagestyle}{%
  \fancyhf{}%
  \renewcommand{\headrulewidth}{0pt}%
  \renewcommand{\footrulewidth}{0pt}%
}
\fancyfoot{}%
\thispagestyle{firstpagestyle}
\makeatother

\section{Introduction}
3D characters are fundamental elements in digital content such as films, animation, and games, where their motions serve as the primary means of conveying actions, emotions, and narrative progression. Traditionally, animating these characters has relied on manual keyframing or motion capture techniques, both of which become time-consuming and labor-intensive when applied to a large number of characters. Motion retargeting addresses this challenge by enabling the automatic transfer of existing motion data, originally created for one character, to others with different skeletal structures or body shapes. As a result, motion retargeting has become a foundational technique in the field of computer graphics. 

While traditional motion retargeting focuses on aligning skeletal motion, it often overlooks surface-level interactions that are crucial for conveying semantic intent. The meaning of a motion is not solely determined by joint trajectories, but is significantly shaped by interaction relationships between body parts, such as self-contacts or near-body proximity. In particular, self-contact constitutes a fundamental component of human motion, as it encodes deliberate and semantically meaningful interactions between specific body parts that are crucial for maintaining balance or expressing intent. These spatial patterns are inherently dependent on a character’s geometry, including body shape and topology, even when the underlying skeletal motion remains identical. As a result, transferring motion without accounting for such geometric dependencies can lead to missing intended contacts, the emergence of unintended contacts, or artifacts such as interpenetration, ultimately compromising both the semantic accuracy and visual quality of the retargeted motion.

To address these challenges, recent motion retargeting methods have progressively evolved to incorporate character geometry and interaction inherent in the source motion. For instance, CAR \cite{villegas2021contact} identifies vertices involved in self-contact in the source motion and encourages the corresponding vertices on the target character to remain in contact, thereby enabling contact-aware retargeting. However, this approach struggles to generalize to motions involving close but non-contact interactions. 

Another line of work, \textit{proximity matching}-based approaches, addresses non-contact interactions by preserving spatial relationships between predefined geometrically corresponding regions. These regions are defined as either volumetric spaces surrounding the character mesh \cite{jin2018aura} or as sparse point sets sampled on the character surface \cite{ho2010spatial, jang2024geometry, ye2024skinned, cheynel2025reconform, liu2018surface}, which we refer to as \textit{anchors} in this work, enabling unified handling of both contact and non-contact interactions across varying character geometries. Unfortunately, these methods often struggle when the target character exhibits significant morphological differences, such as exaggerated body proportions as shown in Figure \ref{fig:tesasr}. This limitation arises from the use of predefined static geometric correspondences, which do not consider whether the target character can accommodate the intended interactions. As a result, motion retargeting may fail to preserve contact semantics when the corresponding region on the target is unreachable or incompatible with the target character's body structure and proportions. 

In this work, we propose a geometry-aware motion retargeting framework that transfers motion between skinned characters by leveraging proximity matching over spatially adaptive anchors. Unlike prior proximity matching-based methods that rely on predefined static anchors, our method dynamically repositions anchors on the target mesh to semantically plausible locations that lie within the kinematically reachable regions of the target character. 
This is achieved by an \textbf{Adaptive Anchor Sampling} module, which refines initial anchor positions using a Transformer-based residual predictor and constrains them to the mesh surface through a differentiable soft projection. This module incorporates deformed source anchors, which implicitly encode the source character’s pose-dependent spatial structure, into a learnable direction-aware attention bias that captures pairwise geometric relationships. The resulting anchors are reachable and structurally coherent, providing effective guidance for proximity-based motion retargeting. 

Conditioned on these adapted anchors, a \textbf{Proximity-based Retargeting} module predicts target skeletal motion that preserves the spatial configurations observed in the source motion through a graph-based encoder-decoder architecture. To facilitate task-aligned coordination between anchor repositioning and motion retargeting, we adopt an alternating optimization strategy in which each module is iteratively optimized while the other remains fixed. This decoupled training scheme allows each component to specialize in its respective objective, enabling robust transfer of self-contact and near-body interactions across characters with varying body shapes, while avoiding over-constrained or physically implausible results.

Our technical contributions can be summarized as follows:
\begin{itemize} [leftmargin=*] 
    \item We propose a \textit{spatially adaptive anchor representation} for motion retargeting, wherein anchors are dynamically repositioned on the target mesh through a learned sampling process conditioned on both source motion and target geometry. 
    \item We develop an \textit{Adaptive Anchor Sampling} module that repositions interaction-related anchors within kinematically reachable and geometrically compatible regions of the target character, while suppressing interpenetration. This regularization facilitates accurate transfer of self-contact and near-body interactions without inducing poses that are incompatible with the geometry of the target character. 
    \item We adopt a task-aligned training scheme for spatially adaptive anchors, enabling anchor repositioning to be learned with respect to the downstream motion retargeting objective while supporting coordinated learning between anchor adaptation and motion retargeting.
\end{itemize}
\section{Related Work}

\subsection{Skeleton-aware Motion Retargeting} 
Early approaches on motion retargeting formulated the problem as constrained non-linear optimization, aiming to preserve the salient characteristics of the source motion while accommodating differences in bone lengths and proportions \cite{gleicher1998retargetting, choi2000online, lee1999hierarchical, shin2001computer, ho2010spatial}. With the advent of deep learning, data-driven methods have significantly improved the scalability and generalizability of motion retargeting by leveraging large-scale motion datasets assuming an identical skeleton structure between source and target characters. For instance, \citeN{uk2020variational} proposed a variational autoencoder-based framework trained on paired motions with different bone lengths under direct supervision. NKN \cite{villegas2018neural} addressed motion retargeting in an unsupervised manner by enforcing cycle consistency, requiring the retargeted motion to be transferred back to reconstruct the original motion. Similarly, PMnet \cite{lim2019pmnet} introduced a perceptual pose loss that aligns source and retargeted motions in a learned pose embedding space. 

Beyond variations in bone lengths and proportions, some research have focused on retargeting to skeletons with different structure, including varying number of joints and their hierarchy. SAN \cite{aberman2020skeleton} embeds skeletons with different numbers of joints into a shared latent space by pooling them to a common primal skeleton. In a more recent effort, SAME \cite{lee2023same} proposed a skeleton-agnostic motion embedding space that disentangles structural information from input motion while preserving its semantics. Building on the graph-based retargeting framework \cite{lee2023same} that supports kinematic retargeting between diverse skeletons in a unified framework, we extend motion retargeting to a geometry-aware setting by incorporating anchors, surface-based proximity representations that capture spatial relationships on the character mesh.
\pagebreak
\subsection{Geometry-aware Motion Retargeting} 
To address the limitations of skeleton-based approaches in handling surface-level interactions, several studies have proposed methods that incorporate target geometry to mitigate interpenetration artifacts. By incorporating an additional geometric correction stage on meshes deformed by skeleton-aware retargeting, R$^2$ET \cite{zhang2023skinned} introduced a distance field-based motion refinement that softly pushes vertices penetrating the body mesh, whereas MoMa \cite{martinelli2024moma} adjusted joint configurations to resolve mesh intersections at the face level. Building upon R$^2$ET \cite{zhang2023skinned}, STaR \cite{yang2025star} proposed a method to address inter-limb penetrations in a temporally consistent manner. \citeN{lee2024learning} proposed a shape-aware refinement network that explicitly conditions pose correction on mesh-derived features, such as signed distance fields. 

In addition to reducing penetration artifacts, prior work has emphasized modeling the semantic structure of source motions and transferring it to target characters to preserve meaningful interactions. CAR \cite{villegas2021contact} proposed a method that identifies pairs of vertices on the source character involved in self-contact and transfers these relationships to the target by enforcing positional constraints on the corresponding vertices. \citeN{basset2020contact} addressed this using a shape transfer-based approach, in which the target character shape is transferred onto the deformed source pose. During shape transfer, contact pairs identified in the source pose are preserved by penalizing increases in their distances beyond a fixed threshold.

Alternatively, other studies have introduced intermediate geometric representations that can encode both contact and near-body interactions, thereby enabling generalization beyond explicit contact. \citeN{jin2018aura} proposed a volumetric mesh that surrounds the character to capture spatial relationships around the body. Other methods abstract the character geometry using sparse point sets sampled on the surface \cite{ho2010spatial, ye2024skinned, cheynel2025reconform, jang2024geometry, liu2018surface}, whereas \citeN{guo2025ultrafast} introduced a representation based on collections of spheres that approximate the character’s body volume. These abstractions are designed to establish semantically consistent correspondences across characters, such that each sampled point or sphere refers to a geometrically similar region across different body shapes. During motion retargeting, these correspondences enable proximity matching, which preserves the relative spatial relationships of body parts from the source character in the corresponding regions of the target character. 

Extending this line of research, which relies on predefined geometric correspondences that remain static during motion retargeting, we introduce \textit{spatially adaptive anchors}, designed to account for both the semantic intent of the source motion and the morphological affordances of the target character. By jointly considering the spatial structure of the source motion and whether the target character can physically reproduce the intended interactions in regions corresponding to the source, our method achieves robust and semantically faithful motion retargeting under large morphological variations, such as an enlarged head or a bulky torso.
\subsection{Interaction Transfer to New Geometries}
The proximity matching strategy has also been widely adopted to transfer interactions with surrounding environments or objects whose shapes or topologies differ from those of the original interacting geometry. \citeN{al2013relationship} introduced the relationship descriptor, a static set of points sampled on the surface of the interacting geometry, and retargeted motion by preserving relative displacements between body parts and these descriptor points under geometric deformation. Subsequent work has adopted spatial mapping strategies that establish bijective correspondences between the source and target geometries over surfaces and surrounding volumes \cite{kim2016retargeting, choi2023online} to determine target joint positions that remain consistent with the original interaction patterns. Building on these proximity-based formulations, several learning-based approaches have augmented motion capture datasets by adapting interactions originally captured with template shapes to diverse geometries, enabling data-driven models to learn interaction semantics from proximity cues \cite{starke2019neural, hassan2021stochastic, zhang2022couch, jin2023dafnet, jin2025interfacerays}.

To improve robustness on large geometric deviations between the source and target objects, \citeN{tonneau2016character} introduced a \textit{contact repositioning} strategy that searches for alternative contact locations when the contacts defined by the static descriptors \cite{al2013relationship} do not yield physically plausible motion or faithfully reproduce the original contact. Inspired by this idea, which was originally developed for character-environment interactions, we generalize this concept to intra-body interactions, where the contact target becomes other moving body parts rather than static surfaces. 
\section{Method}
\begin{figure*}[t]
    \centering \includegraphics[width=\textwidth]{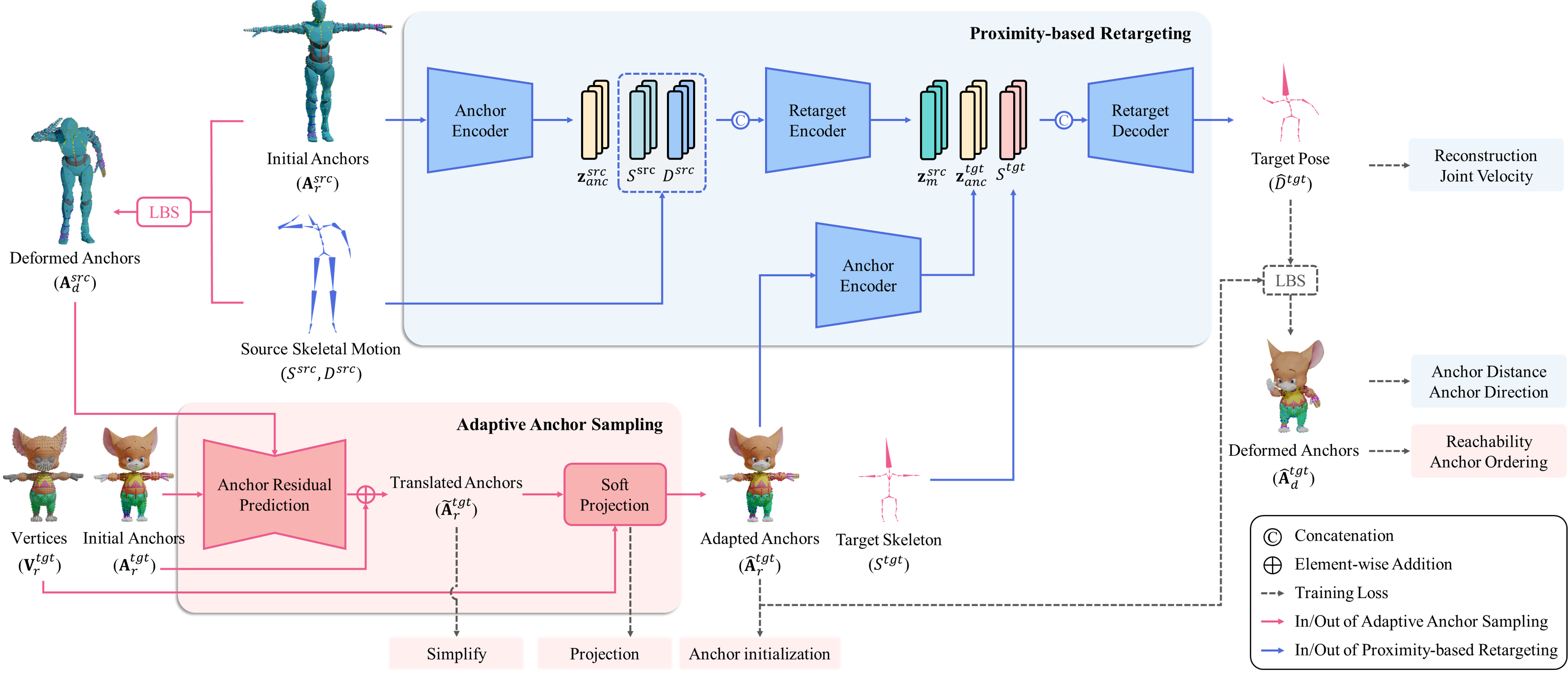}
    \caption{Overview of our method. Pink lines indicate the input-output flow of the Adaptive Anchor Sampling module, whereas blue lines denote that of the Proximity-based Retargeting module. Gray dashed lines connect the components on which each specified loss is computed. Please note that the superscript $t$ is omitted from each symbol for brevity.}
    \label{fig:overview}
\end{figure*}
Our goal is to retarget a source pose to a target skeleton while accounting for geometric differences between characters, such that the retargeted motion preserves spatial relationships between interacting surface regions and remains geometrically plausible on the target character. As shown in Figure \ref{fig:overview}, our method comprises two components: \textbf{Adaptive Anchor Sampling} and \textbf{Proximity-based Retargeting}. The Adaptive Anchor Sampling module begins with a sparse set of predefined anchors on the target character that represent coarse geometric correspondences to the source. It then refines their positions within kinematically reachable regions to better support proximity-based constraints that enforce surface interactions consistent with those in the source motion. Conditioned on the adapted anchors, the Proximity-based Retargeting module subsequently transfers the source pose to the target skeleton by aligning the relative configurations of source and adapted target anchors. By leveraging both skeletal configuration and geometric constraints provided by the anchors, this module generates target motions that preserve the spatial structure implied by the source skinned motion. In the following sections, we first describe the data representations employed in our framework (Section \ref{sec:Data_Representation}). We then detail the architectures of Adaptive Anchor Sampling (Section \ref{sec:Adaptive_Anchor_Sampling}) and Proximity-based Retargeting (Section \ref{sec:Motion_Retargeting}). Finally, we present the training objectives of each module, followed by the alternating optimization strategy used to coordinate the training of two modules (Section \ref{sec:Training_Procedures}). 

\subsection{Data Representation} \label{sec:Data_Representation}
This section describes the data representation used consistently throughout our framework for both training and inference. For each character, we construct a pair of $\{M, G\}$, where $M$ represents the skeletal motion data comprising a skeleton and a set of motion sequences, and $G$ denotes the geometry data of the character in its rest pose. By deforming the geometry data $G$ to follow the skeletal motion $M$, we obtain proximity representations that capture the time-varying spatial relationships induced by the skinned motion.

\paragraph{Skeletal data} Following ~\citeN{lee2023same}, we define the skeletal motion data $M$ as follows:
\begin{equation}
    M=({S, D^{1:N_T}})
\end{equation}
where $S$ and $D^{1:N_T}$ denote the skeleton in the rest pose and a sequence of motion data over $N_T$ frames, respectively. 
The skeleton $S$ is defined as $S=\{ \mathbf{g}_{1:N_J}, \mathbf{o}_{1:N_J} \}$, where $N_J$ indicates the number of joints. Each joint $j$ is associated with its global position $\mathbf{g}_j \in \mathbb{R}^{3}$ and local position $\mathbf{o}_j \in \mathbb{R}^3$, defined relative to its parent joint. For each timestep $t$, the motion data is defined as 
\begin{equation}
    D^t=\{\mathbf{q}^t_{1:N_J}, \mathbf{p}^t_{1:N_J}, \mathbf{p}^{t-1}_{1:{N_J}}, \dot{\mathbf{p}}^t_{1:N_J}, \mathbf{r}^t,\mathbf{c}^t_{1:N_J}\},
\end{equation}
where $\mathbf{q}^t_{j}$ denotes the local joint rotation of joint $j$ with respect to the parent joint, represented using the 6D rotation format \cite{zhou2019continuity}. $\mathbf{p}^t_{j} \in \mathbb{R}^3$ denotes the joint position expressed in the character's facing frames \cite{holden2017phase} and  $\dot{\mathbf{p}}^t_j \in \mathbb{R}^3$ is the linear velocity between consecutive frames computed by $\dot{\mathbf{p}}^t_j=(\mathbf{p}^t_j-\mathbf{p}^{t-1}_j)/\Delta t$, where $\Delta t$ denotes the fixed time interval between adjacent frames. The root movement $\mathbf{r}^t=(\Delta x, \Delta z, \Delta \theta, h)$ represents the root displacement between frames, where $(\Delta x, \Delta z)$ are the translational velocities on the ground, $\Delta \theta$ denotes the rotational velocity around the up-axis, and $h$ is the absolute height of the root joint above the ground. Unless otherwise specified, $(\Delta x, \Delta z, \Delta \theta)$ are measured with respect to the facing frame of the previous time step. Finally, $\mathbf{c}^t_{j}$ is a binary contact label indicating whether joint $j$ is in contact with the ground or not. 

\begin{figure} [t!]
    \centering \includegraphics[width=\columnwidth]{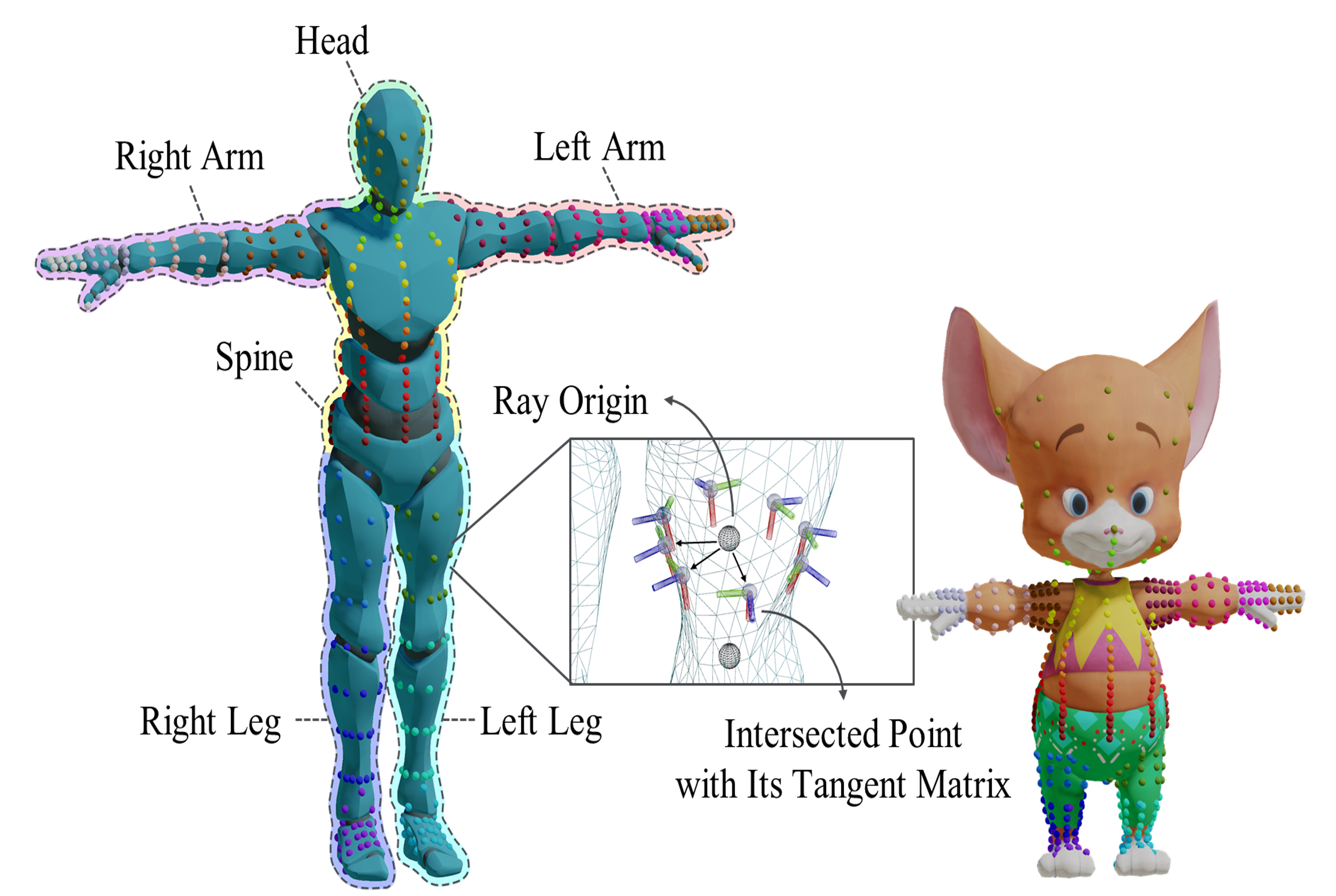}
    \caption{Visualizations of initial anchors and their corresponding body parts. Anchors sharing the same color are derived from the same skeletal bone, demonstrating consistent semantic regions across varying mesh geometries. The center illustration depicts the anchor extraction process, where rays are cast from sampled origins to compute anchor positions along with their corresponding tangent matrices on the mesh surface.}
    \label{fig:initial_anchors} 
\end{figure}

\paragraph{Geometry data} We define the geometry data $G$ based on the character mesh in the rest pose as follows:
\begin{equation}
    G = \{ \mathbf{V}_{r}, \mathbf{A}_{r}, \mathbf{T}_{r}\},
\end{equation}
where $\mathbf{V}_{r}\in \mathbb{R}^{N_V \times 3}$ denotes the vertex positions of the mesh in the rest pose, where $N_V$ is the number of vertices. $\mathbf{A}_{r} \in \mathbb{R}^{N_A \times 3}$ denotes the anchor positions in the rest pose, where $N_A$ is the number of anchors. Because anchors serve as an abstract representation that defines geometrically corresponding locations across characters, $N_A$ remains constant across the characters, while $N_V$ may vary depending on the mesh resolution of each character. $\mathbf{T}_{r} \in \mathbb{R}^{N_A \times 3\times3}$ represents the tangent matrices defined at each anchor. To extract a fixed-size anchor set from character meshes with varying topologies, we adopt the sampling strategy introduced by \citeN{ye2024skinned}. 
Using the rest-pose skeleton $S$ as a canonical reference, this procedure produces a bone-indexed anchor set, with anchors placed in functionally analogous regions across characters, despite differences in body shape and mesh topology as illustrated in Figure \ref{fig:initial_anchors}.
Detailed procedures for extracting the anchors and their corresponding tangent matrices are provided in Section 1.1 of the supplementary document.

\paragraph{Proximity representation} Given a motion frame $D^t$ at timestep $t$, we obtain the deformed anchor positions $\mathbf{A}^t_{d}\in \mathbb{R}^{N_A \times 3}$ by applying Linear Blend Skinning (LBS) to the rest pose anchors $\mathbf{A}_{r}$. Throughout this paper, the subscripts $r$ and $d$ are used to distinguish quantities defined in the rest pose and deformed pose spaces, respectively. Based on $\mathbf{A}^t_{d}$, we compute pairwise proximity cues at each time step to model the spatial relationships between anchors. Each proximity cue consists of the inter-anchor distance, which measures the closeness between anchors, and the relative direction vector, encoding their directional relationships in the tangent frame created by the tangent matrix of the anchor serving as the origin.

We compute a pairwise Euclidean distance matrix $\mathbf{D}^{t}_{dist} \in \mathbb{R}^{{N_A} \times {N_A}}$, where each element $\mathbf{D}_{dist}^t(i, j)$ denotes the distance between deformed anchors $i$ and $j$ at timestep $t$: 
\begin{equation}
    \mathbf{D}^{t}_{dist}(i, j) =  \lVert \mathbf{A}^{t}_{d,i} - \mathbf{A}^{t}_{d, j} \rVert_2.
\end{equation}
In addition, for each pair of anchors $(i, j)$, the direction from anchor $i$ to anchor $j$ expressed in the tangent frame of anchor $i$ is defined following \citeN{ye2024skinned}:
\begin{equation}
    \mathbf{D}_{dir}^t(i, j) = {\mathbf{T}^{t}_{d,i}}^{-1} ({\mathbf{A}_{d,j}^t - \mathbf{A}_{d,i}^t}),
\end{equation}
where $\mathbf{T}^{t}_{d,i} \in \mathbb{R}^{3\times3}$ denotes the deformed tangent frame at anchor $i$ following $D^t$. These pairwise features serve as geometry-aware constraints for modeling proximity-driven interactions. 

To emphasize  anchor pairs when modeling interactions, we further define a distance-based weighting matrix $\mathbf{W}^{t}_{dist} \in \mathbb{R}^{N_A \times N_A}$ using an exponential decay function:
\begin{equation}
    \mathbf{W}^{t}_{dist}(i, j) = \exp \left( -\alpha \cdot \frac{\max(\mathbf{D}^{t}_{dist}(i, j) - d_{\min}, 0)}{d_{\max} - d_{\min}} \right)
\end{equation}
where $\alpha$ is a scaling factor that controls the rate of decay, and $d_{\min}$ and $d_{\max}$ are distance thresholds defined relative to the character height. Following ReConForm \cite{cheynel2025reconform}, we set $d_{\min}$ and $d_{\max}$ to 5$\%$ and 15$\%$ of the character height, respectively. This formulation assigns higher weights to spatially close anchor pairs while suppressing the influence of distant ones. For brevity, we omit $t$ in the following paragraphs. 
\begin{figure*}[t]
    \centering \includegraphics[width=\textwidth]{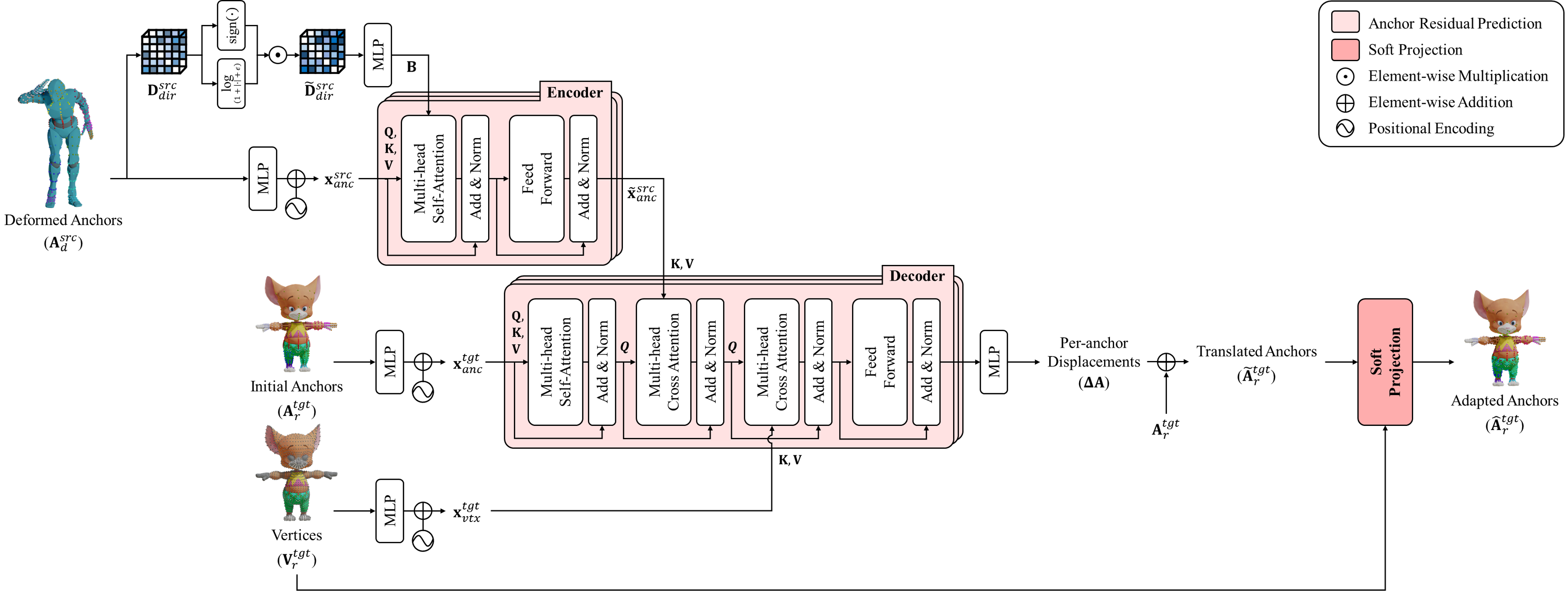}
    \caption{Overview of Adaptive Anchor Sampling module.}
    \label{fig:anchor_sampling_module} 
\end{figure*}
\subsection{Adaptive Anchor Sampling} \label{sec:Adaptive_Anchor_Sampling}
The Adaptive Anchor Sampling module is introduced to address the limitation of the predefined static correspondences under large morphological variations, where preserving interaction semantics often requires identifying alternative yet feasible surface regions on the target character. Therefore, given the initial anchors of the target character, which are defined based on coarse geometric correspondences to the source character, this module aims to reposition the initial anchors of the target character such that resulting anchors provide optimal reference points for subsequent proximity-based retargeting. Rather than enforcing a unique anchor configuration for each source and target character pair, the adapted anchors are designed to serve as learned intermediate guidance that better preserves interaction semantics as well as pose plausibility under shape variation. As illustrated in Figure \ref{fig:anchor_sampling_module}, the Adaptive Anchor Sampling module consists of two components: an \textbf{Anchor Residual Prediction} module, which predicts anchor-wise displacements, and a \textbf{Soft Projection} module, which projects the translated anchors onto the target mesh surface in a differentiable manner.

\subsubsection{Anchor Residual Prediction}
The Anchor Residual Prediction module predicts displacements for the initial anchors of the target character $\mathbf{A}^{tgt}_{r}$, producing translated anchor positions $\mathbf{\tilde{A}}^{tgt}_{r} \in \mathbb{R}^{N_A \times 3}$, which serve as intermediate representations toward obtaining the final adapted anchors. To ensure that anchor repositioning is guided by both the source motion and the target character geometry, we adopt an attention mechanism \cite{vaswani2017attention} that models intra- and inter-modality relationships. Specifically, the encoder operates on the deformed source anchors, which implicitly encode the current source pose, to capture pose-specific spatial relationships among anchors. Conditioned on the encoded source features, the decoder estimates per-anchor offsets for the target character, which are then applied to translate the target anchors. For clarity, the anchor offsets are predicted in the rest pose space, and pose-dependent deformation is handled later in the retargeting stage. 

The encoder models pose-conditioned spatial relationships among source anchors using a stack of Transformer encoder blocks with geometric relative bias. The deformed source anchor positions $\mathbf{A}^{src}_{d} \in \mathbb{R}^{N_A\times3}$ are first embedded through 1D convolution layers to produce per-anchor features $\mathbf{x}^{src}_{anc} \in \mathbb{R}^{N_A \times N_G}$, where $N_G$ is the feature dimension. To encode correspondences between source and target anchors, we assign identical positional encodings to anchor pairs that are assumed to match across characters, thereby encouraging consistent association despite geometric variation. The resulting features are then linearly projected into query, key, and value embeddings, denoted as $\mathbf{Q}$, $\mathbf{K}$ and $\mathbf{V} \in \mathbb{R}^{N_A \times N_G}$, respectively. These tokens are subsequently processed by a stack of Transformer encoder blocks, producing source anchor embeddings $\mathbf{\tilde{x}}^{src}_{anc} \in \mathbb{R}^{N_A \times N_G}$. The self-attention operation is defined as follows:
\begin{equation} \label{eq:Adaptive_Anchor_Sampling_Encoder_Attention}
    Attn(\mathbf{x}^{src}_{anc}) = \text{softmax}\left(\frac{\mathbf{Q} {\mathbf{K}}^{\top}}{\sqrt{N_G}} + \mathbf{B} \right)\mathbf{V}, 
\end{equation}
where $\mathbf{B} \in \mathbb{R}^{N_A \times N_A}$ denotes a learnable relative positional bias that encodes pairwise geometric relationships between deformed source anchors. Inspired by \citeN{hong2023attention}, this bias is constructed from relative displacements between anchors represented in their respective tangent frames, allowing the model to explicitly capture spatial dependencies induced by the source pose. For details on constructing the bias, please refer to Section 1.2 of the supplementary document.

The decoder takes the initial anchors of the target character $\mathbf{A}^{tgt}_r$ as input and predicts their displacements $\Delta \mathbf{A} \in \mathbb{R}^{N_A \times 3}$, which are subsequently applied to obtain the translated anchors $\mathbf{\tilde{A}}^{tgt}_{r}$. The rest pose target anchors $\mathbf{A}^{tgt}_{r}$ are first embedded using 1D convolution layers to produce per-anchor features $\mathbf{x}^{tgt}_{anc} \in \mathbb{R}^{N_A \times N_G}$. Similarly, target mesh vertices in the rest pose $\mathbf{V}^{tgt}_{r}$ are embedded into per-vertex features $\mathbf{x}^{tgt}_{vtx} \in \mathbb{R}^{N_V \times N_G}$. After applying positional encoding, $\mathbf{x}^{tgt}_{anc}$ are iteratively refined through a series of decoder blocks. The attention operations proceed as follows:
\begin{enumerate} [leftmargin=*] 
    \item \textbf{Self-attention among target anchors:} This layer captures inter-anchor relationships by allowing each target anchor to attend to all others through self-attention over $\mathbf{x}^{tgt}_{anc}$.
    \item \textbf{Cross-attention to source anchors:} To inject pose-dependent information and establish anchor-level semantic correspondences, the updated target anchor features attend to the encoded source anchor embeddings $\tilde{\mathbf{x}}^{src}_{anc}$.
    \item \textbf{Cross-attention to target mesh vertices:} To ensure that the predicted anchor displacements are consistent with the target mesh geometry, the target anchor queries attend to per-vertex features $\mathbf{x}^{tgt}_{vtx}$ derived from the rest-pose target mesh.
\end{enumerate}
Through these three attention operations, the representation of each target anchor is iteratively refined by aggregating information from other target anchors, the source pose, and the target mesh geometry. The final output embedding is passed through an MLP to predict displacements $\Delta \mathbf{A}$, which are added to the initial rest pose anchors to obtain the translated target anchors:
\begin{equation}
\mathbf{\tilde{A}}^{tgt}_{r} = \mathbf{A}^{tgt}_{r} + \Delta \mathbf{A}.
\end{equation}

\subsubsection{Soft Projection}
\begin{figure} [!t]
    \centering \includegraphics[width=\columnwidth]{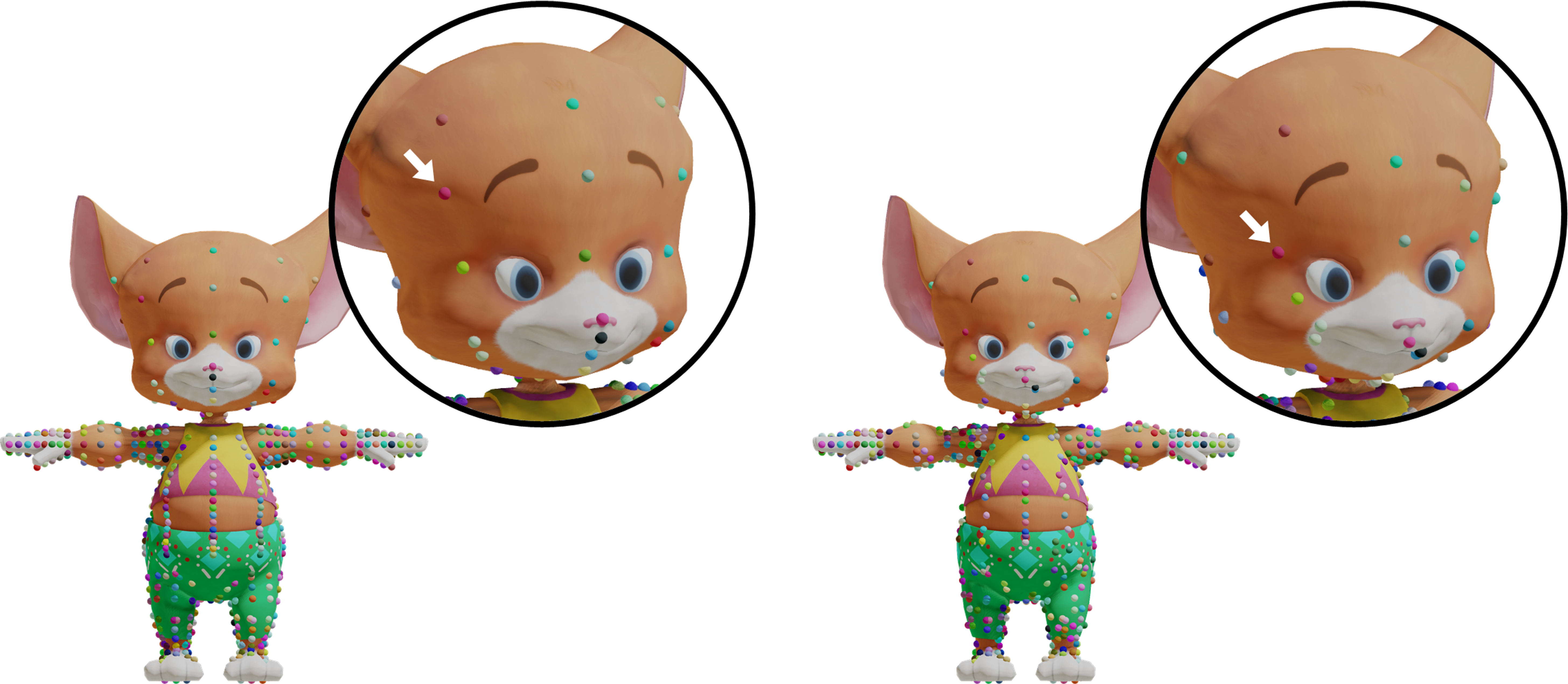}
    \caption{Visualization of the initial (left) and adapted (right) anchors on the target character. Anchors with identical colors represent corresponding anchor identities before and after the adaptive sampling process. A white arrow highlights an anchor pair to illustrate the repositioning induced by the adaptation process.}
    \label{fig:anchor_adapted}
\end{figure}
Directly adding the predicted displacements to the initial target anchors does not guarantee that the resulting anchors lie on the surface of the target character mesh. A common strategy is to apply nearest neighbor (NN) selection, which assigns each translated anchor to its closest mesh vertex. However, such NN selection is inherently non-differentiable, thereby preventing end-to-end gradient-based optimization. To address this, we adopt the Soft Projection operation \cite{lang2020samplenet}, which constrains each translated anchor to the target mesh in a differentiable manner by expressing it as a weighted combination of nearby mesh vertices. 

Specifically, each translated target anchor is expressed as a distance-weighted combination of its $k$-nearest vertices on the target rest mesh $\mathbf{V}^{tgt}_{r}$, yielding the adapted anchors $\hat{\mathbf{A}}^{tgt}_{r} \in \mathbb{R}^{N_A \times 3}$. The smoothness of the distance-based weighting is controlled by a learnable temperature parameter $\tau$. Figure~\ref{fig:anchor_adapted} illustrates the transition from the initial target anchors to the adapted anchors after applying the predicted displacements and the subsequent Soft Projection onto the target mesh surface. For details on the soft projection operation, please refer to Section 1.3 of the supplementary document.
\begin{figure}
    \centering
    \begin{minipage}{0.23\columnwidth}
        \centering
        \includegraphics[width=\columnwidth]{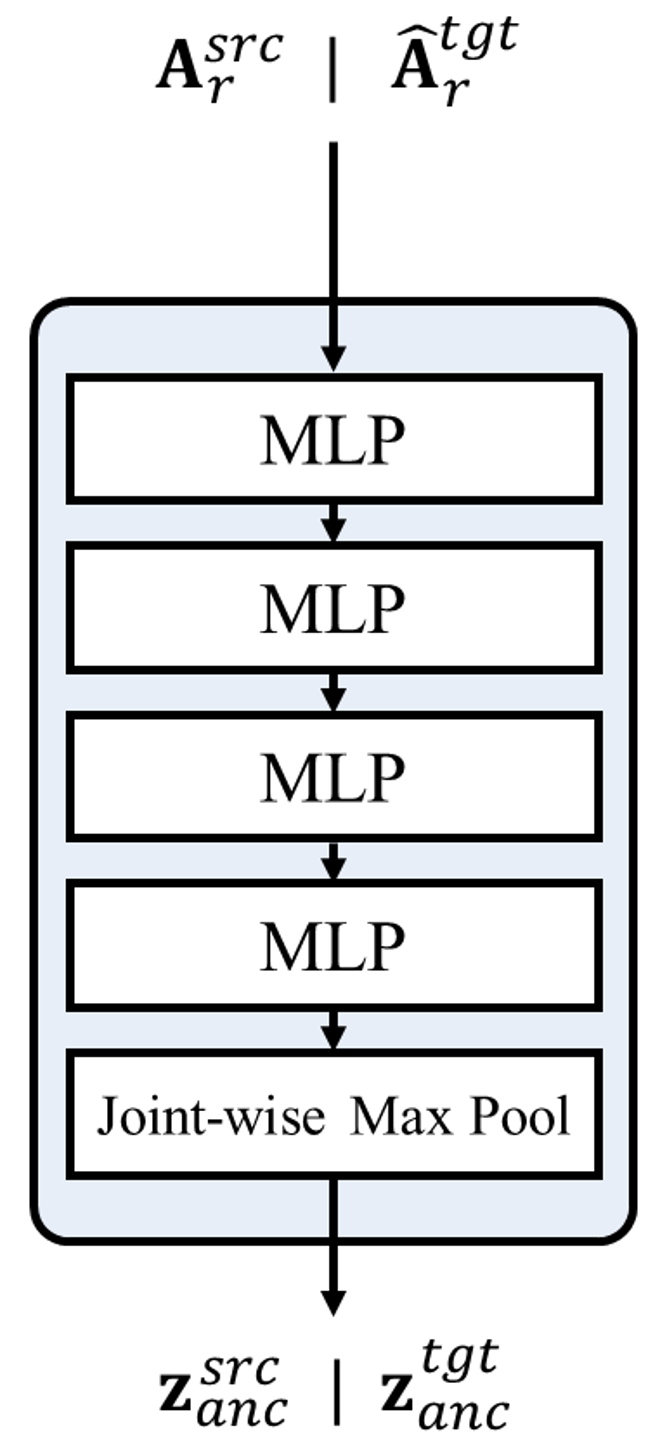}
        \subcaption{\footnotesize \textit{Anchor Encoder}}        
        \label{fig:anchor_encoder}
    \end{minipage}
    \hspace{2.5em}
    \begin{minipage}{0.23\columnwidth}
        \centering
        \includegraphics[width=\columnwidth]{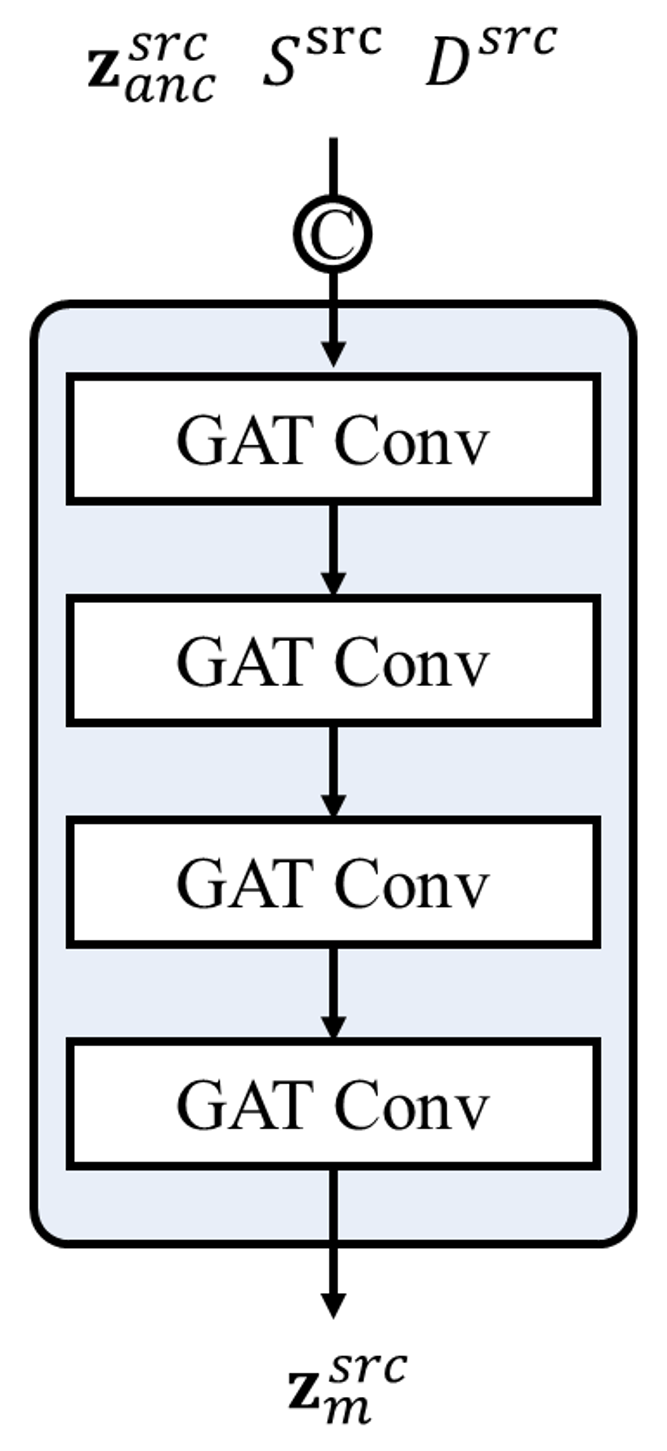}
        \subcaption{\footnotesize \textit{Retarget Encoder}}        
        \label{fig:retarget_encoder}
    \end{minipage}
    \hspace{2.5em}
    \begin{minipage}{0.23\columnwidth}
        \centering 
        \includegraphics[width=\columnwidth]{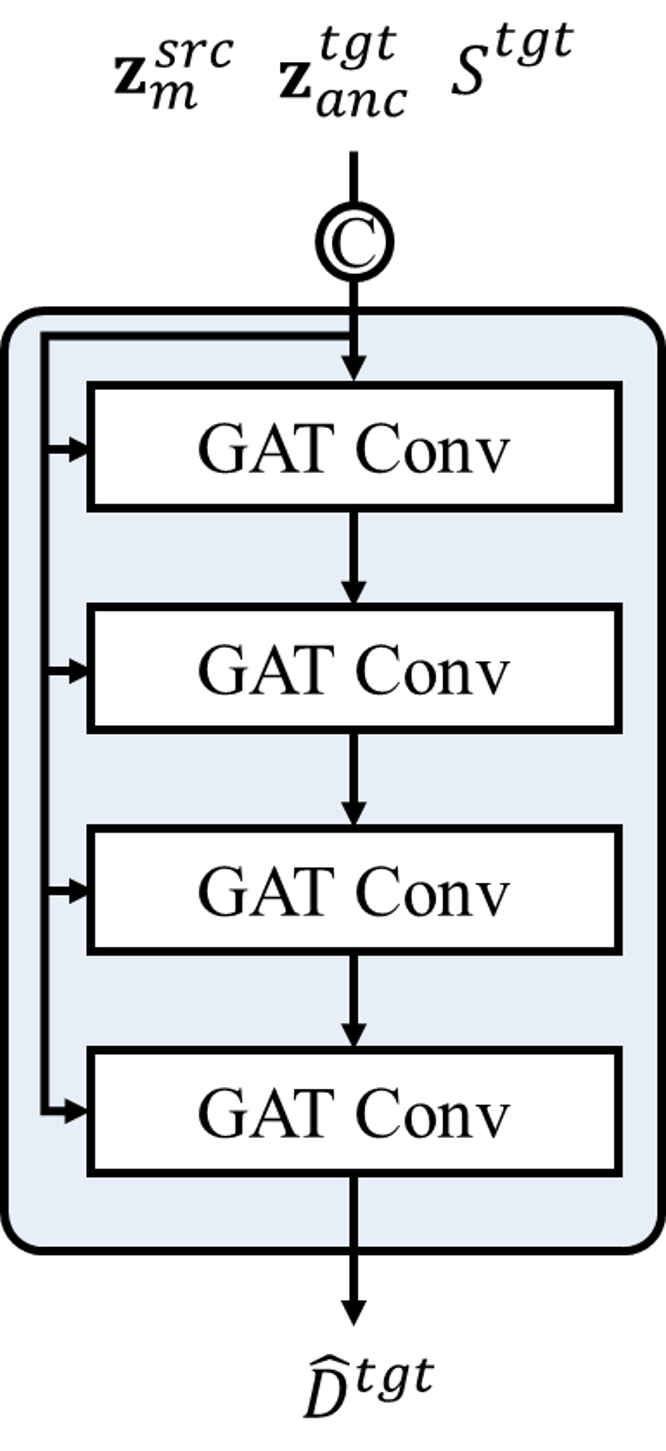}
        \subcaption{\footnotesize \textit{Retarget Decoder}}        
        \label{fig:retarget_decoder}
    \end{minipage}
    \caption{Overview of (a) anchor encoder, (b) retarget encoder, and (c) retarget decoder.}
    \label{fig:retargetnet_architecture}
\end{figure}
\subsection{Proximity-based Retargeting} \label{sec:Motion_Retargeting}
The Proximity-based Retargeting module transfers a source pose sequence to a target skeleton by performing proximity matching in the anchor space, thereby ensuring that the retargeted motion remains geometrically plausible with respect to the target character. To this end, this module utilizes the deformed source anchors $\mathbf{A}^{src}_{d}$ as a geometric reference for defining proximity objectives, and predicts the target motion conditioned on the adapted target anchors $\hat{\mathbf{A}}^{tgt}_{r}$. By aligning the relative configurations between source and target anchors through proximity objectives, the model preserves pose-dependent interaction patterns, such as contact and near-body proximity exhibited in the source motion while adapting them to the geometry of the target character. The module comprises two components: an \textbf{Anchor Encoder}, which aggregates spatial features from anchors into joint-wise embeddings and a \textbf{Retargeting Network}, which predicts target joint motions conditioned on both the skeletal input and the encoded anchor features.

\subsubsection{Anchor Encoder} 
To align anchor features with joint representations, we employ a PointNet-like architecture \cite{qi2017pointnet} as shown in Figure \ref{fig:anchor_encoder}. The encoder operates on the rest pose anchor positions: for the source character, we use the initial rest pose anchors $\mathbf{A}^{src}_{r}$, and for the target character, we use the adapted rest pose anchors $\hat{\mathbf{A}}^{tgt}_{r}$ produced by the Adaptive Anchor Sampling module. Each anchor is treated as an individual point and processed through a stack of MLPs, producing per-anchor feature embeddings. Subsequently, we aggregate anchor embeddings by applying a max-pooling operation over all anchors associated with the same joint, resulting in joint-wise anchor features for the source and target characters, denoted as $\mathbf{z}^{src}_{anc},\, \mathbf{z}^{tgt}_{anc} \in \mathbb{R}^{N_J \times N_R}$, where $N_R$ is the feature dimension. These joint-wise features are used as geometry-aware conditioning signals that can be directly integrated into the following retargeting network. 

\subsubsection{Retargeting Network}
% Use of GCN AutoEncoder -> Encoder/Decoder Input/Output -> Effect 
Inspired by \citeN{lee2023same}, we adopt a graph-based autoencoder architecture to predict skeletal motion for the target character. The skeleton is modeled as a graph, in which joints correspond to nodes and bones define the edges. Specifically, we employ Graph Attention Networks (GATs) \cite{velivckovic2017graph}, which extend graph convolutions by learning attention weights over neighboring nodes. Although we assume a consistent skeletal structure across characters, we adopt a graph-based representation in which each node is extended to jointly encode a skeleton joint and its corresponding anchors. This formulation allows the network to capture not only structural motion patterns but also the local geometric context around each joint, thereby facilitating the faithful transfer of interaction semantics beyond skeletal kinematics.

Given a source skeleton and a sequence of poses, the encoder processes each frame independently. For a single frame, the encoder takes the source pose $D^{src}$, the source skeleton $S^{src}$, and the encoded source anchors in the rest pose $\mathbf{z}^{src}_{anc}$ as input as shown in Figure \ref{fig:retarget_encoder}. These features are concatenated and passed through a stack of GAT layers, resulting in a source motion embedding $\mathbf{z}^{src}_{m} \in \mathbb{R}^{N_J \times N_R}$. This embedding captures the spatial structure of the source pose while being conditioned on both the skeleton and the anchors located on the character geometry. The decoder then generates the target motion conditioned on the target character as shown in Figure \ref{fig:retarget_decoder}. Specifically, it takes the target skeleton $S^{tgt}$, the embedded target anchors $\mathbf{z}^{tgt}_{anc}$, and the source motion embedding $\mathbf{z}^{src}_{m}$ as input and reconstructs the target pose $\hat{D}^{tgt}=\{\mathbf{\hat{q}}^{tgt}_{1:N_J}, \mathbf{\hat{r}}^{tgt}, \mathbf{\hat{c}}^{tgt}_{1:N_J}\}$. The decoder adopts the same architecture as the encoder, but additionally incorporates residual connections for each GAT layer.

\subsection{Training} \label{sec:Training_Procedures}
In this section, we describe the training objectives and optimization strategy used to train the proposed method. We first introduce the loss functions for the Adaptive Anchor Sampling and Proximity-based Retargeting modules, each designed to promote geometrically plausible and interaction-aware motion transfer. We then present the alternating optimization strategy adopted to coordinate the training of the two modules.

\subsubsection{Training Adaptive Anchor Sampling} 
The adapted anchors are intended to serve as intermediate variables that are learned to facilitate subsequent proximity-based retargeting. Because the final target pose is determined by proximity constraints aggregated over multiple weighted anchor pairs, we do not enforce a unique optimal anchor layout for every source pose and target character pair. Instead, we define a set of desirable properties for adapted anchors: they should lie on interaction-relevant surface regions, remain reachable under the predicted pose, and maintain a spatially distributed arrangement rather than collapsing into a small region. The following objective terms are designed to encourage these properties:
\begin{equation}
    \begin{aligned}
    \mathcal{L}_{anc} &= \lambda_{simp}\mathcal{L}_{simp} + \lambda_{proj}\mathcal{L}_{proj} 
    \\ &+ \lambda_{reach}\mathcal{L}_{reach} + \lambda_{ord}\mathcal{L}_{ord} + \lambda_{init}\mathcal{L}_{init},   
    \end{aligned}
\end{equation}
where each $\lambda$ denotes the weighting coefficient for the corresponding loss term. While $\mathcal{L}_{simp}$ and $\mathcal{L}_{init}$ are defined on the translated and adapted anchor positions in the rest pose, denoted as $\tilde{\mathbf{A}}^{tgt}_{r}$ and $\hat{\mathbf{A}}^{tgt}_{r}$ respectively, $\mathcal{L}_{reach}$ and $\mathcal{L}_{ord}$ are computed on the deformed adapted anchors $\hat{\mathbf{A}}^{tgt}_{d}$, which are obtained by deforming $\hat{\mathbf{A}}^{tgt}_{r}$ according to the target character pose produced by the Proximity-based Retargeting.

\paragraph{Simplification Loss} To ensure that the adapted anchors of the target character remain close to the original mesh surface and avoid anchor collapse to a small region, we employ a simplification loss \cite{dovrat2019learning}. Given the rest pose vertices $\mathbf{V}^{tgt}_{r}$ and the adapted anchors $\hat{\mathbf{A}}^{tgt}_{r}$ of the target character, this loss measures the average and maximum nearest-neighbor distances between the two sets. For two arbitrary point sets $\mathbf{X}$ and $\mathbf{Y}$, the average and maximum nearest-neighbor distances denoted as $\mathcal{L}_a$ and $\mathcal{L}_m$, are defined as follows: 
\begin{align}
    \mathcal{L}_a(\mathbf{X}, \mathbf{Y}) &= \frac{1}{|\mathbf{X}|} \sum_{\mathbf{x} \in \mathbf{X}} \min_{\mathbf{y} \in \mathbf{Y}} || \mathbf{x} - \mathbf{y} ||^2_2, \\
    \mathcal{L}_m(\mathbf{X}, \mathbf{Y}) &= \max_{\mathbf{x} \in \mathbf{X}} \min_{\mathbf{y} \in \mathbf{Y}} || \mathbf{x} - \mathbf{y} ||^2_2.
\end{align}
The simplification loss is then formulated as:
\begin{equation}
    \begin{aligned}
        \mathcal{L}_{simp}(\hat{\mathbf{A}}^{tgt}_{r}, \mathbf{V}^{tgt}_{r}) & =
        \mathcal{L}_a(\hat{\mathbf{A}}^{tgt}_{r}, \mathbf{V}^{tgt}_{r}) +
        \mathcal{L}_m(\hat{\mathbf{A}}^{tgt}_{r}, \mathbf{V}^{tgt}_{r}) \\ &+
        \mathcal{L}_a(\mathbf{V}^{tgt}_{r}, \hat{\mathbf{A}}^{tgt}_{r}),
    \end{aligned}  
    \label{eq:proj_loss}
\end{equation} 
where the overall formulation captures a bidirectional geometric discrepancy between $\hat{\mathbf{A}}^{tgt}_{r}$ and $\mathbf{V}^{tgt}_{r}$.
The first two terms encourage the adapted anchors to lie close to the mesh vertices by minimizing both the average and maximum distances from anchors to vertices ($\hat{\mathbf{A}}^{tgt}_{r} \rightarrow \mathbf{V}^{tgt}_{r}$). The last term enforces sufficient coverage of the mesh vertices by penalizing vertices that are far from any anchor ($\mathbf{V}^{tgt}_{r} \rightarrow \hat{\mathbf{A}}^{tgt}_{r}$).

\paragraph{Projection Loss} Following \citeN{lang2020samplenet}, we additionally employ a projection loss to regularize the temperature parameter $\tau$ used in the Soft Projection operation. In the Soft Projection, each adapted anchor is represented as a weighted combination of its $k$-nearest vertices. $\tau$ , which is initialized to $1.0$, controls the smoothness of the interpolation: larger values produce smoother averaging over multiple vertices, whereas smaller values yield a sharper weighting that concentrates on the nearest vertex. During training, the projection loss encourages $\tau$ to gradually decrease as follows: 
\begin{equation}
    \mathcal{L}_{proj} = \tau^2. 
\end{equation}
This allows the projection to transition from smooth interpolation to near-discrete matching, which is required at inference. 

\paragraph{Reachability Loss} When the source motion involves self-contact, such as a hand touching the torso, it implicitly indicates that the end-effector is intended to interact with a nearby region on another body part. Because anchors are initially defined in one-to-one correspondence across characters, such interactions can be identified based on the proximity between source anchors and transferred to the target configuration. In this process, target anchors that are expected to interact with an end-effector should lie within a reachable range of the associated limb. To address this, we introduce a reachability loss that penalizes violations of reach constraints, ensuring that target anchors associated with end-effector interactions remain within the maximum extension of the relevant limb.

We consider the set of end-effectors $\mathcal{E}=\{e \mid e \in \{\textit{lh}, \textit{rh}, \textit{lf}, \textit{rf}\}\}$, where \textit{lh}, \textit{rh}, \textit{lf} and \textit{rf} denote the left hand, right hand, left foot, and right foot, respectively. For each end-effector $e$, we associate a corresponding ball joint $b(e)$, defined as the shoulder joint for hands and the pelvis joint for feet. The limb length $\ell(e)$ is defined as the total length of the kinematic chain from $b(e)$ to the end-effector $e$ in the rest pose, representing the maximum straightened reach of the limb. Following \citeN{shin2001computer}, we define a reachability penalty between an end-effector $e$ and an adapted target anchor $j$ as follows:
\begin{equation}
    \mathcal{P}(e, j)= [\text{max}(0, \|{\hat{\textbf{p}}_{b(e)}}-{\hat{\mathbf{A}}_{d, j}} \|_2 - {\ell(e)})]^2,
\end{equation}
where $\hat{\mathbf{p}}_{b(e)} \in \mathbb{R}^3$ denotes the position of the ball joint $b(e)$ on the target character in the retargeted pose, and $\hat{\mathbf{A}}_{d, j} \in \mathbb{R}^3$ is the position of the adapted anchor $j$ after being deformed under that retargeted pose. This term penalizes anchors that lie outside the straightened reach of the end-effector.

Let $\mathcal{N}_e \subset \{1, ..., N_A\}$ be the index set of anchors associated with $e$, and $\mathcal{N}_{limb(e)} \subset \{1, ..., N_A\}$ the index set of anchors located on the same limb as $e$. We first define the set of anchor pairs that contribute to the reachability loss, parameterized by an end-effector $e$ as follows:
\begin{equation}
    \mathcal{R} = \{(e, i, j) \mid e \in \mathcal{E}, i \in \mathcal{N}_e, j \notin \mathcal{N}_{limb(e)} \}.
\end{equation}
That is, for each end-effector $e$, we consider pairs consisting of an end-effector anchor $i$ and a candidate interaction anchor $j$ located outside the corresponding limb, thereby focusing on cross-limb interactions. The reachability loss is then defined as follows:
\begin{equation}
    \mathcal{L}_{reach} = 
            \frac{1}{|\mathcal{R}|}
            \sum_{(e, i, j) \in \mathcal{R}}
            \mathbf{W}^{src}_{dist}(i, j) \ \mathcal{P}(e, j), 
\end{equation}
where $|\mathcal{R}|$ denotes the total number of valid anchor pairs used for normalization, and $\mathbf{W}^{src}_{dist}(i, j)$ is a proximity-based interaction weight computed from the source motion, indicating the likelihood that anchor $j$ participates in an interaction with an end-effector anchor $i$. By measuring the distance from the ball joint to each candidate interaction anchor $j$ and penalizing when $j$ lies beyond the limb length, $\mathcal{L}_{reach}$ encourages the anchor adaptation stage to yield target anchor configurations that are reachable by the corresponding limb while the proximity weights $\mathbf{W}^{src}_{dist}(i, j)$ suppress constraints from distant anchors that are unlikely to contribute to interactions.

\paragraph{Anchor Ordering Loss} While the reachability loss constrains anchors to remain within the feasible kinematic range of the corresponding limb, it does not explicitly regulate the relative spatial arrangement between anchors. As a result, anchors may satisfy reach constraints while drifting to configurations that distort interaction-related spatial structure and potentially induce local penetrations. To explicitly regularize inter-anchor arrangement during optimization, we employ an anchor ordering loss. Following ReConForm \cite{cheynel2025reconform}, we characterize pairwise anchor relationships using a signed distance matrix $\mathbf{D}_{ord} \in \mathbb{R}^{N_A \times N_A}$ with respect to the deformed anchors. Each element $\mathbf{D}_{ord}(i, j)$ represents the signed displacement of anchor $j$ along the surface normal at anchor $i$:
\begin{equation}
    \mathbf{D}_{ord}(i, j) = \mathbf{n}_{d, i} \cdot (\mathbf{A}_{d, j} - \mathbf{A}_{d, i}), 
\end{equation}
where $\mathbf{A}_{d, i}$ and $\mathbf{n}_{d, i}$ denote the position and normal of anchor $i$, respectively. A negative value of $\mathbf{D}_{ord}(i, j)$ indicates that anchor $j$ lies behind the tangent plane at anchor $i$, capturing relative ordering of anchor pairs along the local surface normal direction. 
Rather than explicitly resolving penetration artifacts at the geometric level, these signed values are treated as relative geometric signals that optimize anchors. 

Given the source and target signed distance matrices $\mathbf{D}^{src}_{ord}$ and $\mathbf{D}^{tgt}_{ord}$, we encourage the adapted anchors to preserve source-consistent ordering patterns by minimizing their discrepancy over interacting anchor pairs as follows:
\begin{equation}
    \mathcal{L}_{ord} = 
    \frac{1}{|\mathcal{M}|} \sum_{(i,j)\in \mathcal{M}} 
    \mathbf{W}^{src}_{dist}(i,j) \, 
    \lVert 
        \mathbf{D}^{src}_{ord}(i, j) - \mathbf{D}^{tgt}_{ord}(i, j)
    \rVert^{2},
\end{equation}
where $\mathcal{M}$ denotes the set of valid anchor pairs defined by a body-part mask that excludes anchor pairs belonging to the same body part as illustrated in Figure \ref{fig:initial_anchors}. The weighting matrix $\mathbf{W}^{src}_{dist}$ assigns higher values to spatially close anchor pairs, emphasizing their relationships. Importantly, this loss does not aim to correct penetration artifacts present in the source motion, but instead focuses on preserving relative inter-anchor ordering that characterizes the interaction structure. Although penetration reduction is not an explicit objective, optimizing anchor positions to maintain consistent spatial ordering naturally discourages interpenetration in the target configuration, given the source motion that does not exhibit penetration artifacts.

\paragraph{Anchor Initialization Loss.} To stabilize anchor adaptation and prevent excessive deviation from the initial anchor configuration, we penalize the deviation of the adapted target anchors $\hat{\mathbf{A}}^{tgt}_{r}$ from their original positions $\mathbf{A}^{tgt}_{r}$ using a mean squared error: 
\begin{equation}
    \mathcal{L}_{\text{init}}
    = \frac{1}{N_A}\sum_{i=1}
    \lVert\
        \hat{\mathbf{A}}^{tgt}_{r, i} - \mathbf{A}^{tgt}_{r, i}
    \rVert^2_2.
\end{equation}

\subsubsection{Training Proximity-based Retargeting Module} 
The objective terms to train the Proximity-based Retargeting module are as follows:
\begin{equation}
    \mathcal{L}_{retarget} = \lambda_{rec}\mathcal{L}_{rec} + \lambda_{vel}\mathcal{L}_{vel}+ \lambda_{dist}\mathcal{L}_{dist} + \lambda_{dir}\mathcal{L}_{dir},
\end{equation}
where each $\lambda$ denotes the weighting coefficient for the corresponding loss term.

\paragraph{Reconstruction Loss} 
We adopt a reconstruction loss following \citeN{lee2023same}, which stabilizes motion initialization by measuring discrepancies between the ground-truth motion $D^{tgt}$ and the predicted motion $\hat{D}^{tgt}$:
\begin{equation}
    \begin{aligned}
        \mathcal{L}_{rec} = 
        & \lambda_{q} \lvert| \mathbf{q}^{tgt} - \hat{\mathbf{q}}^{tgt} \rvert|^2_2 + 
        \lambda_{p} \lvert| \mathbf{p}^{tgt} - \hat{\mathbf{p}}^{tgt} \rvert|^2_2 
        \\ &+ \lambda_{r} \lvert| \mathbf{r}^{tgt} - \hat{\mathbf{r}}^{tgt} \rvert|^2_2 + 
        \lambda_{c} \lvert| \mathbf{c}^{tgt} - \hat{\mathbf{c}}^{tgt} \rvert|^{2} , 
    \end{aligned}    
\end{equation}
where $\hat{\mathbf{p}}^{tgt}$ denotes the predicted joint positions in the world coordinate, calculated by forward kinematics from the predicted joint rotations $\hat{\mathbf{q}}_{tgt}$.

\paragraph{Joint Velocity Loss} 
To ensure smooth transitions between adjacent frames, we employ a joint velocity loss: 
\begin{equation}
    \mathcal{L}_{vel} = \lvert| \dot{\mathbf{p}}^{tgt} - \hat{\dot{\mathbf{p}}}^{tgt} \rvert|^2_2,
\end{equation}
where $\dot{\mathbf{p}}^{tgt}$ and $\hat{\dot{\mathbf{p}}}^{tgt}$ denote the ground truth and the predicted joint linear velocities, respectively. By minimizing the discrepancy between the predicted and ground-truth joint velocities, which inherently exhibit temporal smoothness, the model encourages the retargeted motion to maintain smooth temporal transitions.

\paragraph{Anchor Distance Loss} 
We encourage the target character to follow the spatial configuration of the source motion by applying an anchor distance loss that preserves inter-anchor distances observed in the source. Let $\mathbf{A}^{src}_{d}$ denote the source anchors deformed by the source skeletal pose $D^{src}$, and $\hat{\mathbf{A}}^{tgt}_{d}$ denote the adapted target anchors deformed by the retargeted pose $\hat{D}^{tgt}$. From these anchor positions, we compute the pairwise Euclidean distance matrices $\mathbf{D}^{src}_{dist}$ and $\hat{\mathbf{D}}^{tgt}_{dist}$. The loss is formulated as follows: 
\begin{equation}
    \mathcal{L}_{dist} = 
    \frac{1}{|\mathcal{M}|} \sum_{(i,j)\in \mathcal{M}} 
    \mathbf{W}^{src}_{dist}(i,j) \, 
    \| \mathbf{D}^{src}_{dist}(i,j) - \hat{\mathbf{D}}^{tgt}_{dist}(i,j) \|^2,
    \label{eq:anchor_dist_loss}
\end{equation} 
By minimizing $\mathcal{L}_{dist}$, the model encourages the target anchors to maintain the relative spatial layout observed in the source motion, which is essential for reproducing interactions such as self-contacts or near-body proximities.

\paragraph{Anchor Direction Loss}
While the anchor distance loss enforces local spatial consistency, it does not account for directional relationships between anchors. Because different directional configurations can yield similar pairwise distances, relying solely on distance regularization may lead to interpenetration between body parts. To address this, we employ an anchor direction loss that encourages alignment between the normalized displacement vectors of source and target anchor pairs:
\small \begin{equation}
    \mathcal{L}_{\text{dir}} = \frac{1}{|\mathcal{M}|} \sum_{(i,j)\in \mathcal{M}} 
    \mathbf{W}^{src}_{dist}(i,j) \left( 1 - 
    \frac{\mathbf{D}^{src}_{dir}(i,j)}{\|\mathbf{D}^{src}_{dir}(i,j)\|} \cdot
    \frac{\hat{\mathbf{D}}^{tgt}_{dir}(i,j)}{\|\hat{\mathbf{D}}^{tgt}_{dir}(i,j)\|}
    \right)
\end{equation} \normalsize
where $\mathbf{D}^{src}_{dir}(i, j)$ and $\hat{\mathbf{D}}^{tgt}_{dir}(i, j)$ denote the relative displacement vectors between the $i$-th and $j$-th anchors, computed from the initial anchors of the source character and the adapted anchors of the target character, respectively. The inner product measures the cosine similarity between the two vectors, encouraging alignment in direction. By minimizing $\mathcal{L}_{dir}$, the model encourages directional alignment of corresponding anchor pairs, thereby preserving the relative orientations of body parts and preventing interpenetration artifacts that may arise from distance-based constraints alone.
\subsubsection{Alternating Optimization Strategy}
Our method involves two optimization variables: the adapted anchors $\hat{\mathbf{A}}^{tgt}_{r}$ and the target pose $\hat{D}^{tgt}$. Because the proximity objectives, such as $\mathcal{L}_{dist}$ and $\mathcal{L}_{dir}$, are evaluated on anchors deformed by the final pose, their errors can be reduced either by relocating anchors on the target mesh or by adjusting the target pose. Under joint training, pose updates provide a more immediate means of reducing the error, while anchor adaptation is additionally constrained by a regularization term such as $\mathcal{L}_{init}$. As a result, the anchors tend to remain close to their initial positions, weakening the intended role of the Adaptive Anchor Sampling module and often leading to distorted postures. 

To mitigate this ambiguity, we adopt a fixed two-step alternating optimization strategy. The Adaptive Anchor Sampling module is first updated to adjust anchor positions based on the downstream retargeting result. The Proximity-based Motion Retargeting module is then updated using the resulting adapted anchors. During each step, all loss terms including $\mathcal{L}_{anc}$ and $\mathcal{L}_{retarget}$ remain active, but only the parameters of the selected module are updated. This decoupled training allows each module to focus on its respective objective while being progressively refined. In particular, it enables the Adaptive Anchor Sampling module to first adapt anchors toward reachable regions without being overridden by pose changes, generating anchor configurations that better support subsequent retargeting. Simultaneously, it allows the Proximity-based Retargeting module to more effectively transfer motion conditioned on the given anchors. Consequently, our method effectively preserves self-contact and near-body interactions across diverse body shapes, while avoiding implausible postures. For more details on adopting the alternating optimization strategy, please refer to Section 2 of the supplementary document.
\section{Experiments}
This section provides an evaluation of our method. We first describe the dataset and implementation details, followed by quantitative and qualitative comparisons with state-of-the-art methods. We then present a series of ablation studies to validate the impact of each component in our framework.

\paragraph{Dataset}
To construct our dataset, we randomly selected 8 characters for training and 9 characters for testing from Mixamo \cite{Mixamo}. Among the test characters, 5 overlapped with the training set, while the remaining 4 characters were unseen during training. For each training character, we collected 276 non-overlapping motion sequences of varying lengths, resulting in a total of 2,208 sequences sampled at 30 fps and comprising 161,448 frames. From each sequence, 8 consecutive frames were randomly sampled to construct the training input. An additional set of 275 motion sequences with varying lengths was collected for evaluation. Following \citeN{villegas2018neural}, we organized the test set into four splits based on whether the source motion and target character were seen during training: seen character with seen motion (SC+SM), seen character with unseen motion (SC+UM), unseen character with seen motion (UC+SM), and unseen character with unseen motion (UC+UM). For more details on the dataset preparation used in both training and evaluation, please refer to Section 3.1 of the supplementary document.

\paragraph{Implementation Details}
Our method was implemented in PyTorch and trained on a single NVIDIA RTX A6000 GPU with 48GB of VRAM. We used the Adam optimizer \cite{kingma2014adam} with a fixed learning rate of $10^{-3}$. Our model was trained for 200 epochs with a batch size of 16, which required approximately 48 hours of training. The hyperparameters used to train the Adaptive Anchor Sampling Module were set to $\lambda_{simp}=0.01$, $\lambda_{proj}=0.01$, $\lambda_{reach}=1000.0$, $\lambda_{ord}=1.0$, $\lambda_{init}=1.0$, and $k=10$. For the Proximity-based Retargeting module, the loss weights were set to $\lambda_{rec}=1.0$, $\lambda_{q}=15.0$, $\lambda_{p}=0.01$, $\lambda_{r}=10.0$, $\lambda_{c}=1.0$, $\lambda_{vel}=1.0$, $\lambda_{dist}=1.0$, and $\lambda_{dir}=1500.0$. For details on the design of the loss weights, please refer to Section 3.2 of the supplementary document.
\subsection{Comparisons with State-of-the-art Methods}
\paragraph{Baselines} 
We compared our method with existing motion retargeting approaches. As a reference, we included results obtained by MotionBuilder \cite{MotionBuilder}, an off-the-shelf retargeting tool, using its default solver. While MotionBuilder provides advanced control options, such as auxiliary effectors and reach targets, that can be manually configured to better accommodate target geometry, these features are primarily intended for post-processing. Because they are not designed to automatically infer contact semantics from the source motion or determine corresponding reach targets on the target character, we restrict the comparison to its automated retargeting pipeline. We further included results obtained using several learning-based approaches: SAME \cite{lee2023same}, R$^2$ET \cite{zhang2023skinned}, and MeshRet \cite{ye2024skinned}. SAME is a skeleton-aware method that operates on joint-based representations, while R$^2$ET and MeshRet incorporate character geometries to improve the plausibility of retargeted motions. MeshRet is most closely related to our method in that it performs proximity matching between source and target characters, but differs in relying on a predefined static set of anchors. 

\paragraph{Metric}
To evaluate the effectiveness of our method in preserving interaction semantics as well as ensuring geometric plausibility, we employed two metrics: penetration rate and contact preservation. Penetration rate (Pen) quantifies how well the retargeted motion mitigates interpenetration artifacts, defined as the percentage of limb vertices that intersect with other body parts, including other limbs. For evaluating contact preservation with respect to the source motion, we adopted the evaluation scheme proposed by~\citeN{jang2024geometry}, which measures precision (Prec), recall (Rec), and accuracy (Acc), computed from frame-wise true/false positives and negatives. For detailed definitions, please refer to Section 3.3 of the supplementary document. Because our goal is to generate geometrically plausible motion of the target character while preserving the interaction semantics induced by the source, we do not evaluate kinematic fidelity, such as joint positional errors, which may not reflect perceptual or geometric plausibility, especially when minor deviations are necessary to avoid collisions or to maintain self-contact.

\paragraph{Quantitative Evaluation} 
\begin{table}
    \caption{Quantitative comparison between our method and baselines. The best result for each metric is highlighted in bold.}
    \centering
    \setlength{\tabcolsep}{14pt}
    \resizebox{\columnwidth}{!}{%
    \begin{tabular}{l|c|cll}
    \hline
    \multicolumn{1}{c|}
    {\multirow{2}{*}{Methods}} 
    & \multirow{2}{*}{\begin{tabular}[c]{@{}c@{}}Pen (\%) ↓ \end{tabular}} 
    & \multicolumn{3}{c}{Contact Preservation}\\ \cline{3-5} 

    \multicolumn{1}{c|}{} 
    & & Prec ↑ & Rec ↑ & Acc ↑ \\ \hline 
    MotionBuilder      
    & 18.647 & 0.254 & 0.311& 0.927 \\
    SAME      
    & 17.409 & 0.263 & 0.289& 0.931 \\
    R2ET      
    & 14.926 & 0.250 & 0.195 & 0.937 \\
    MeshRet   
    & 18.783 & 0.295 & 0.282 & 0.937 \\ 
    Ours   
    & \textbf{14.730} & \textbf{0.415} &\textbf{0.350} & \textbf{0.948}
    \\ \hline
    \end{tabular}%
    }
    \label{tab:quan_results}
\end{table}
As shown in Table \ref{tab:quan_results}, our method outperformed all baselines, achieving the lowest penetration rate, as well as the highest precision, recall, and accuracy in contact preservation. MotionBuilder and SAME exhibited relatively high penetration rates and lower contact preservation scores. R$^2$ET achieved the second-best performance in penetration avoidance; however, its limited consideration of interaction-related spatial structures led to low contact preservation scores. In contrast, MeshRet better preserved contact by leveraging proximity matching through static correspondences, but its relatively high penetration rate suggests that such static mappings may be insufficient to handle geometric variation across characters.

\paragraph{User Study}
\begin{table} [!t]
    \caption{Results of user study. The best result for each metric is highlighted in bold.}
    \setlength{\tabcolsep}{13pt}
    \resizebox{\columnwidth}{!}{%
    \begin{tabular}{l|ccc}
    \hline
    \multicolumn{1}{c|}{Methods}      
    & \begin{tabular}[c]{@{}c@{}}Semantic \\ Preservation ↑\end{tabular} 
    & \begin{tabular}[c]{@{}c@{}}Artifact \\ Avoidance ↑\end{tabular} \
    & \begin{tabular}[c]{@{}c@{}}Overall  \\ Quality ↑\end{tabular} \\ \hline
    MotionBuilder & 3.342 & 2.388 & 2.625 \\
    SAME          & 2.875 &	2.454 &	2.413 \\
    R$^2$ET       & 3.438 &	3.263 &	3.079 \\
    MeshRet       & 2.300 &	1.721 &	1.683 \\ 
    Ours          & \textbf{4.679} & \textbf{4.646} & \textbf{4.592} \\ \hline
    \end{tabular}%
    } \label{tab:user_study}
    \vspace{-1.0em}
\end{table}
We conducted a user study to assess the visual quality of retargeted motions produced by our method compared to existing retargeting approaches. We constructed 50 evaluation tasks by applying each method to 10 motion sequences sampled from the test set, with each task comprising a 4–5 second animation clip. The resulting motions were presented in a randomized order to mitigate potential bias. After watching each clip, participants were asked to evaluate the following three criteria: Semantic Preservation (whether the retargeted motion conveys the intent of the source motion, including intended contact), Artifact Avoidance (absence of visual artifacts such as interpenetration), and Overall Quality (overall naturalness and plausibility of the motion). All questions were rated on a 5-point Likert scale, ranging from 1 (Strongly disagree) to 5 (Strongly agree). We recruited 24 participants (13 males and 11 females, ages ranging from 23 to 35). As summarized in Table \ref{tab:user_study}, our method achieved the highest average scores across all three criteria, with 4.679 in Semantic Preservation, 4.646 in Artifact Avoidance, and 4.592 in Overall Quality, demonstrating its effectiveness in producing visually plausible and semantically faithful retargeted motions.

\paragraph{Qualitative Evaluation} 
\begin{figure*} 
    \centering \includegraphics[width=\textwidth]{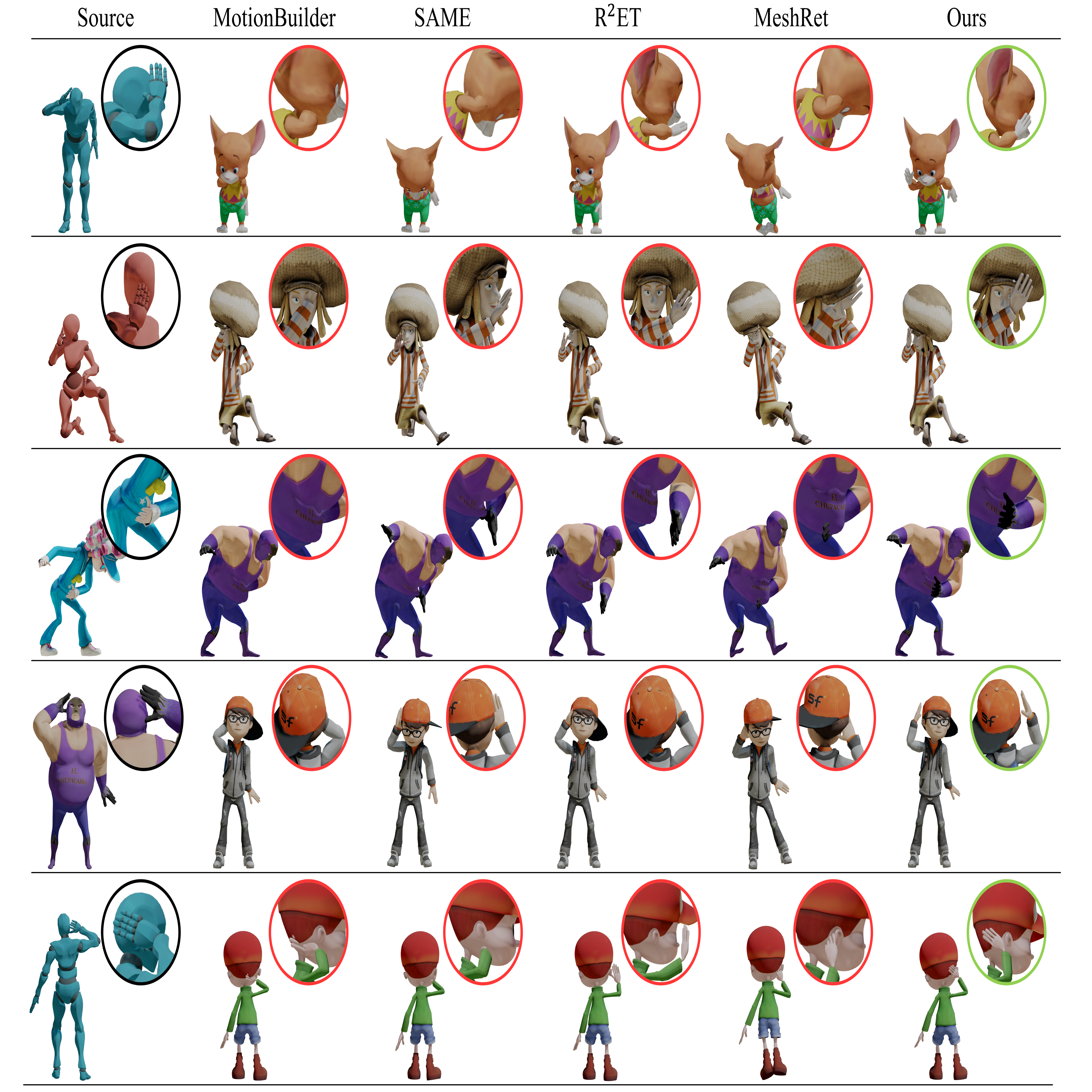}
    \caption{Qualitative comparison with baselines on skinned motion retargeting. Each row presents a case where the source character performs a motion involving self-contact, while the corresponding body part of the target character exhibits enlarged proportions. Results enclosed in circles provide a closer view of the corresponding outputs. Red highlights indicate cases with penetration artifacts or a failure to preserve the intent of the original motion, whereas green highlights indicate successful preservation of both semantic intent and geometric plausibility.}
    \label{fig:qual_baselines}
\end{figure*}
Figure \ref{fig:qual_baselines} presents qualitative comparisons of skinned motion retargeting involving self-contacts or near-body interactions across characters with exaggerated body proportions, such as enlarged heads or bulky torsos. MotionBuilder and SAME, which do not account for character geometry, frequently produced penetration artifacts or failed to preserve contact present in the source motion. R$^2$ET effectively mitigated penetration artifacts, but its correction primarily enforced geometric validity on the target character without considering semantic context of the source motion, yielding target motion often deviated from the intent of the original motion. This aligns with the observation in the quantitative comparison, highlighting that incorporating source-driven spatial relationships in addition to geometric consistency is essential for semantically coherent skinned motion retargeting.
MeshRet, which relies on proximity matching using predefined static correspondences, generally maintained end-effectors in regions consistent with the source motion. However, when the target character exhibited enlarged or bulky body parts, interactions intended by the source were mapped to unreachable regions on the target mesh, resulting in penetration artifacts. 
In contrast, our method not only effectively mitigated penetration but also faithfully preserved contact semantics, demonstrating the effectiveness of spatially adaptive anchors that account for both the source motion and the geometric feasibility of the target character in supporting the intended interactions. For animation results, please see the supplementary video.
\subsection{Ablation Studies}
This section presents a comprehensive evaluation to validate the design choices of our method, analyzing the contribution of each component in enhancing contact preservation as well as penetration avoidance of retargeted motions. The evaluation is organized into three parts: (1) ablations on the Adaptive Anchor Sampling module, (2) ablations on the Proximity-based Retargeting module, and (3) the effectiveness of the alternating optimization strategy compared to joint training. Each experiment includes quantitative and qualitative evaluations to highlight its contribution to overall performance. 

\subsubsection{Adaptive Anchor Sampling Module}
To validate the design choices of the Adaptive Anchor Sampling module, we conducted a series of ablation studies. First, we assessed the effectiveness of the spatially adaptive anchors by comparing our full model with a variant that uses only the initial target anchors without anchor adaptation. Second, we evaluated the contribution of each loss term used to train the Adaptive Anchor Sampling module. Third, we examined the architectural components of the Anchor Residual Prediction module, including the relative positional bias and attention operations. Finally, we investigated the effect of the hyperparameter $k$, which determines the number of neighboring vertices used in the Soft Projection operation.

\begin{figure*} 
    \centering
    \begin{minipage}{0.12\textwidth}
        \centering
        \includegraphics[width=\textwidth]{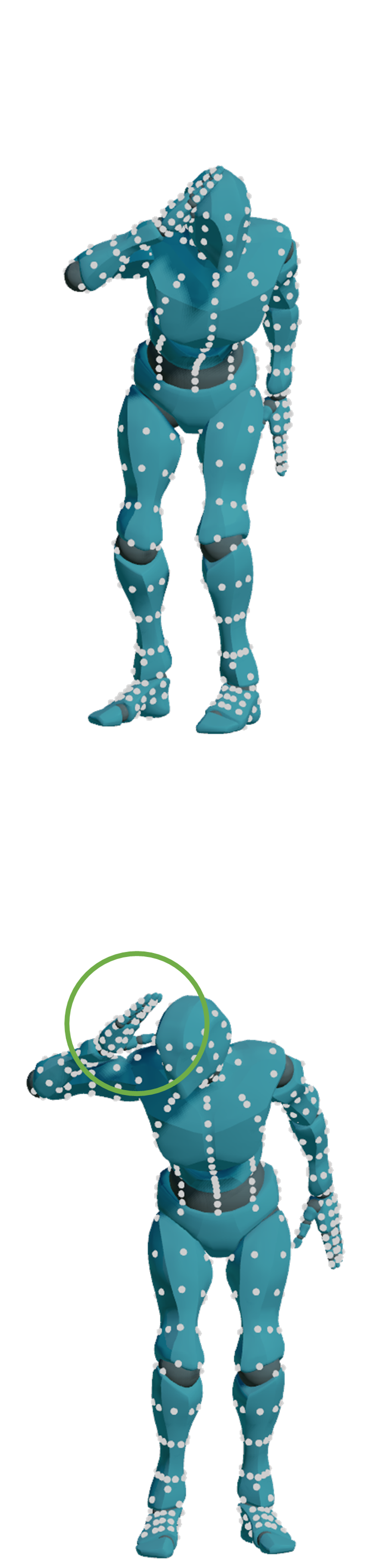}
        \caption*{\textit{Source Pose}} \label{fig:ablation_k_source}
    \end{minipage}
    \begin{minipage}{0.12\textwidth}
        \centering
        \includegraphics[width=\textwidth]{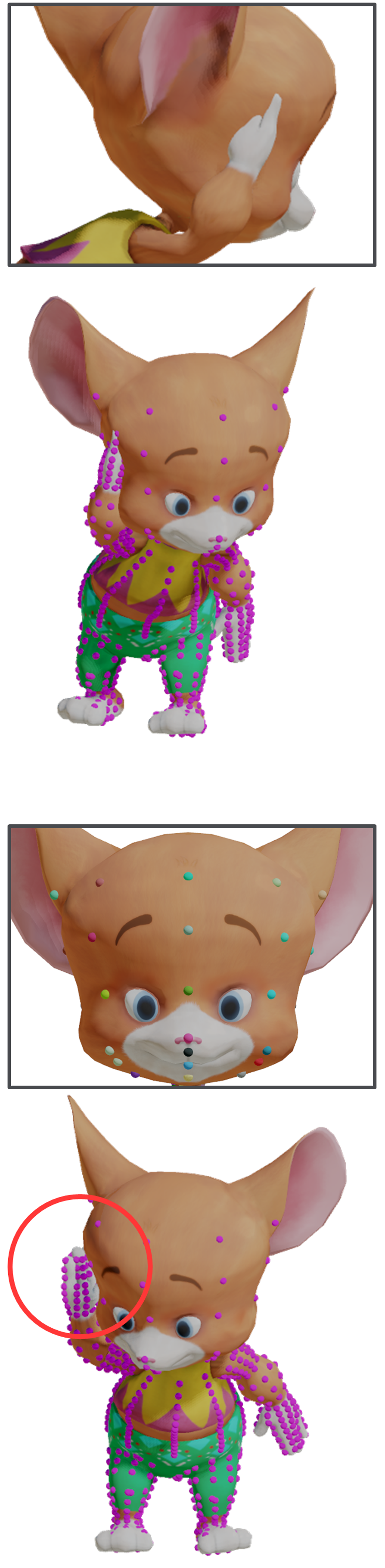}
        \caption*{(a)} \label{fig:ablation_aas_loss_static}
    \end{minipage}
    \begin{minipage}{0.12\textwidth}
        \centering 
        \includegraphics[width=\textwidth]{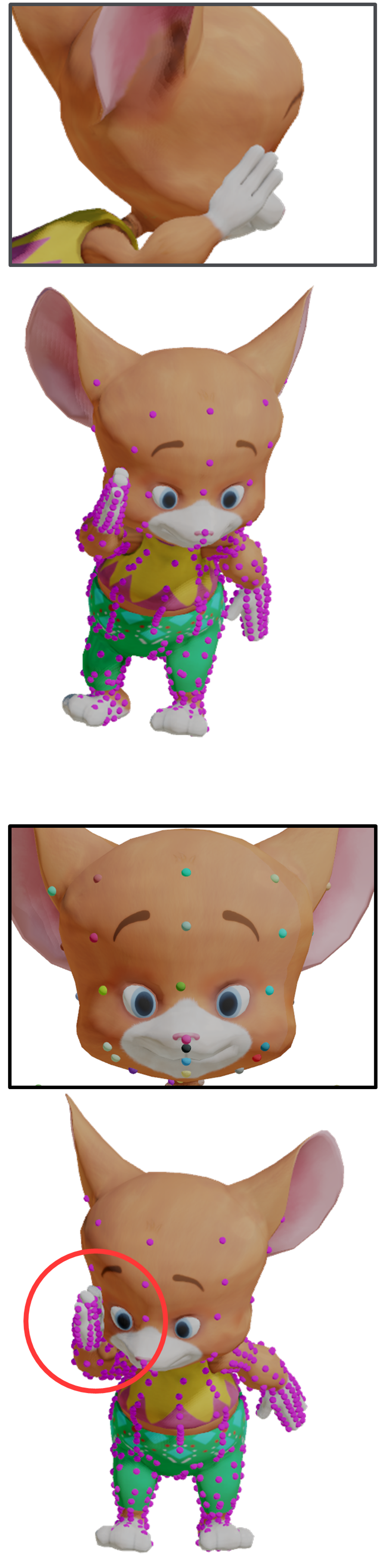}
        \caption*{(b)} \label{fig:ablation_k_5}
    \end{minipage}
    \begin{minipage}{0.12\textwidth}
        \centering 
        \includegraphics[width=\textwidth]{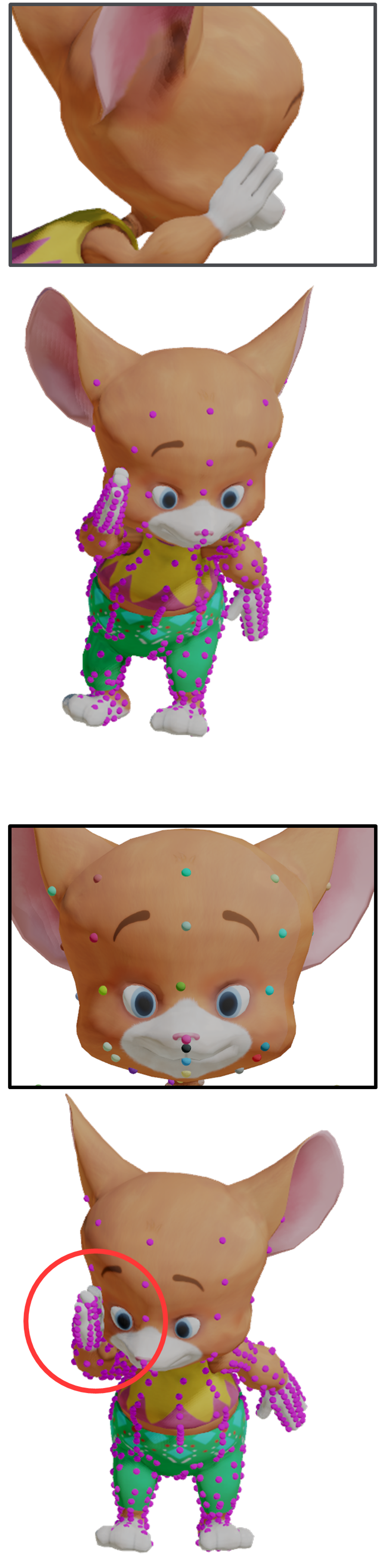}
        \caption*{(c)} \label{fig:ablation_aas_loss_woproj}
    \end{minipage}
    \begin{minipage}{0.12\textwidth}
        \centering 
        \includegraphics[width=\textwidth]{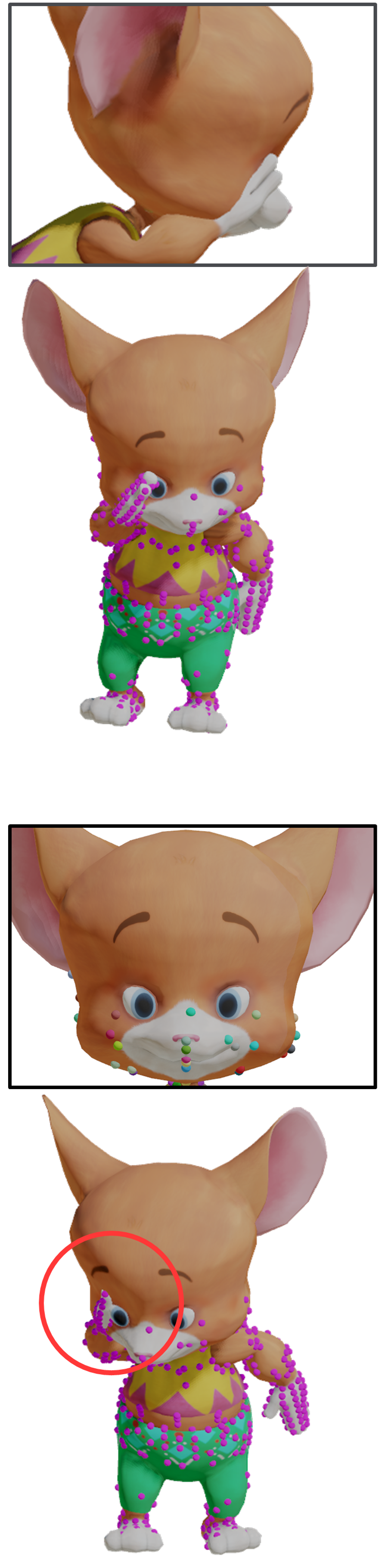}
        \caption*{(d)} \label{fig:ablation_aas_loss_woinit}
    \end{minipage}
    \begin{minipage}{0.12\textwidth}
        \centering 
        \includegraphics[width=\textwidth]{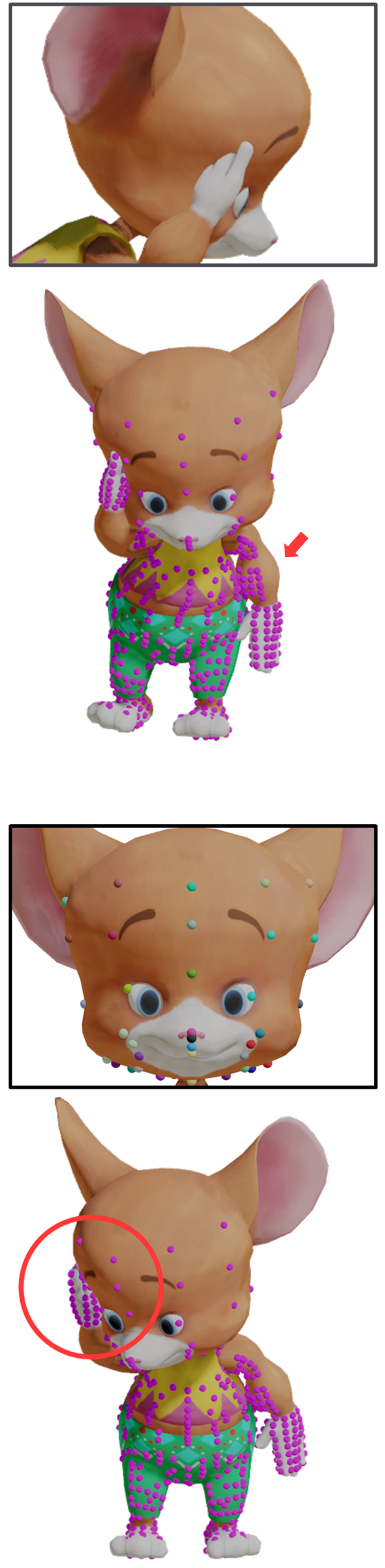}
        \caption*{(e)} \label{fig:ablation_aas_loss_woord}
    \end{minipage}
    \begin{minipage}{0.12\textwidth}
        \centering 
        \includegraphics[width=\textwidth]{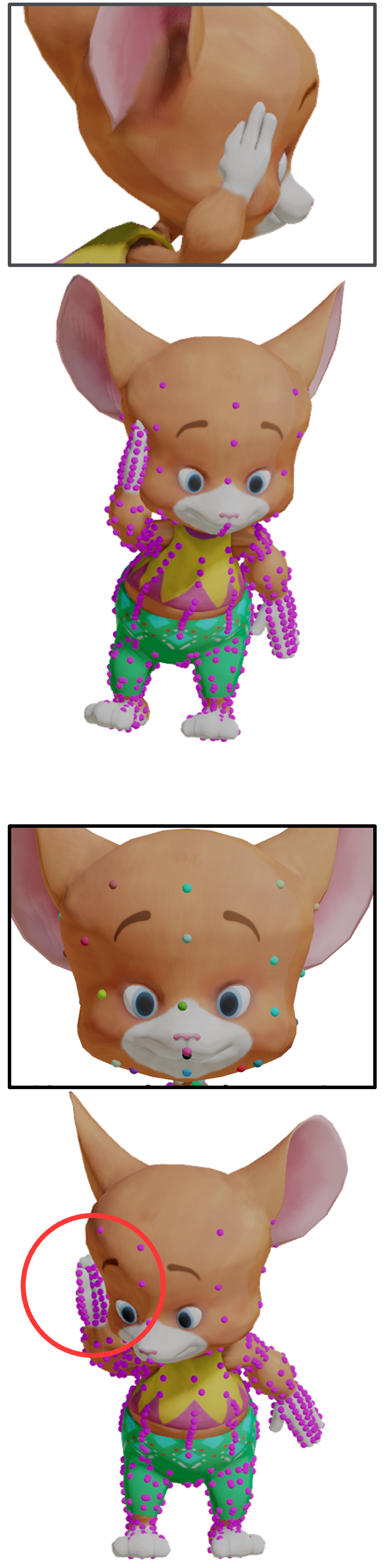}
        \caption*{(f)} \label{fig:ablation_aas_loss_woreach}
    \end{minipage}
    \begin{minipage}{0.12\textwidth}
        \centering 
        \includegraphics[width=\textwidth]{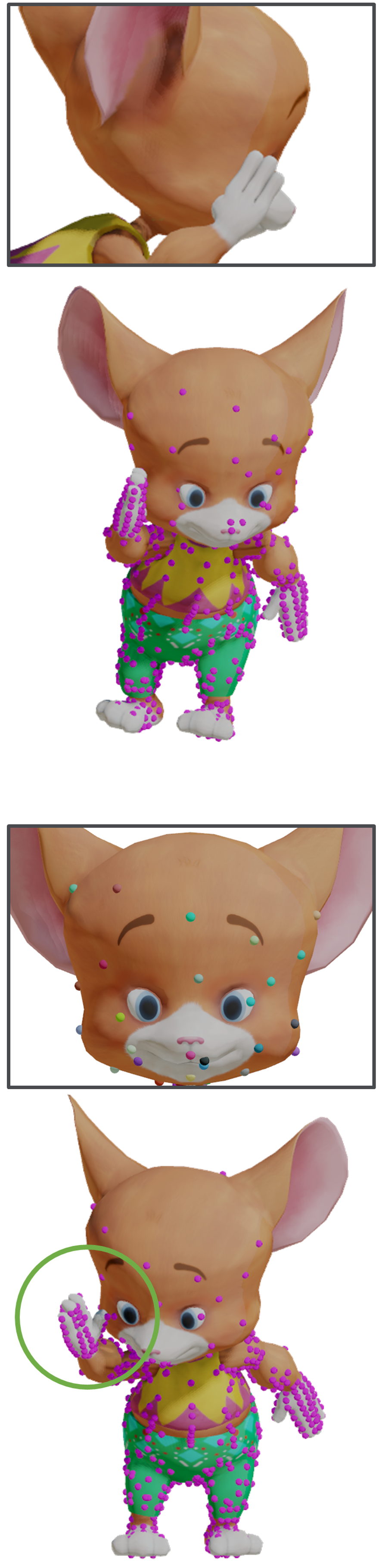}
        \caption*{(g)} \label{fig:ablation_aas_loss_ours}
    \end{minipage}
    \caption{Qualitative comparison of the ablation study on the Adaptive Anchor Sampling module. The leftmost figure shows the source pose with its initial anchors visualized in white. Each subfigure (a)–(g) shows retargeted results produced by the corresponding variants listed in Table~\ref{tab:ablation_aas_loss}, with adapted anchors visualized in pink. Results enclosed in boxes depict the corresponding results of a closer view.}
    \label{fig:ablation_loss_aas}
\end{figure*}
\begin{table}[]
    \caption{Quantitative results of the ablation study on the use of adaptive anchors and loss terms to train the Adaptive Anchor Sampling module. The best result for each metric is highlighted in bold.}
    \resizebox{\columnwidth}{!}{
    \begin{tabular}{l|c|ccc}
    \hline
    \multicolumn{1}{c|}
    {\multirow{2}{*}{Methods}} 
    & \multirow{2}{*}{\begin{tabular}[c]{@{}c@{}}Pen (\%) ↓\end{tabular}} 
    & \multicolumn{3}{c}{Contact Preservation} \\ \cline{3-5} 

    & & Prec ↑ & Rec ↑ & Acc ↑ \\ \hline
    (a) \ w/o Adaptive Anchor Sampling 
    & 16.080  & 0.383   & \textbf{0.392} & 0.944 \\ \hline
    (b) \ w/o $\mathcal{L}_{simp}$ 
    & 15.573 & 0.371 & 0.358 & 0.944 \\
    (c) \ w/o $\mathcal{L}_{proj}$ 
    & 16.061 & 0.368 & 0.378 & 0.943 \\
    (d) \ w/o $\mathcal{L}_{init}$ 
    & 17.160 & 0.319 & 0.327 & 0.938 \\
    (e) \ w/o $\mathcal{L}_{ord}$ 
    & 15.926 & 0.347 & 0.349 & 0.941 \\
    (f) \ w/o $\mathcal{L}_{reach}$ 
    & 15.707 & 0.371 & 0.351 & 0.944 \\ \hline
    (g) Ours 
    & \textbf{14.730} & \textbf{0.415} & 0.350 & \textbf{0.948} \\ \hline
    \end{tabular}
    }\label{tab:ablation_aas_loss}
\end{table}

\paragraph{The use of adaptive anchors} 
We evaluate the effectiveness of the spatially adaptive anchors on the quality of retargeted motion, by comparing our full model against a variant that uses predefined static anchors without anchor adaptation. The results are shown in Table \ref{tab:ablation_aas_loss}. The variant (a) corresponds to a setting where only the Proximity-based Retargeting module is trained, and the initial target anchors remain fixed. This variant exhibited higher penetration rates and lower performance in contact preservation compared to our model (g), which incorporates spatially adaptive anchors. Although variant (a) achieved higher recall, the increased penetration rate suggests that many of the detected contacts arose from near-surface configurations immediately preceding interpenetration, rather than from physically valid interactions.

Qualitative results in Figure~\ref{fig:ablation_loss_aas} further support these findings by illustrating a case where the source motion brings the right hand toward the upper region of the head. The variant (a) attempted to reproduce this motion by guiding the target character's hand to the corresponding static anchors located near the top of the head. However, due to the limited reach of the target character's limb, these anchors became unreachable, resulting in noticeable penetration artifacts. In contrast, our method (g), which leveraged adaptive anchors that are dynamically repositioned based on both source motion and target geometry, guided the hand to a reachable location that avoided penetration while faithfully preserving the intent of the original motion. These results collectively highlight the importance of anchor adaptation in preserving interaction semantics and ensuring the geometric plausibility of the retargeted motion.

\paragraph{Loss terms} 
To evaluate the contribution of each loss term, we conducted an ablation study by removing individual components from the full objective. As shown in (b) to (g) of Table \ref{tab:ablation_aas_loss}, the model trained with full loss terms (g) outperformed all ablated variants, achieving the lowest penetration rate along with the highest contact precision and accuracy. The moderate recall observed in (g) likely stems from the same rationale discussed for variant (a): our model tends to suppress near-surface contacts that precede penetration, thereby favoring physically valid interactions over recall.

\begin{figure} [!t]
    \centering \includegraphics[width=\columnwidth]{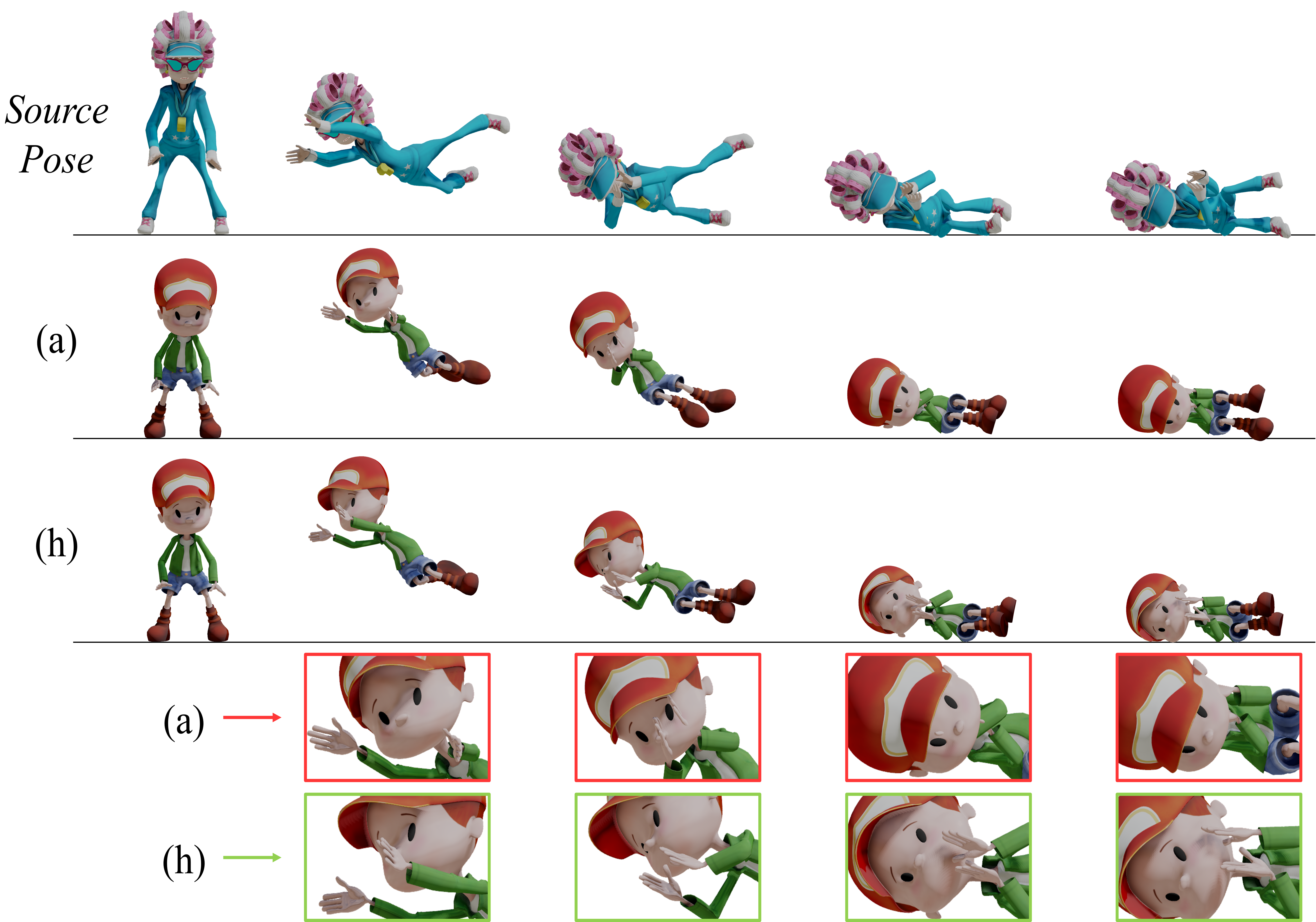}
    \caption{Qualitative results of the ablation study on the use of $\mathbf{B}$. Results enclosed in boxes depict the corresponding results of a closer view. Subfigures (a) and (h) correspond to retargeted poses produced by the variants described in Table~\ref{tab:ablation_aas_alter_design}.}
    \label{fig:qual_ablation_b}
\end{figure}

This observation aligns with the qualitative comparisons in Figure \ref{fig:ablation_loss_aas}, which illustrates how the complete loss objective enables the model to predict structurally coherent anchor configurations that effectively guide the subsequent retargeting. 
Removing the anchor initialization loss $\mathcal{L}_{init}$ (d) led to anchor drifting toward geometrically irrelevant regions, as the model lacked guidance to preserve coarse correspondences with the source. Such drift may produce suboptimal performance compared to variant (a), in which the anchors remain fixed. This suggests that $\mathcal{L}_{init}$ serves to preserve the stable initialization provided by the initial correspondences while still allowing necessary adaptation. Without the anchor ordering loss $\mathcal{L}_{ord}$ (e), anchors tended to collapse onto a single side of each body part, disrupting their spatial distribution across the surface as indicated by the red arrow. Both cases resulted in significant deviations from the source motion, ultimately degrading the quality of the retargeted results. Removing the simplification $\mathcal{L}_{simp}$ (b), projection $\mathcal{L}_{proj}$ (c), and reachability $\mathcal{L}_{reach}$ (f) losses prevented effective anchor adaptation. As shown in the frontal views highlighted in the boxed regions, the anchors remained close to their initial configuration, which can be identified by comparing anchors of the same color with those in variant (a), demonstrating limited spatial adjustment. With limited adaptation, the target anchors failed to retain the near‑contact distances observed in the source motion, leading to unintended contact and minor penetration. In contrast, the anchors produced by the full objective (g) helped generate target motion that preserves the close-proximity relationships present in the source motion, even under substantial geometric discrepancies between characters.
\paragraph{Alternative architectures}
We conducted an ablation study to evaluate the design choices of the Anchor Residual Prediction module by comparing our full model against four variants, with the quantitative results summarized in Table 4. First, to assess the impact of the relative positional bias $\textbf{B}$, which serves as an additional input to the encoder and captures pairwise geometric relationships among deformed source anchors, we trained a variant without it, denoted as (a). In addition, because the decoder consists of three attention operations across different modalities, we trained three variants by independently removing each operation, denoted as (b) to (d). As shown in (a) to (d) and (h) of Table \ref{tab:ablation_aas_alter_design}, our model (h) consistently achieved the lowest penetration rate, as well as the highest precision and accuracy in contact preservation. The effect of incorporating the positional bias $\textbf{B}$ is further illustrated in Figure \ref{fig:qual_ablation_b}. The model trained without $\textbf{B}$ often struggled to preserve the directional relationships inherent in the source motion, leading to implausible retargeted poses with noticeable interpenetration. These results demonstrate the importance of both the geometric encoding in the encoder and the multi-modal attention structure in the decoder for achieving robust and reliable anchor adaptation.

\begin{table} [!t]
    \caption{Quantitative results of the ablation study on the alternative architectures for the Adaptive Anchor Sampling module. The best result for each metric is highlighted in bold.}
    \centering
    \setlength{\tabcolsep}{12pt}
    \resizebox{\columnwidth}{!}{
    \begin{tabular}{l|c|cll}
    \hline
    \multicolumn{1}{c|}{\multirow{2}{*}{Methods}} 
    & \multirow{2}{*}{\begin{tabular}[c]{@{}c@{}}Pen (\%) ↓\end{tabular}} 
    & \multicolumn{3}{c}{Contact Preservation}                                                            
    \\ \cline{3-5} 
    \multicolumn{1}{c|}{}                  
    & & Prec ↑ & Rec ↑ & Acc ↑ \\ \hline
    (a) \ w/o $\mathbf{B}$         
    & 15.725 & 0.395 & 0.364 & 0.946 \\ 
    (b) \ w/o Self-Attn            
    & 15.376 & 0.374 & 0.325 & 0.945 \\ 
    (c) \ w/o 1st Cross-Attn       
    & 14.925 & 0.376 & 0.345 & 0.944 \\ 
    (d) \ w/o 2nd Cross-Attn       
    & 15.553 & 0.385 & 0.368 & 0.945 \\ \hline
    (e) \ $k=1$                    
    & 16.471 & 0.390 & \textbf{0.404} & 0.945 \\
    (f) \ $k=5$                    
    & 15.830 & 0.370 & 0.348 & 0.944 \\ 
    (g) \ $k=20$                   
    & 15.908 & 0.370 & 0.362 & 0.943 \\ \hline
    (h) \ Ours                  
    & \textbf{14.730} & \textbf{0.415} & 0.350 & \textbf{0.948} \\ \hline
    \end{tabular}%
    } 
    \label{tab:ablation_aas_alter_design}
\end{table}
\begin{figure} 
    \centering
    \begin{minipage}{0.18\columnwidth}
        \centering
        \includegraphics[width=\columnwidth]{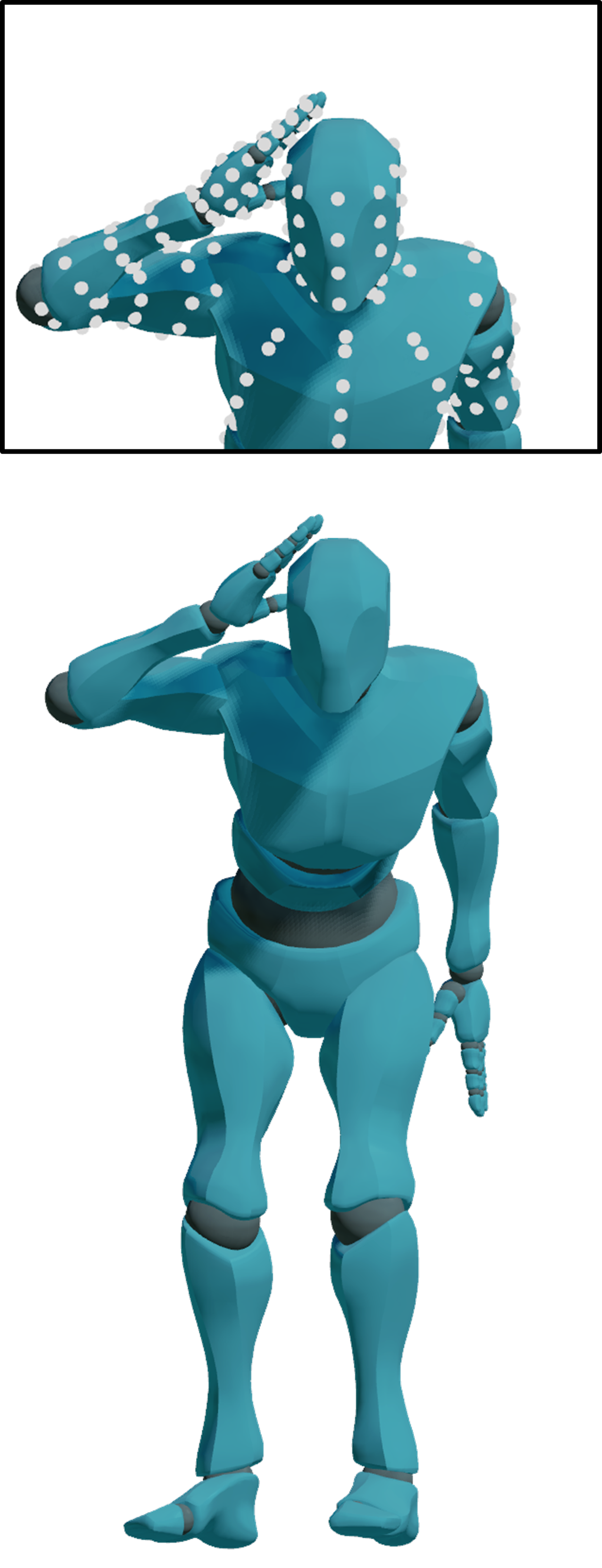}
        \caption*{\textit{Source Pose}}        
        \label{fig:ablation_k_source}
    \end{minipage}
    \begin{minipage}{0.18\columnwidth}
        \centering
        \includegraphics[width=\columnwidth]{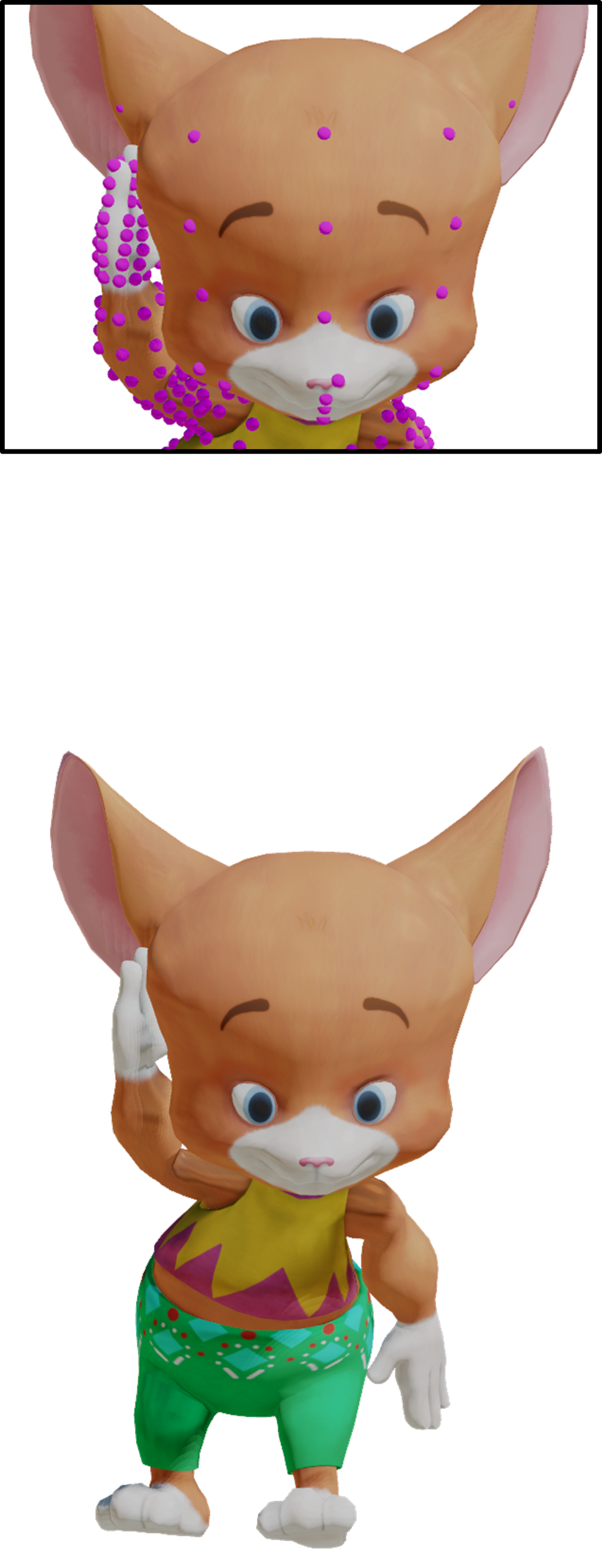}
        \caption*{(e)} 
        \label{fig:ablation_k_1}
    \end{minipage}
    \begin{minipage}{0.18\columnwidth}
        \centering 
        \includegraphics[width=\columnwidth]{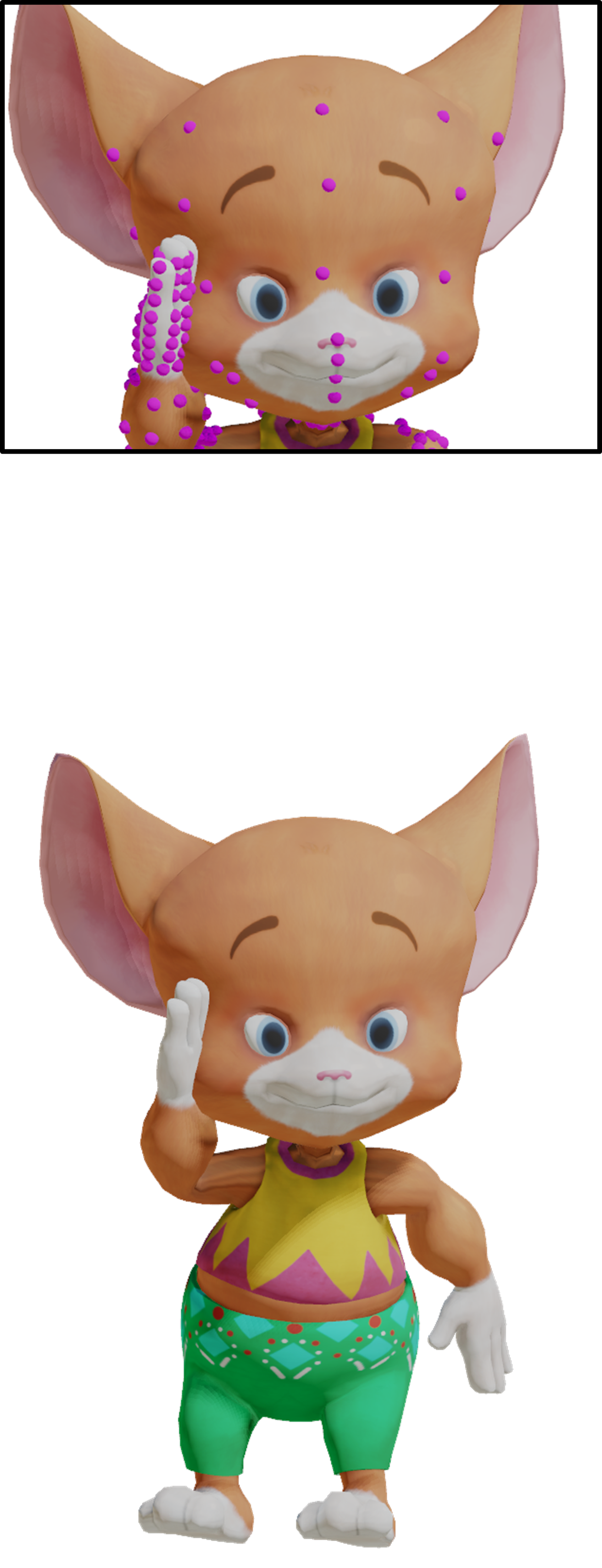}
        \caption*{(f)} 
        \label{fig:ablation_k_5}
    \end{minipage}
    \begin{minipage}{0.18\columnwidth}
        \centering 
        \includegraphics[width=\columnwidth]{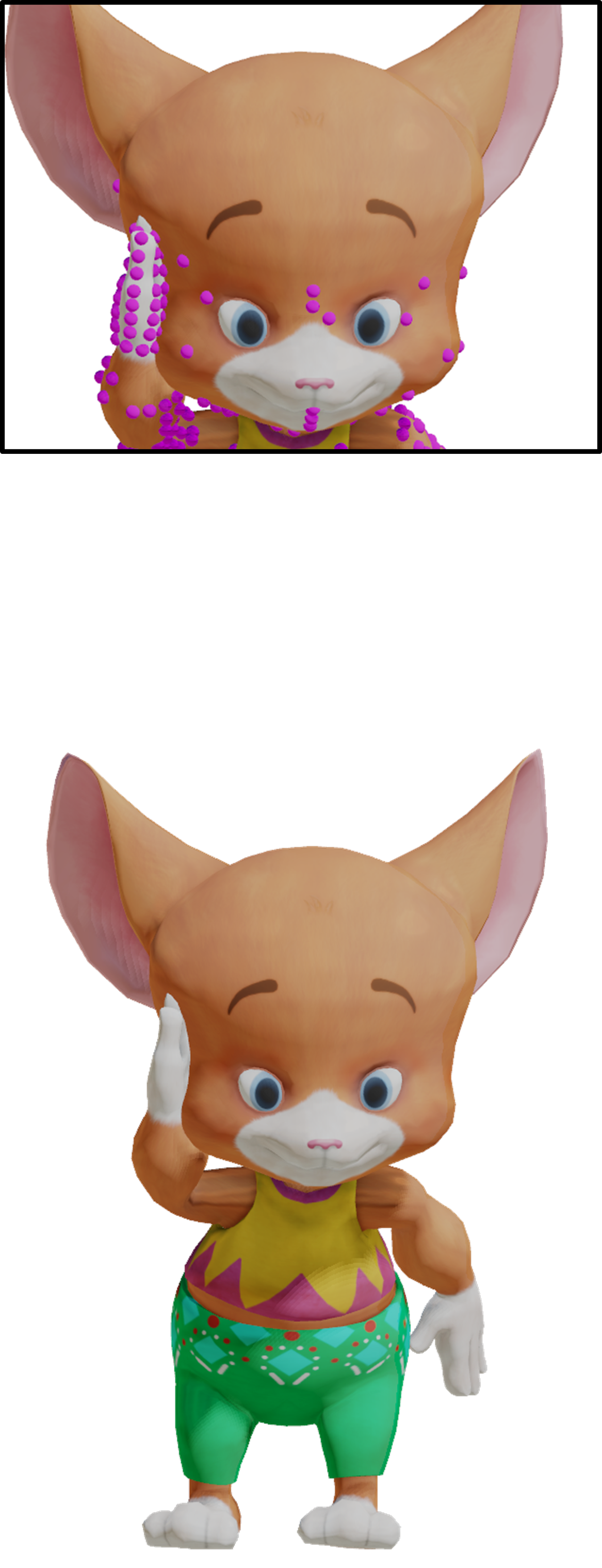}
        \caption*{(g)} 
        \label{fig:ablation_k_20}
    \end{minipage}
    \begin{minipage}{0.18\columnwidth}
        \centering 
        \includegraphics[width=\columnwidth]{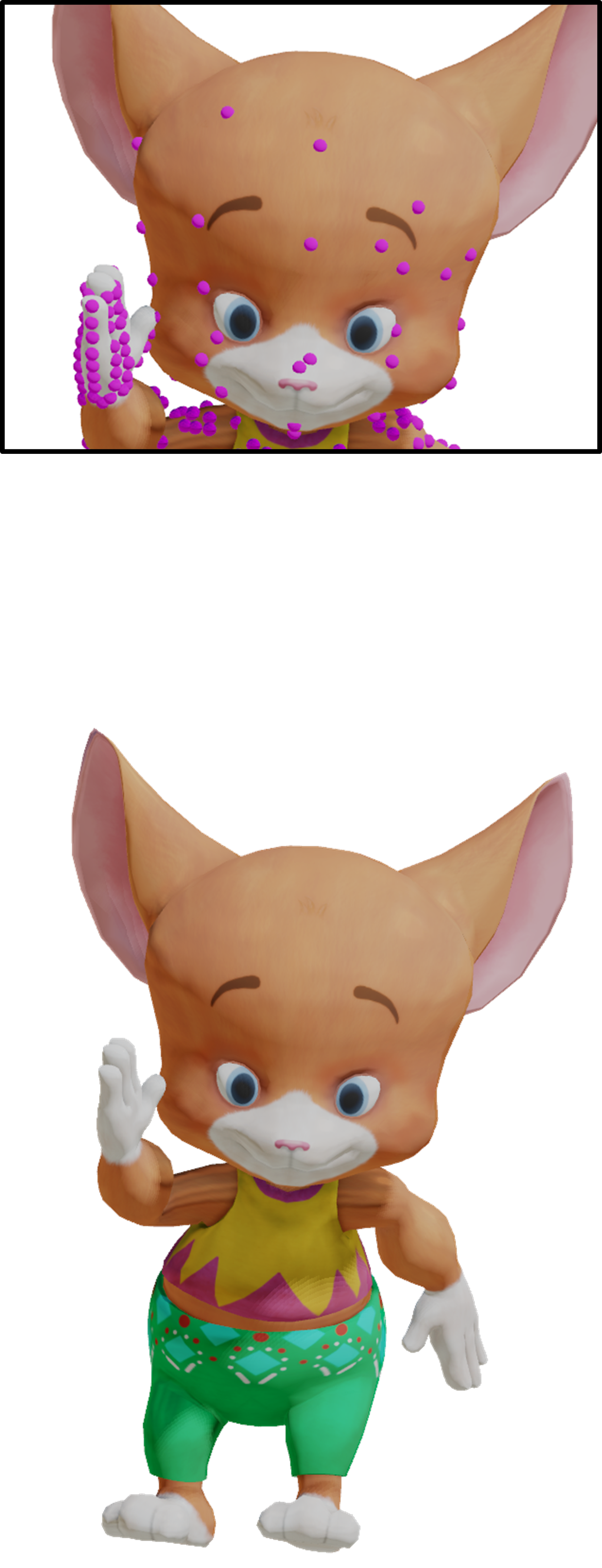}
        \caption*{(h)} 
        \label{fig:ablation_k_10}
    \end{minipage}
    \caption{Qualitative results of the ablation on the parameter $k$. Figures enclosed in boxes provide closer views for detailed comparison with initial anchors (white) and adapted anchors (pink). Subfigures (e) to (h) correspond to retargeted poses produced by the variants described in Table~\ref{tab:ablation_aas_alter_design}.}
    \label{fig:ablation_k}
\end{figure}
We further investigated the effect of the hyperparameter $k$, which controls the number of neighboring vertices considered during the Soft Projection process. To assess the influence of this parameter, we evaluated the quality of retargeted motions under various $k$, including smaller values ($k=1$ (e), $k=5$ (f)) and a larger value ($k=20$ (g)) against our choice of $k=10$ (h). As shown in Figure~\ref{fig:ablation_k}, smaller values of $k$ produced minimal anchor displacements, which frequently caused penetration artifacts in regions where the target character has enlarged body parts and the source motion involves close interactions, similar to the use of static anchors. On the other hand, a larger value of $k$ introduced excessive deviation from the initial anchor positions, because it incorporates unnecessary vertices that were spatially distant and thus irrelevant to the local geometric and semantic context of the anchor. This often led to semantically inconsistent anchor placements, ultimately causing the model to fail to preserve the spatial configuration of the source motion. These results suggest that our choice of $k=10$ provides a balanced point, offering sufficient flexibility to adapt to local geometry while preserving semantic alignment with the source motion. 

\subsubsection{Proximity-based Retargeting Module}
\begin{figure}
    \centering
    \begin{minipage}{0.18\columnwidth}
        \centering
        \includegraphics[width=\columnwidth]{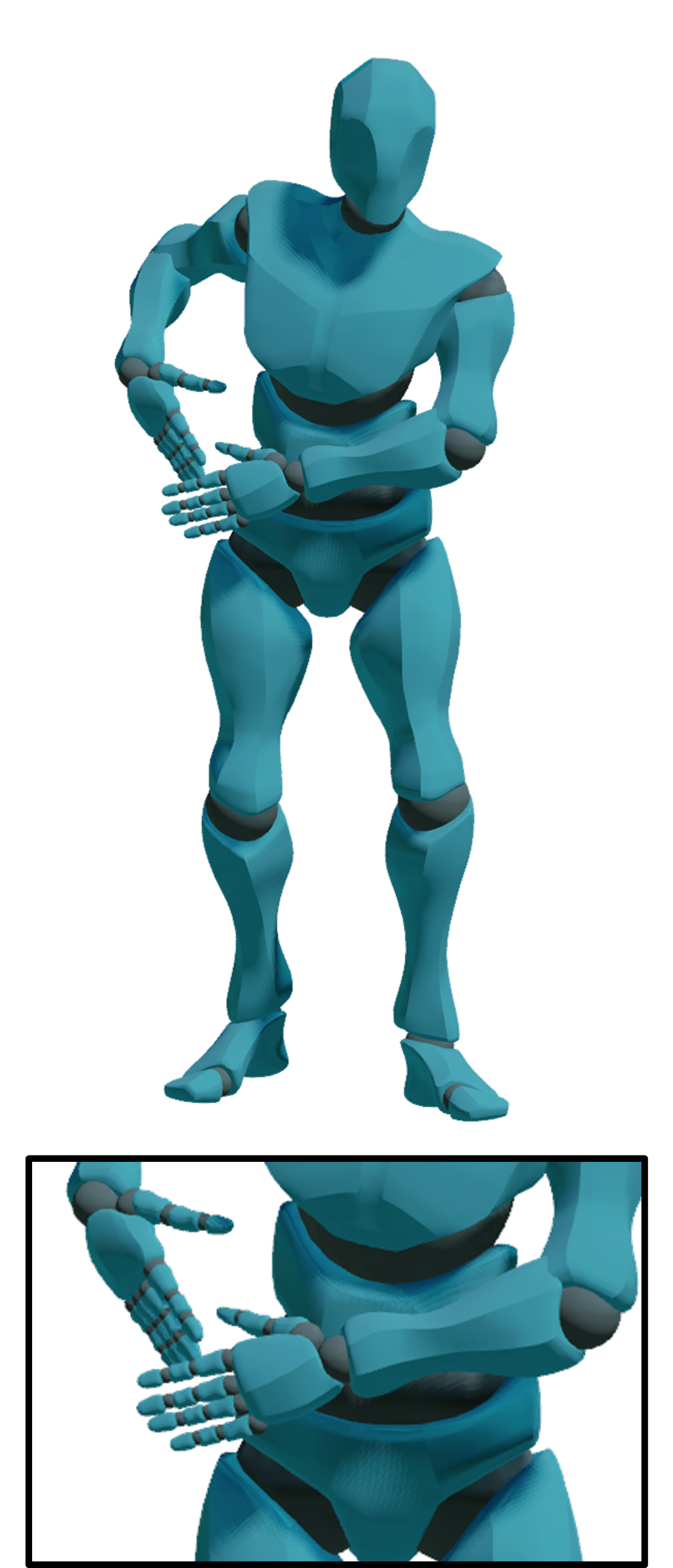}
        \caption*{\textit{Source Pose}}        
        \label{fig:ablation_loss_retarget_source}
    \end{minipage}
    \begin{minipage}{0.18\columnwidth}
        \centering
        \includegraphics[width=\columnwidth]{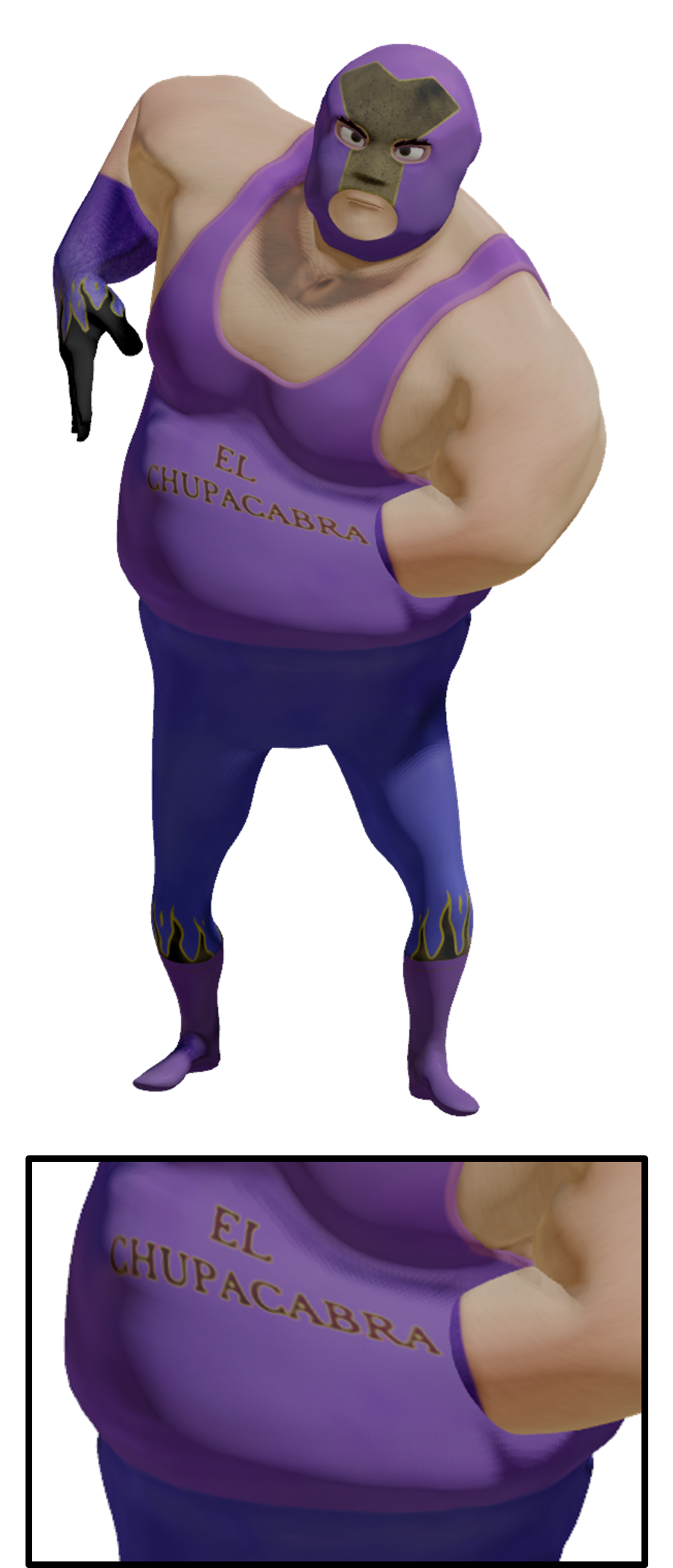}
        \caption*{(a)} 
        \label{fig:ablation_loss_retarget_tgt_a}
    \end{minipage}
    \begin{minipage}{0.18\columnwidth}
        \centering 
        \includegraphics[width=\columnwidth]{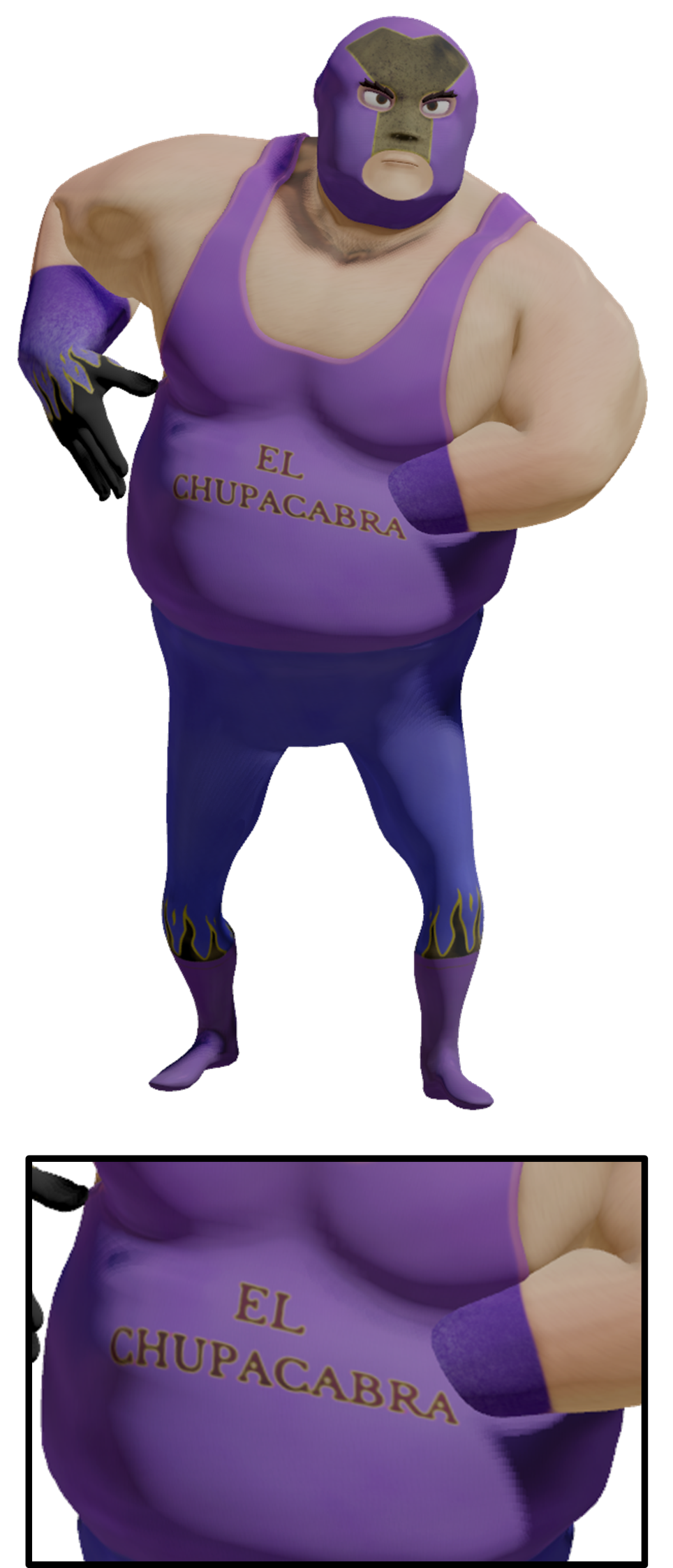}
        \caption*{(b)} 
        \label{fig:ablation_loss_retarget_tgt_b}
    \end{minipage}
    \begin{minipage}{0.18\columnwidth}
        \centering 
        \includegraphics[width=\columnwidth]{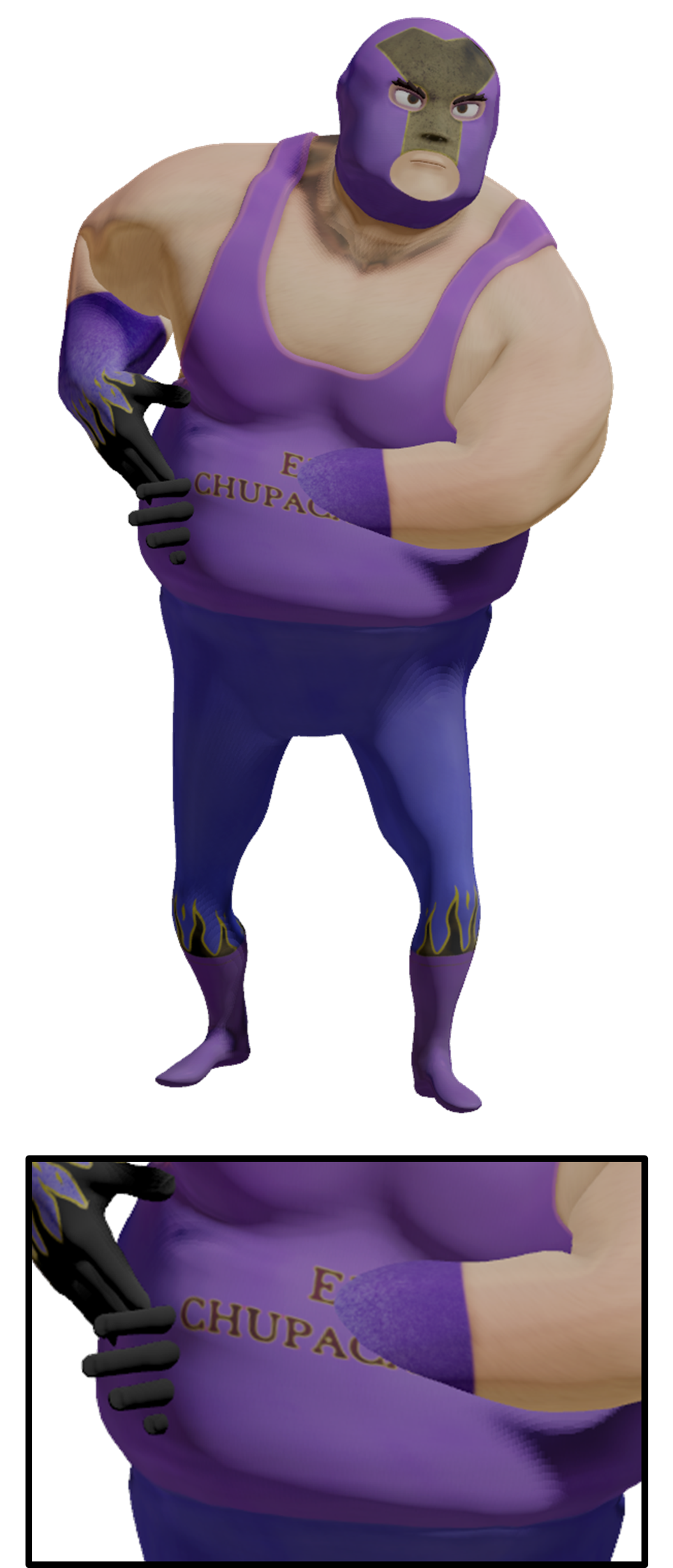}
        \caption*{(c)} 
        \label{fig:ablation_loss_retarget_tgt_c}
    \end{minipage}
    \begin{minipage}{0.18\columnwidth}
        \centering 
        \includegraphics[width=\columnwidth]{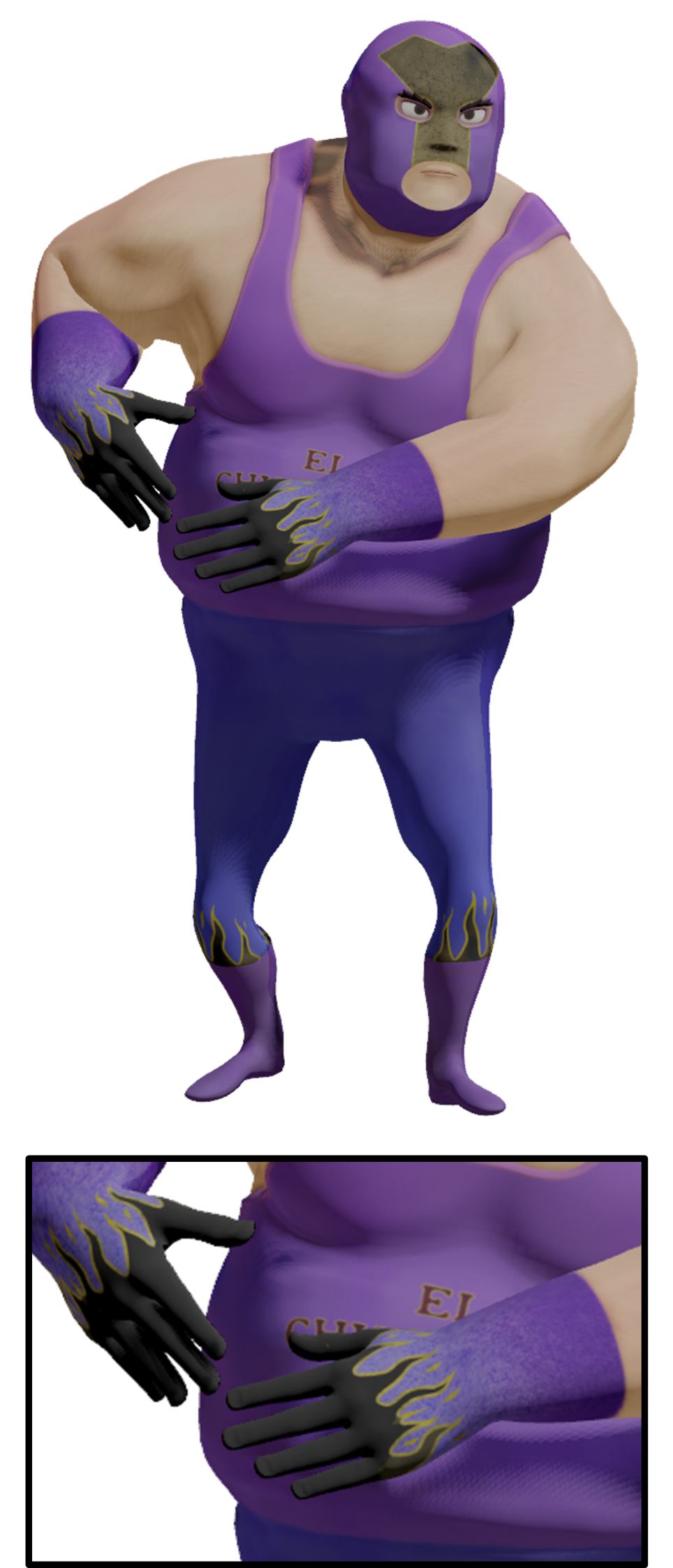}
        \caption*{(d)} 
        \label{fig:ablation_loss_retarget_tgt_d}
    \end{minipage}
    \caption{Qualitative results of the ablation study on the Proximity-based Retargeting module. Subfigures (a) to (d) correspond to retargeted poses produced by the variants described in Table~\ref{tab:ablation_loss_retarget}.}
    \label{fig:ablation_loss_retarget}
\end{figure}

To evaluate the contribution of each loss term in the Proximity-based Retargeting module, we conducted an ablation study by progressively removing individual terms. The quantitative and qualitative results are summarized in Table~\ref{tab:ablation_loss_retarget} and Figure~\ref{fig:ablation_loss_retarget}, respectively. Starting from the baseline (a) that uses only the reconstruction loss, we first added the joint velocity loss (b), which provided temporal regularization but showed limited improvement in preserving contact semantics. Incorporating the anchor distance loss $\mathcal{L}_{dist}$ (c) led to a noticeable gain in semantic consistency, because it encourages the spatial configuration of the adapted anchors to remain similar to that of the source motion. However, because different spatial layouts can share similar pairwise distances, relying solely on distance constraints still resulted in noticeable penetration artifacts. This is addressed by incorporating the anchor direction loss $\mathcal{L}_{dir}$, which completes the full objective of our model (d). By promoting the alignment of relative displacement vectors between anchor pairs, this term encourages consistent spatial orientation and mitigates unnatural intersections that may arise when relying solely on distance preservation. As shown in the final row (d) of Table~\ref{tab:ablation_loss_retarget}, including $\mathcal{L}_{dir}$ resulted in the lowest penetration rate and the highest contact preservation scores across all settings. These results confirm that the full loss configuration is crucial for achieving robust and interaction-aware motion transfer.
\begin{table} [!t]
    \caption{Quantitative results of the ablation study on loss terms to train the Proximity-based Retargeting module. The best result for each metric is highlighted in bold.}
    \setlength{\tabcolsep}{10pt}
    \resizebox{\columnwidth}{!}{
    \begin{tabular}{l|c|cll}
    \hline
    \multicolumn{1}{c|}
    {\multirow{2}{*}{Methods}} 
    & \multirow{2}{*}{\begin{tabular}[c]{@{}c@{}}Pen (\%) ↓\end{tabular}} 
    & \multicolumn{3}{c}{Contact Preservation}                                                        
    \\ \cline{3-5} 

    \multicolumn{1}{c|}{} 
    &   &Prec ↑ &Rec ↑ &Acc ↑ \\ \hline
    (a) \ $\mathcal{L}_{rec}$                                                             
    & 17.693 & 0.275 & 0.307 & 0.932 \\ 
    (b) \ $\mathcal{L}_{rec} + \mathcal{L}_{vel}$                                           
    & 17.385 &0.256 &0.256 &0.933 \\ 
    (c) \ $\mathcal{L}_{rec} + \mathcal{L}_{vel} + \mathcal{L}_{dist}$                      
    & 18.110 &0.309 &0.338 &0.936 \\
    (d) \ $\mathcal{L}_{rec} + \mathcal{L}_{vel} + \mathcal{L}_{dist} + \mathcal{L}_{dir}$ (Ours) 
    & \textbf{14.730} &\textbf{0.415} &\textbf{0.350} &\textbf{0.948} \\ \hline
    \end{tabular}%
    }
    \label{tab:ablation_loss_retarget}
\end{table}

\pagebreak
\subsubsection{Training Strategies}
\begin{figure} [!t]
    \centering
    \begin{minipage}{0.15\columnwidth}
        \centering
        \includegraphics[width=\columnwidth]{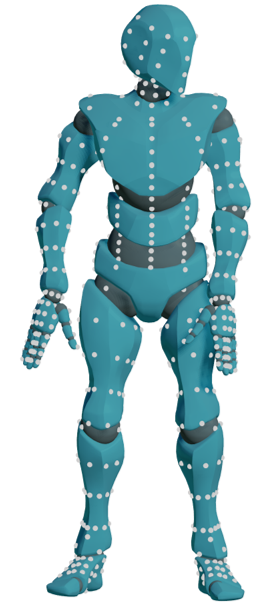}
        \caption*{\textit{Source Pose}}        
        \label{fig:ablation_alter_source}
    \end{minipage}
    \hspace{2.5em}
    \begin{minipage}{0.15\columnwidth}
        \centering
        \includegraphics[width=\columnwidth]{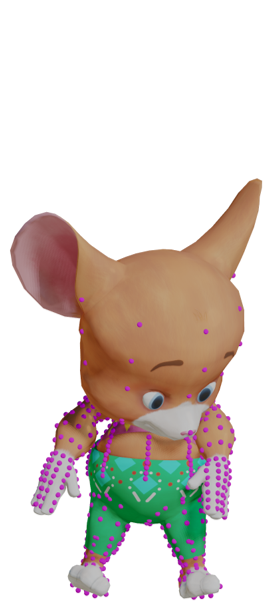}
        \caption*{(a)} 
        \label{fig:ablation_alter_joint}
    \end{minipage}
    \hspace{2.5em}
    \begin{minipage}{0.15\columnwidth}
        \centering 
        \includegraphics[width=\columnwidth]{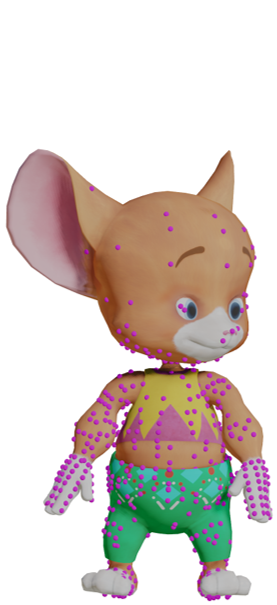}
        \caption*{(b)} 
        \label{fig:ablation_alter_alter}
    \end{minipage}
    \caption{Qualitative comparison of training strategies. The leftmost figure shows the source pose with its initial anchors visualized in white. Subfigures (a) and (b) correspond to the retargeted results obtained by the variants described in Table~\ref{tab:ablation_training_procedures}, with the adapted anchors visualized in pink.}
    \label{fig:ablation_alternative_optimization}
    \vspace{-1.0em}
\end{figure}

We conducted an ablation study to validate the effectiveness of the alternating optimization strategy by comparing it against a variant that jointly trained the Adaptive Anchor Sampling and Proximity-based Retargeting modules. As shown in Table \ref{tab:ablation_training_procedures}, the joint training variant (a) resulted in noticeably higher penetration rates and lower contact preservation scores, whereas the alternating optimization strategy (b) substantially improved performance across all metrics. In addition, as shown in Figure \ref{fig:ablation_alternative_optimization}, the anchors refined with the alternating scheme exhibit noticeable spatial displacements, whereas those from the jointly trained model remain largely unchanged. This suggests that joint training tends to satisfy proximity constraints primarily by adjusting skeletal poses rather than repositioning anchors, which often results in overly distorted postures. These findings demonstrate that our alternating optimization strategy allowed each module to better focus on its designated objective and improved overall performance in a complementary manner. For additional details on the training curves for each training strategy, please refer to Section 2 of the supplementary material.
\section{Limitations and Future Work}
% Limitations + Corresponding future work
While our method effectively preserves interaction-driven spatial relationships, several limitations remain. First, while the reachability loss encourages plausible retargeting in cases where the original interaction is kinematically infeasible, it serves as a soft constraint that only approximates the reachable ranges without explicitly modeling joint-level articulation limits. As a result, adapted anchors may be located near the boundary of the feasible region without accurately reflecting the actual kinematic capabilities of the target character. Future work could incorporate more precise representations such as per-joint range of motion and character-specific affordances to improve the physical plausibility of retargeted motions \cite{tonneau2014using}. 

Second, while our method effectively preserves surface-level interactions, it implicitly assumes that the source motion does not contain geometric artifacts, such as interpenetration, to ensure the generation of clean target motions. Because proximity constraints are transferred without explicit correction, any artifacts present in the source motion may propagate to the retargeted result, resulting in undesirable penetrations in the target character. Accordingly, integrating a refinement module that detects and corrects such artifacts in the source motion, while preserving its intended contact semantics can be a future direction. 

\begin{table} [!t]
    \caption{Quantitative results of the ablation study on training procedures. The best result for each metric is highlighted in bold.}
    \centering
    \setlength{\tabcolsep}{14pt}
    \resizebox{\columnwidth}{!}{%
    \begin{tabular}{l|c|cll}
    \hline
    \multicolumn{1}{c|}
    {\multirow{2}{*}{Methods}} 
    & \multirow{2}{*}{\begin{tabular}[c]{@{}c@{}}Pen (\%) ↓\end{tabular}} 
    & \multicolumn{3}{c}{Contact Preservation}                                                        
    \\ \cline{3-5} 
    \multicolumn{1}{c|}{}    
    &   & Prec ↑   & Rec ↑   & Acc ↑ \\ \hline
    
    (a) \ Joint training           
    & 21.987 & 0.262 & 0.349 & 0.926 \\    
    (b) \ Ours                     
    & \textbf{14.730} & \textbf{0.415} & \textbf{0.350} & \textbf{0.948} \\ \hline    
    \end{tabular}%
    } 
    \label{tab:ablation_training_procedures}
\end{table}

Third, although $\mathcal{L}_{vel}$ is employed to reduce local jitter, our method primarily relies on frame-wise spatial correspondences and does not explicitly model long-term motion dynamics. This can become problematic under large geometric discrepancies, where temporally evolving interactions may be truncated once the target character reaches its feasible limit, potentially causing the motion to stall and disrupting its original temporal progression. For instance, when a hand is supposed to slide continuously upward along the head, a target character with a larger head may cause the hand to stop near a reachable boundary rather than maintaining the intended upward progression. Incorporating motion-level modeling through intermediate key-joint trajectories \cite{tonneau2014using, hwang2025motion, hwang2025goal} prior to proximity enforcement or introducing timing-aware motion adjustment \cite{goel2025generative}, would be a promising direction for improving temporal continuity.

Our method could be extended in several directions to improve its generality. While it allows variation in joint locations, it assumes a consistent number of joints and shared skeletal connectivity across characters, due to the Anchor Encoder that constructs joint-level features by pooling anchors associated with the same joint for proximity-based matching. This could be alleviated by replacing joint-wise aggregation with body-part-wise aggregation over semantically related regions, enabling more flexible matching across heterogeneous skeletons. In addition, the current formulation considers only ground contact through $c^t_j$ and does not explicitly model interactions with external objects during retargeting. With sufficiently annotated data \cite{jin2025interfacerays}, $c^t_j$ could be generalized to indicate whether each joint is in contact with elements of the surrounding environment, such as sitting on a chair or resting the elbows on a table. Such extensions would broaden the applicability of the framework to a wider range of character rigs and more complex motions involving object interactions.

Lastly, motion retargeting is inherently subjective, as the aspects of motion that should be preserved may vary depending on the context. In some cases, preserving contact may be desirable, whereas in others, maintaining joint configurations despite sacrificing contact may be more appropriate. Our method explicitly adopts the former perspective, prioritizing contact preservation over joint-angle preservation by relocating anchors. Addressing this trade-off more explicitly would likely require higher-level semantic priors, such as vision-language models or human feedback. Moreover, while our user study indicates that the retargeted motions are perceptually acceptable, a more comprehensive evaluation could better characterize this subjectivity by comparing our results with artist-produced results and examining whether the outputs of our method lie within the distribution of results made by artists.

\section{Conclusion}
In this work, we present a novel motion retargeting framework that transfers motion between skinned characters by leveraging spatially adaptive anchors. Given a set of initial anchors that define coarse correspondences between source and target characters, the key insight of our method is to reposition the anchors on the target character by jointly considering the spatial structure of the source motion and the feasibility of reproducing the intended interactions on the target character. To this end, we propose an Adaptive Anchor Sampling module that refines anchor positions within kinematically reachable regions to better support proximity-based constraints, ensuring surface interactions consistent with those in the source motion. The resulting anchors then serve as effective guidance for the Proximity-based Retargeting module, which predicts target skeletal motion while preserving both geometric plausibility and intended interactions. To facilitate task-aligned coordination between anchor adaptation and motion retargeting, we adopt an alternating optimization strategy that allows each module to specialize in its designated objective. Through extensive experiments, we demonstrated that our method achieves robust and semantically faithful motion transfer across characters with diverse body shapes.

\begin{acks}
This work was supported by the National Research Foundation of Korea (NRF) grant funded by the Korea government (MSIT) (RS-2024-00333478).
\end{acks}

\bibliographystyle{ACM-Reference-Format}
\bibliography{99.Ref}

\clearpage

% Restart section numbering from 1
\setcounter{section}{0}
\setcounter{subsection}{0}
\setcounter{subsubsection}{0}
\setcounter{figure}{0}
\setcounter{table}{0}

\renewcommand{\thesection}{\arabic{section}}
\renewcommand{\thesubsection}{\thesection.\arabic{subsection}}
\renewcommand{\thesubsubsection}{\thesubsection.\arabic{subsubsection}}

% Supplementary title spanning both columns, left-aligned like the original ACM title
\makeatletter
\twocolumn[
    \begingroup
        \@twocolumnfalse
        \noindent
        {\@titlefont Skinned Motion Retargeting with Spatially Adaptive Interaction Guidance\par}
        \vspace{1.0em}
        
        % \noindent
        % {\@titlefont\large Supplementary Material\par}
        \begin{center}
            {\@titlefont\Large Supplementary Material\par}
        \end{center}
        \vspace{2.0em}
    \endgroup
]
\makeatother

\section{Details on Adaptive Anchor Sampling}
\subsection{Anchor Extraction} \label{sec:Supp_Anchor_Extraction}
To extract a fixed-size set of anchors from character meshes with varying topologies, we adopt the sampling strategy introduced by \citeN{ye2024skinned}, which employs the rest pose skeleton $S$ as a canonical reference. For each skeletal bone, defined by a parent and child joint pair, a fixed number of intermediate points are uniformly sampled along the bone and treated as ray origins. At each origin, multiple rays are cast in evenly spaced directions within the plane perpendicular to the bone axis, and are traced until they intersect the mesh surface. The resulting intersection point, along with its surface normal, is expressed in barycentric coordinates with respect to the triangle where the point intersected. The tangent direction is aligned with the bone axis used for raycasting, and the bi‑tangent is computed as the cross product between the normal and tangent. These three vectors collectively constitute a tangent matrix at each anchor. This procedure yields a fixed-size, bone-indexed anchor set with consistent semantic correspondence across characters, in the sense that anchors are placed in functionally analogous regions with respect to the canonical skeleton, regardless of differences in mesh topology, as illustrated in Figure 3 of the main paper. For more details, we refer the readers to \citeN{ye2024skinned}.

\subsection{Learnable Relative Positional Bias} \label{Appendix_RelativeBias}
Inspired by \citeN{hong2023attention}, which models pairwise spatial relationships within a point cloud using relative Cartesian offsets and relative normal angles, we incorporate a learnable bias term derived by relative displacements between anchors, represented in their respective tangent frame, to model spatial dependencies induced by the source pose. This allows the model to explicitly capture how anchors are spatially arranged relative to one another, in a way that remains consistent under global rigid transformations. Specifically, given the pairwise relative direction vectors $\mathbf{D}^{src}_{dir} \in \mathbb{R}^{N_A \times N_A \times 3}$ between source anchors, we apply a signed logarithmic transformation defined as follows: 
\begin{equation}
    \tilde{\mathbf{D}}^{src}_{dir}=\text{sign}(\mathbf{D}^{src}_{dir}) \odot \text{log}(1+|\mathbf{D}^{src}_{dir}|+\epsilon),
\end{equation}
where $\odot$ denotes element-wise multiplication and $\epsilon$ is a small constant for numerical stability.
This formulation applies logarithmic scaling to relative positional offsets to compress their dynamic range and facilitate stable learning. In our setting, we additionally preserve the sign of each component to retain directional polarity, as the relative directions between anchors encode meaningful geometric relationships that would be lost under magnitude-only representations. The transformed vectors are then projected through a two-layer MLP to produce the relative bias matrix as follows:
\begin{equation}
    \mathbf{B} = \lambda_b \cdot \tanh(\text{{MLP}}(\tilde{\mathbf{D}}^{src}_{dir})),
\end{equation}
where the tanh activation bounds the output range for stable training and $\lambda_b$ is a scaling factor that controls the influence of the bias term relative to attention logits.

\subsection{Soft Projection} \label{Appendix_SoftProjection}
Given the translated target anchors $\tilde{\mathbf{A}}^{tgt}_{r}$, Soft Projection produces the adapted anchors $\hat{\mathbf{A}}^{tgt}_{r} \in \mathbb{R}^{N_A \times 3}$ by projecting each translated anchor onto the target rest mesh $\mathbf{V}^{tgt}_{r}$. Let $\mathcal{N}_i \subset \{1, ..., N_V\}$ represent the set of indices of $k$-nearest vertices to $\tilde{\mathbf{A}}^{tgt}_{r, i}$. The adapted anchor $\hat{\mathbf{A}}^{tgt}_{r, i}$ is computed as follows: 
\begin{equation}
    \hat{\mathbf{A}}^{tgt}_{r, i} = \sum_{j \in \mathcal{N}_i}{\omega}_j \mathbf{V}^{tgt}_{r, j},
\end{equation}
where $\mathbf{V}^{tgt}_{r, j}$ denotes the position of the $j$-th target vertex. The weight $\omega_j$ associated with each neighboring vertex is defined based on its Euclidean distance to the translated anchor $\tilde{\mathbf{A}}^{tgt}_{r, i}$ as follows: 
\begin{equation}
     {\omega}_j = \frac{\exp(-d^2_j / \tau^2)}
    {\sum_{l \in \mathcal{N}_i} 
    \exp(-d^2_l / \tau^2)},
    \label{eq:soft_projection_weighting}
\end{equation}
where $d_j=\lVert \tilde{\mathbf{A}}^{tgt}_{r, i}-\mathbf{V}^{tgt}_{r, j}  \rVert_2$ denotes the distance between the $i$-th translated anchor and the $j$-th vertex while $\tau$ is a learnable temperature parameter that controls the smoothness of the interpolation.
\section{Training Behaviors under Optimization Strategies}
\begin{figure*}[t]
    \centering
    \begin{subfigure}[t]{0.78\textwidth}
        \centering
        \includegraphics[width=\linewidth]{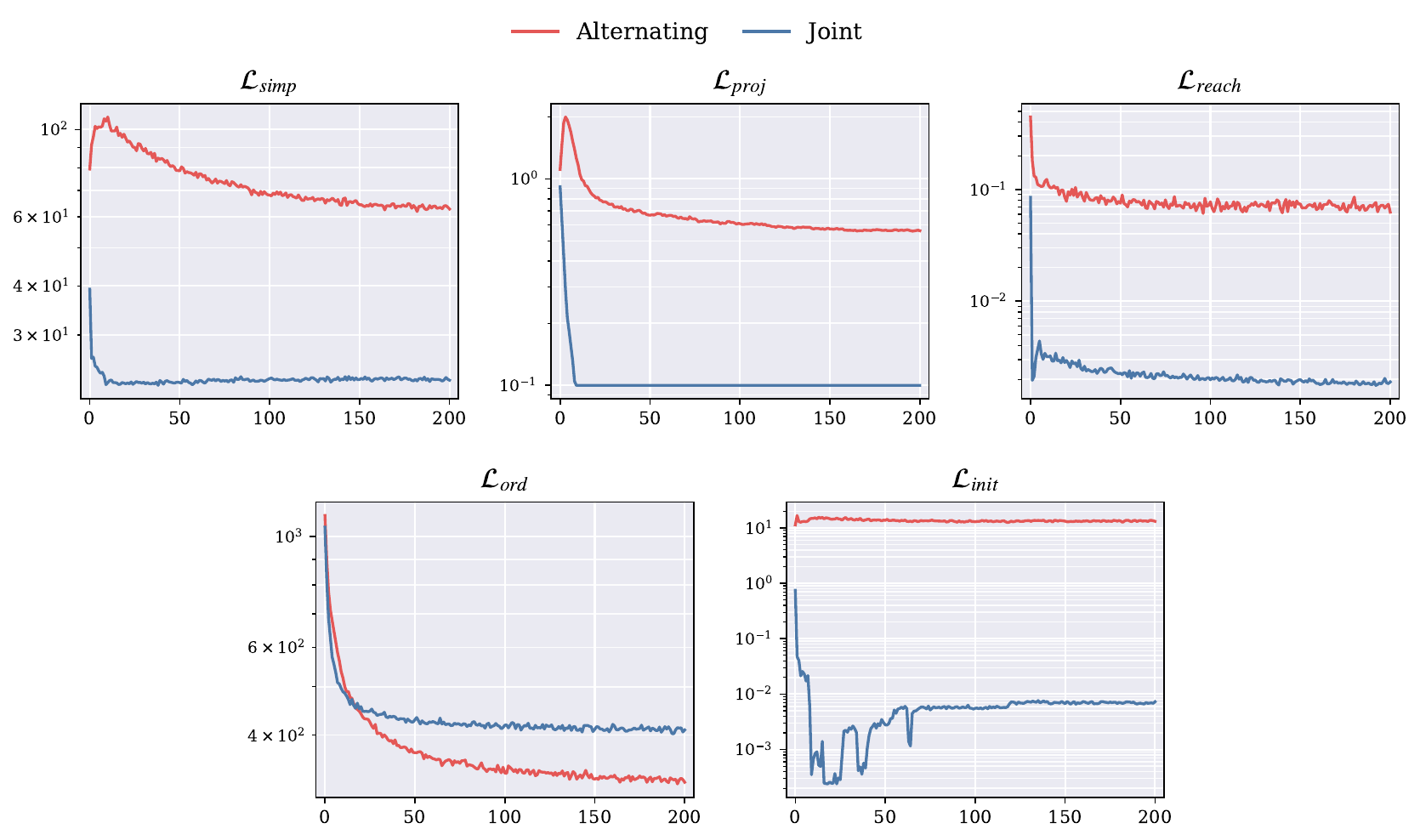}
        \caption{Loss curves for the Adaptive Anchor Sampling module.}
        \label{fig:subfig_a}
    \end{subfigure}

   \par \vspace{1em}
    
    \begin{subfigure}[t]{1.0\textwidth}
        \centering
        \includegraphics[width=\linewidth]{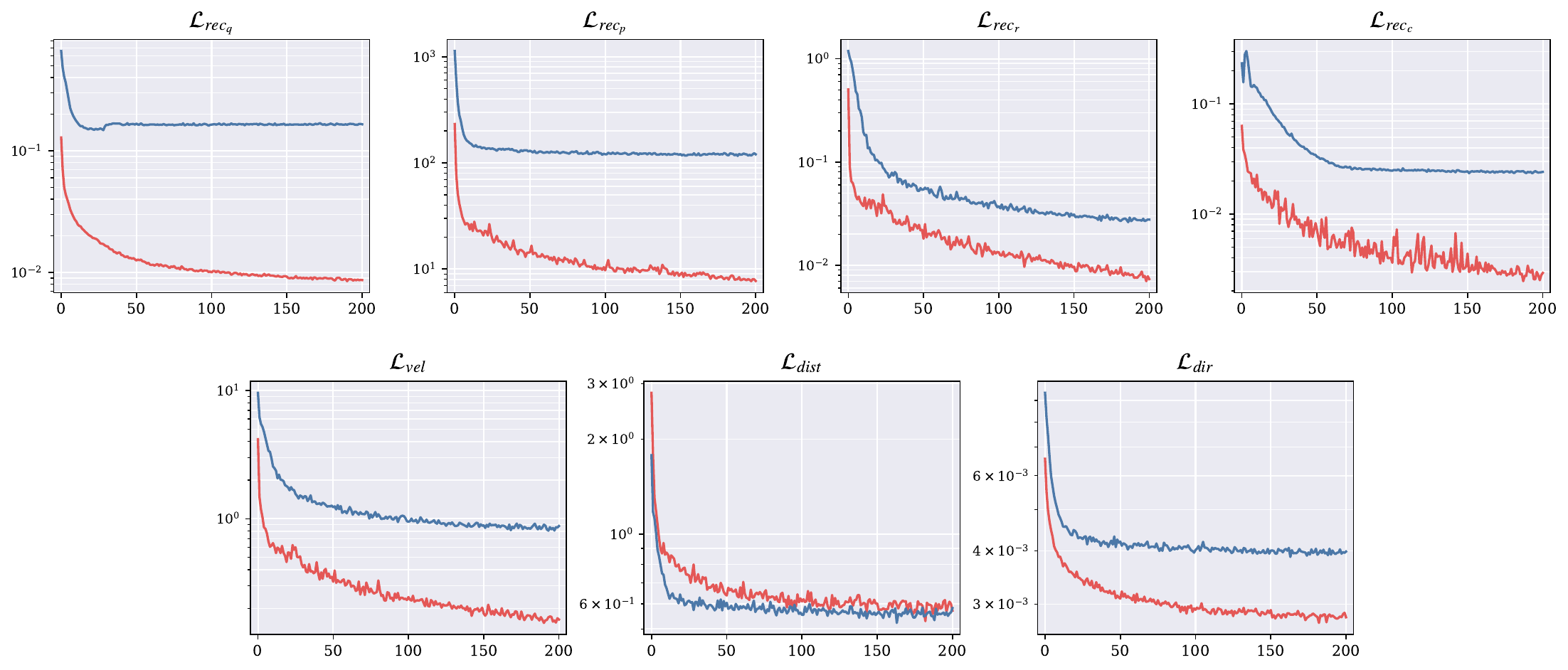}
        \caption{Loss curves for the Proximity-based Retargeting module.}
        \label{fig:subfig_b}
    \end{subfigure}
    \caption{Comparison of training loss curves under alternating optimization (red) and joint optimization (blue).}
    \label{fig:Alternating_Joint_Loss_Compare}
\end{figure*} 
Our method involves two optimization variables: the adapted anchors $\hat{\mathbf{A}}^{tgt}_{r}$ and the target pose $\hat{D}^{tgt}$. A practical difficulty is that the proximity objectives, such as $\mathcal{L}_{dist}$ and $\mathcal{L}_{dir}$, can be affected either by relocating anchors on the target mesh or by adjusting the target pose, thereby introducing ambiguity as to which variable should primarily account for satisfying the optimization objective. This coupling makes joint training problematic, because proximity objectives are evaluated on anchors deformed by the final pose configuration, for which pose updates provide a more immediate means of reducing the error, whereas anchor adaptation is additionally constrained by a regularization term such as $\mathcal{L}_{init}$. In practice, under joint training, the anchors tend to remain close to their initial positions while the proximity objectives are still reduced. This suggests that the reduction is achieved primarily by adjusting skeletal poses, with only marginal adaptation of the anchors. As a result, joint training weakens the intended role of the Adaptive Anchor Sampling module and often leads to excessively distorted postures.

To address this, we adopt an alternating optimization strategy that decouples the training of two modules, optimizing each in turn while keeping the other fixed. The Adaptive Anchor Sampling module is first updated to adjust anchor positions based on the downstream retargeting result, and the updated module predicts the adapted target anchors. The Proximity-based Motion Retargeting module is subsequently updated using these adapted anchors. During each step, all loss terms including $\mathcal{L}_{anc}$ and $\mathcal{L}_{retarget}$ remain active, but only the parameters of the selected module are updated. We use this fixed two-step schedule throughout training without any additional switching criterion. This design allows the anchors to first adapt toward reachable regions without being overridden by pose changes. Once the anchors have been adjusted, the retargeting module is then optimized to predict a target pose that satisfies the proximity constraints induced by the adapted anchors.

To further examine the effect of the training strategy described in Section 3.4.3 of the main paper, we provide an additional comparison of the training loss curves under joint and alternating optimization. As shown in Figure \ref{fig:Alternating_Joint_Loss_Compare}, compared to alternating optimization, joint optimization yielded lower $\mathcal{L}_{init}$ and $\mathcal{L}_{reach}$, indicating that the anchors remain closer to their initial positions while staying within reachable regions from anchors associated with interacting end-effectors. At the same time, $\mathcal{L}_{dist}$ stayed at a comparable level, whereas losses that encourage the predicted target motion $\hat{D}^{tgt}$ to remain close to the reference target motion $D^{tgt}$, particularly $\mathcal{L}_{rec}$ and $\mathcal{L}_{vel}$, remained noticeably higher than those obtained with alternating optimization. This suggests that, under joint training, the proximity objectives were satisfied primarily through pose adjustment, while anchor adaptation remained limited. This behavior reflects an optimization ambiguity, as the same objectives can be achieved either through pose updates or anchor relocation. In contrast, the alternating optimization better preserved consistency with the reference target motion while allowing larger adjustments on anchors, indicating a more balanced use of the two modules. As a result, the two optimization variables contributed more complementarily to the training process, reducing the optimization conflict observed under joint training.
\section{Implementation Details}
\subsection{Dataset Preparation} \label{Appendix_DatasetPreparation}
All skeletons in the dataset share an identical structure, such that the number and connectivity of joints remain consistent across characters, while mesh topologies vary. To improve training efficiency, meshes with a large number of vertices are decimated to fewer than 3,000 vertices. For motion data, we utilized sequences of YBot provided by Mixamo \cite{Mixamo}, which offer clean skinned motion with accurate self-contact. These sequences are retargeted to other character models using MotionBuilder \cite{MotionBuilder}, an off-the-shelf retargeting tool, constructing a shared set of motion sequences for all characters. As a result, the total number of motion frames $N_T$ remains consistent across the entire dataset.

Existing off-the-shelf retargeting tools \cite{MotionBuilder}, as well as large-scale motion repositories such as Mixamo \cite{Mixamo}, do not consistently produce artifact-free skinned motion across diverse characters. As a result, the resulting skeletal motions may exhibit geometric artifacts, such as interpenetration or missing contacts, when applied to different character meshes. For evaluation, we randomly selected motion sequences from a subset with no observable geometric artifacts to ensure reliable assessment. During training, however, we allowed sequences that may contain interpenetration in order to include motions involving self-contact and near-body interactions. Accordingly, although we refer to the target skeletal motion as ground truth for notational convenience, it should not be interpreted as an exact physically or geometrically correct motion. Instead, it provides a structurally plausible initialization, which is subsequently refined through geometry-aware objectives to improve consistency with the target character in an unsupervised manner.

\subsection{Design of Loss Weights} 
The loss weights were determined empirically through a staged process. We first tuned the Proximity-based Retargeting module alone on relatively less challenging cases, where plausible retargeting could already be achieved without anchor adaptation. Starting from $\mathcal{L}_{rec}$ and $\mathcal{L}_{vel}$ with their weights initialized following SAME \cite{lee2023same}, we introduced $\mathcal{L}_{dist}$ and $\mathcal{L}_{dir}$ to extend the model from skeleton-aware to mesh-aware retargeting. While $\mathcal{L}_{rec}$ encourages the predicted motion to remain close to the reference target motion $D^{tgt}$, $\mathcal{L}_{dist}$ and $\mathcal{L}_{dir}$ encourage the retargeted motion to preserve the proximity structure induced by the source motion. Because stronger proximity preservation may require deviations from the reference motion, these terms, together with the rotational term weighted by $\lambda_{q}$ in $\mathcal{L}_{rec}$, were empirically balanced to preserve interactions while avoiding penetration artifacts. A larger weight was assigned to $\mathcal{L}_{dir}$ than to $\mathcal{L}_{dist}$ because $\mathcal{L}_{dir}$ had a relatively smaller numerical scale.

We then incorporated the Adaptive Anchor Sampling module and tuned its weights on more challenging cases, where anchor adaptation was necessary for successful motion transfer on the target character. Following SampleNet \cite{lang2020samplenet}, the weights of $\mathcal{L}_{simp}$ and $\mathcal{L}_{proj}$ were first set to $0.01$. We then initialized the weight of $\mathcal{L}_{init}$ to $1.0$ as a stabilizing prior, and empirically adjusted the weights of $\mathcal{L}_{reach}$ and $\mathcal{L}_{ord}$ to permit sufficient anchor displacement without destabilizing retargeting. In practice, assigning excessive weight to $\mathcal{L}_{ord}$ often over-constrained the relative spatial arrangement of anchors, which in turn led to distorted postures.

\subsection{Evaluation Metric} \label{Appendix_Metric}
\paragraph{Contact Preservation}
Following ~\citeN{jang2024geometry}, we evaluated contact preservation by comparing frame-wise binary contact labels between the source and retargeted motions. A contact is defined as a case where two body parts (e.g., hand and torso) are in close proximity without interpenetration. Such events are detected on a per-frame basis and classified into one of the following categories:
\begin{itemize} [label=\small$\bullet$, leftmargin=1.5em]
    \item \textbf{True Positive (TP)}: Contact exists in both source and target motions (contact correctly preserved). 
    \item \textbf{False Negative (FN)}: Contact exists in the source but not in the target (contact missed).
    \item \textbf{False Positive (FP)}: Contact exists in the target but not in the source (unintended contact).
    \item \textbf{True Negative (TN)}: No contact in both motions (non-contact correctly preserved).
\end{itemize}
Based on these definitions, we compute the following metrics:
\begin{itemize} [label=\small$\bullet$, leftmargin=1.5em]
    \item \textbf{Precision} ($= \frac{TP}{TP + FP}$): Proportion of predicted contacts that are also present in the source, indicating the reliability of predicted interactions.
    \item \textbf{Recall} ($= \frac{TP}{TP + FN}$): Proportion of source contacts correctly recovered in the target, reflecting how well the intended interactions are preserved.
    \item \textbf{Accuracy} ($= \frac{TP + TN}{TP + TN + FP + FN}$): Overall correctness across both contact and non-contact cases.
\end{itemize}
These metrics collectively evaluate how accurately and completely the interaction semantics of the source motion are preserved in the retargeted results.
\pagebreak
\section{Additional Experiments}
\subsection{Experiments on Adaptive Anchor Behavior}
\begin{figure*}
    \centering
    \begin{subfigure}[t]{1.0\textwidth}
        \centering
        \includegraphics[width=\linewidth]{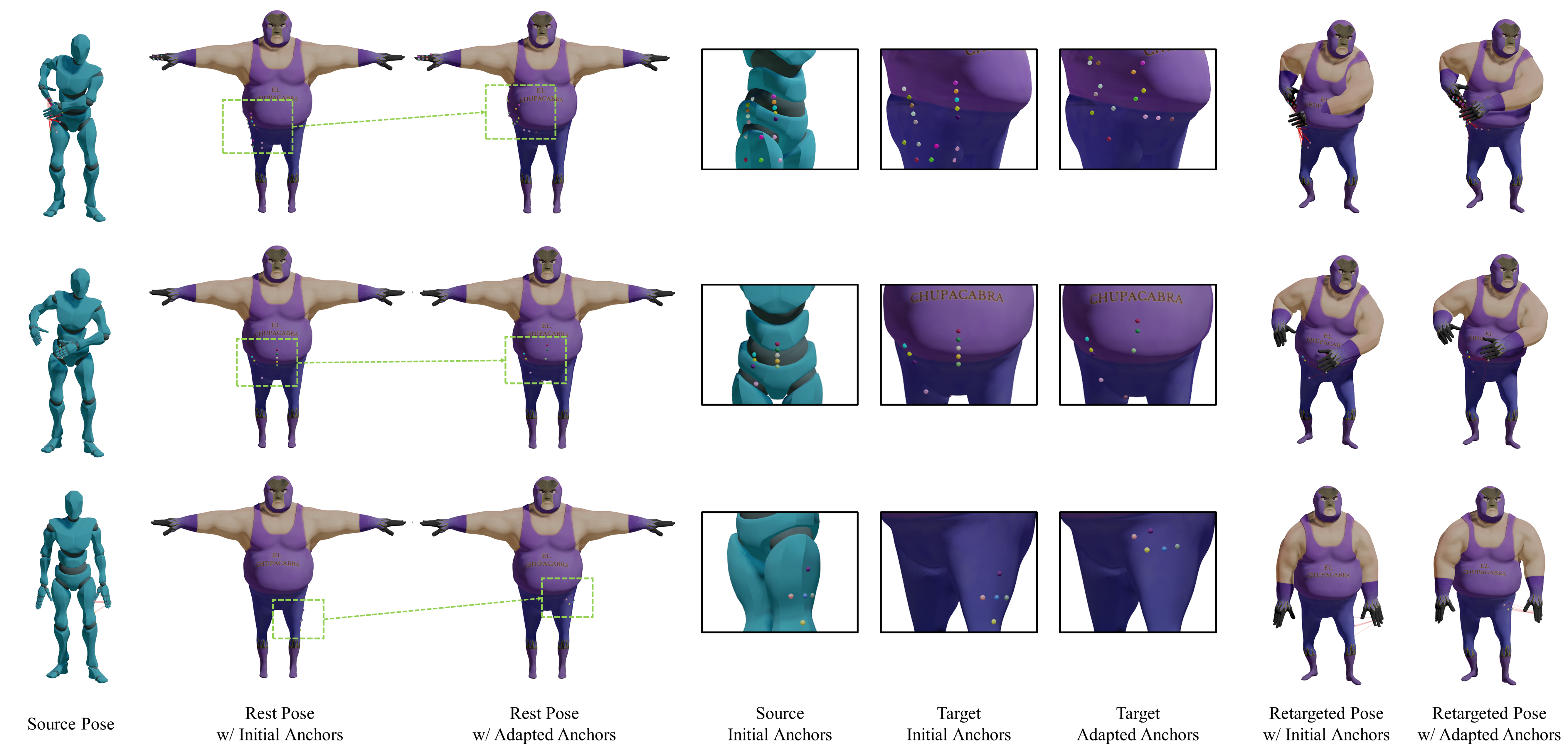}
        \caption{Source pose changes with fixed source and target characters}
        \label{fig:additional_exp_anchor_behavior_sourposechanges}
    \end{subfigure}
   \par \vspace{1em}
    \begin{subfigure}[t]{1.0\textwidth}
        \centering
        \includegraphics[width=\linewidth]{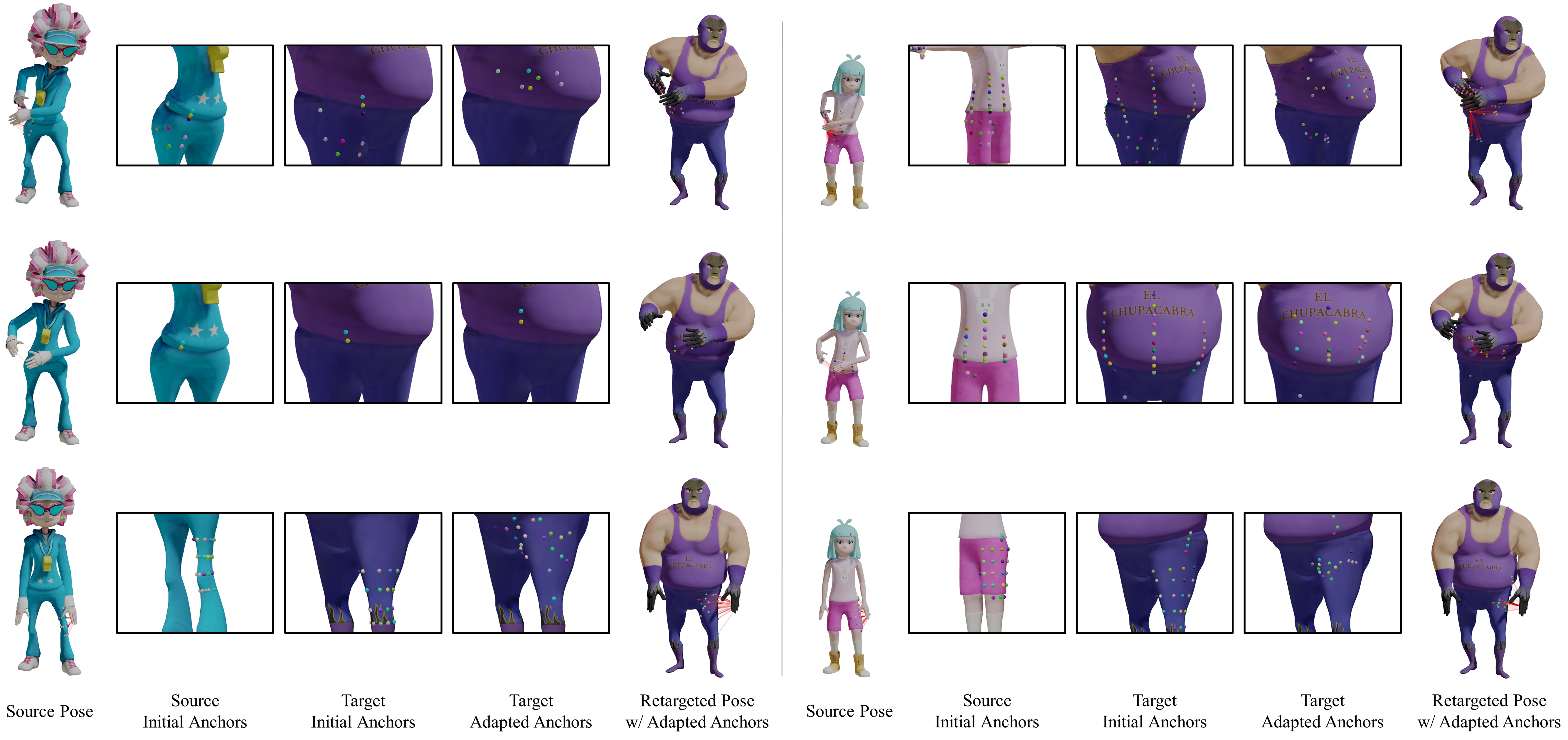}
        \caption{Extension under varying source characters}
        \label{fig:additional_exp_anchor_behavior_sourcharchanges}
    \end{subfigure}
    \caption{Qualitative results of adaptive anchor behavior on a fixed target character under varying source characters and poses. A query anchor is selected on the left hand of the source character, and the corresponding interaction-relevant anchors are illustrated, with the opacity of the red edges representing the corresponding interaction weights relative to the query anchor. On the target character, the corresponding initial and adapted anchors are shown to highlight the displacements induced by the anchor adaptation process. 
    }
    \label{fig:additional_exp_anchor_behavior}
\end{figure*}
We provide additional qualitative results to demonstrate how the predicted anchor locations shift from their initial positions to support proximity-based retargeting. Figure \ref{fig:additional_exp_anchor_behavior} shows the retargeted poses of a fixed target character under varying source characters and poses, along with the corresponding initial and adapted anchors. To effectively demonstrate the anchor adaptation behavior, we randomly selected a query anchor from the left hand of the source character and visualized the anchors involved in its interaction, which is determined by the interaction weighting matrix $\textbf{W}^{src}_{dist}$. The anchors visualized on the source character indicate those with non-zero entries in the weighting matrix, and the opacity of the red edges represents the associated weight values, with brighter edges representing larger weights. The anchors on the target character denote the corresponding anchors of the source. The initial anchor locations are also shown for reference to clearly illustrate how the anchors are repositioned through the Adaptive Anchor Sampling module. Anchors sharing the same color across the initial anchors of the source and both the initial and adapted anchors of the target indicate corresponding anchor pairs.

Figure \ref{fig:additional_exp_anchor_behavior_sourposechanges} illustrates how the anchors are adapted as the source pose changes while the source and target characters remain fixed. From top to bottom, the source pose transitions from configurations in which the hands are located close to the right side of the body to an idle pose. As shown in the first and second rows, when retargeting source poses in which the left hand is placed near the right side and front of the body to a target character with an enlarged belly, the target anchors are repositioned upward from their initial locations. This adaptation enabled the target character's hand to reach the right side of the body without penetration artifacts. A similar tendency is observed as the left hand of the source character moves back toward its original side, as shown in the last row. In this case, the adapted anchors help preserve the target character's body shape preventing the hand from dropping excessively in the idle pose, in contrast to the retargeted results obtained by the initial anchors. 

Figure \ref{fig:additional_exp_anchor_behavior_sourcharchanges} further extends the above observation under varying source characters. Although the source characters assume similar overall poses, slight differences in end-effector locations lead to variations in the precise spatial relationships between body parts. Moreover, differences in character height lead to changes in the interaction ranges defined by the interaction weighting matrix. As a result, both the interacting anchor pairs and their associated weights vary across source characters. Nevertheless, when different source characters perform a similar pose, the adaptation behavior of the target anchors remains largely consistent despite differences in body shape and proportions among the source characters. This suggests that the anchor adaptation is driven primarily by the spatial interaction structure induced by the source pose, rather than by the morphology of the source character itself. Please refer to the supplementary video for the animation results. 

Although the examples illustrate how the anchors interacting with a single query anchor are adapted, the resulting adapted anchors are not determined solely by the query anchor. Instead, they arise from the aggregation of relationships between multiple anchor pairs, conditioned on the source pose configuration. In addition, during adaptation, the influence of such anchor pairs is modulated by the weighting matrix, which quantifies their importance with respect to the interactions present in the source motion. Consequently, anchor pairs with larger interaction weights exert stronger influence when the anchors are adjusted toward reachable regions, which may occasionally alter their relative topology. Accordingly, rather than interpreting the adaptation mechanism on a strictly per-anchor basis, we focus on illustrating the overall adaptation tendency induced by the source interaction structure. Moreover, even when the adapted anchor configurations are not exactly identical across different source characters, the target character can still reproduce similar poses, since proximity matching is governed primarily by highly weighted anchor pairs, and such moderate topology changes therefore have only limited impact on the final retargeted motion.
\subsection{Experiments on Less Extreme Cases}
\begin{figure}
    \centering
    \begin{minipage}{0.18\columnwidth}
        \centering
        \includegraphics[width=\columnwidth]{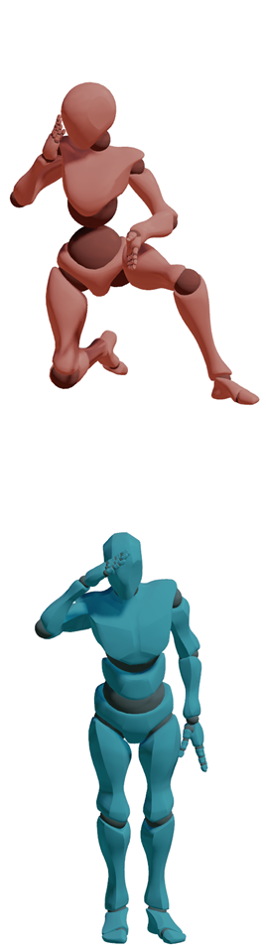}
        \caption*{\textit{Source Pose}}        
        \label{fig:additional_exp_less_extreme_cases_source}
    \end{minipage}
    \hspace{0.5em}
    \begin{minipage}{0.36\columnwidth}
        \centering
        \includegraphics[width=\columnwidth]{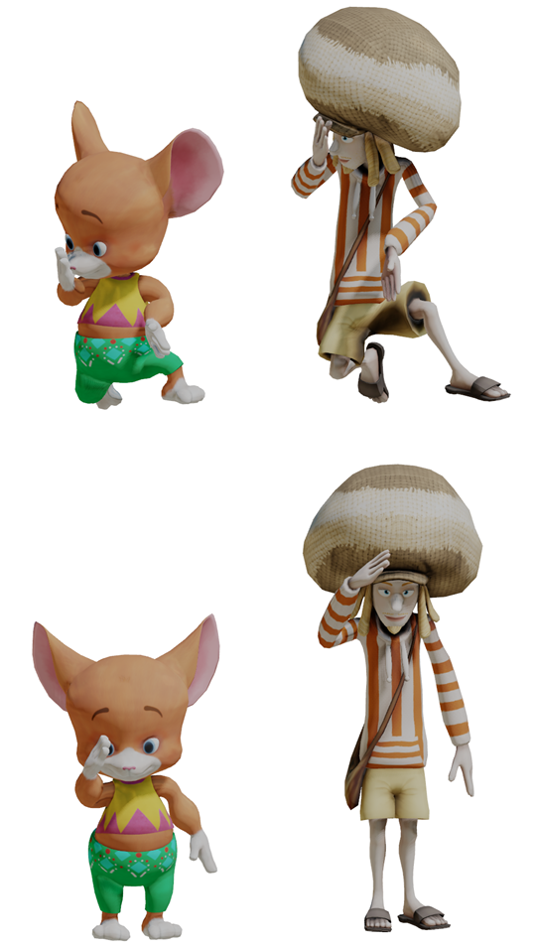}
        \caption*{(a) Extreme cases} 
        \label{fig:additional_exp_less_extreme_cases_extreme}
    \end{minipage}
    \hspace{0.5em}
    \begin{minipage}{0.36\columnwidth}
        \centering 
        \includegraphics[width=\columnwidth]{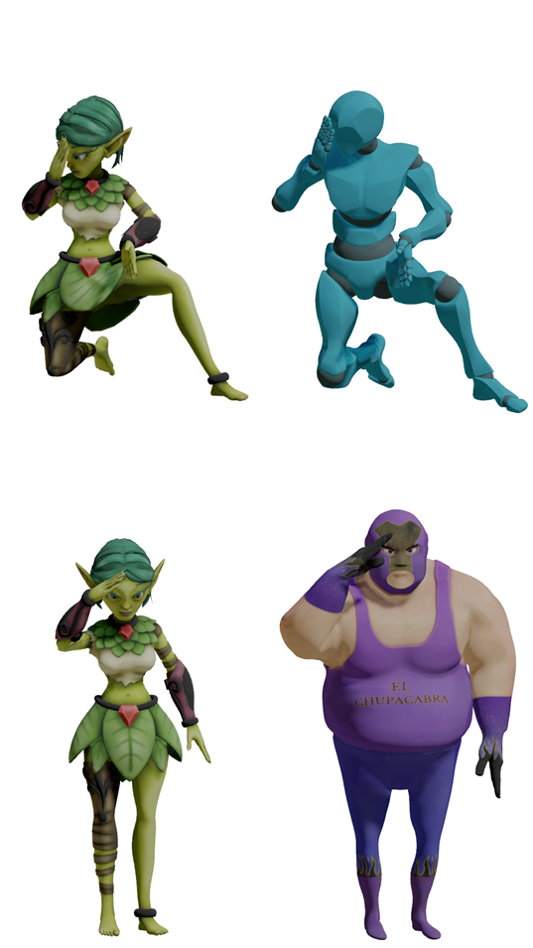}
        \caption*{(b) Less extreme cases} 
        \label{fig:additional_exp_less_extreme_cases_lessextreme}
    \end{minipage}
    \caption{Qualitative results of retargeting to target characters where the morphology of the body part involved in hand interaction is enlarged (a) or similar to that of the source (b). 
    }
    \label{fig:additional_exp_less_extreme_cases}
\end{figure}
While the proposed method primarily focuses on challenging cases in which large morphological discrepancies make exact reproduction of the source interaction inherently ambiguous, we additionally present qualitative results on less extreme cases to provide a clearer assessment of motion fidelity. As shown in Figure \ref{fig:additional_exp_less_extreme_cases}, variant (a) illustrates relatively extreme cases, where the source motion involves a hand-to-head interaction and the target character exhibits a substantially enlarged head compared to the source. In contrast, variant (b) presents less extreme cases, in which the geometry of the interaction region is more similar to that of the source character. Across both settings, our method successfully preserved the intended contact interaction: variant (a) achieved this through substantial yet plausible pose adaptation, whereas variant (b) maintained the interaction while remaining much closer to the source pose. These results indicate that the proposed method is effective not only in highly challenging transfers that require substantial geometric adaptation, but also in less extreme settings where the source interaction can be reproduced more faithfully.
\begin{table*}[!t]
\caption{Quantitative comparison between our method and baselines across different evaluation splits. The best result for each metric is highlighted in bold.}
\centering
\small
\setlength{\tabcolsep}{5pt}
\resizebox{\textwidth}{!}{%
\begin{tabular}{l|c|ccc|c|ccc|c|ccc|c|ccc}
\hline
\multicolumn{1}{c|}{\multirow{3}{*}{Methods}}
& \multicolumn{4}{c|}{SC+SM}
& \multicolumn{4}{c|}{SC+UM}
& \multicolumn{4}{c|}{UC+SM}
& \multicolumn{4}{c}{UC+UM} \\ \cline{2-17}

\multicolumn{1}{c|}{}
& \multirow{2}{*}{Pen (\%) $\downarrow$}
& \multicolumn{3}{c|}{Contact Preservation}
& \multirow{2}{*}{Pen (\%) $\downarrow$}
& \multicolumn{3}{c|}{Contact Preservation}
& \multirow{2}{*}{Pen (\%) $\downarrow$}
& \multicolumn{3}{c|}{Contact Preservation}
& \multirow{2}{*}{Pen (\%) $\downarrow$}
& \multicolumn{3}{c}{Contact Preservation} \\ \cline{3-5} \cline{7-9} \cline{11-13} \cline{15-17}

\multicolumn{1}{c|}{}
&  & Prec $\uparrow$ & Rec $\uparrow$ & Acc $\uparrow$
&  & Prec $\uparrow$ & Rec $\uparrow$ & Acc $\uparrow$
&  & Prec $\uparrow$ & Rec $\uparrow$ & Acc $\uparrow$
&  & Prec $\uparrow$ & Rec $\uparrow$ & Acc $\uparrow$ \\ \hline

MotionBuilder
& 0.211          & 0.266         & 0.323           & 0.901 
& 0.212          & 0.375         & \textbf{0.280}  & 0.925
& 0.170          & 0.216         & 0.328           & 0.932
& 0.181          & 0.234         & 0.305           & 0.938 \\

SAME
& 0.185          & 0.287         & 0.327           & 0.906
& 0.189          & 0.366         & 0.257           & 0.925
& 0.164          & 0.241         & 0.332           & 0.938
& 0.175          & 0.202         & 0.210           & 0.941 \\

R2ET
& \textbf{0.151} & 0.268         & 0.211           & 0.914
& \textbf{0.169} & 0.338         & 0.135           & 0.928
& 0.134          & 0.218         & 0.210           & 0.944
& 0.160          & 0.244         & 0.220           & 0.947 \\

MeshRet
& 0.223          & 0.246         & 0.238           & 0.906
& 0.244          & 0.322         & 0.241           & 0.920
& 0.154          & 0.284         & 0.312           & 0.946
& 0.184          & 0.347         & 0.329           & 0.953 \\

Ours
& 0.152          & \textbf{0.488} & \textbf{0.415} & \textbf{0.935}
& 0.187          & \textbf{0.409} & 0.229          & \textbf{0.930}
& \textbf{0.124} & \textbf{0.387} & \textbf{0.380} & \textbf{0.956}
& \textbf{0.157} & \textbf{0.394} & \textbf{0.364} & \textbf{0.957} \\

\hline
\end{tabular}%
}
\label{tab:splitwise_quan_results}
\end{table*} 
\subsection{Experiments on Generalization Capabilities}
\paragraph {Details on Evaluation Splits}
We additionally include split-wise quantitative results corresponding to Table 1 in the main paper, which summarizes performance over the full test set. While the aggregated results in Table 1 already show that our method outperforms all baselines, the split-wise analysis provides a more detailed view of its behavior under different combinations of seen and unseen characters and motions. As shown in Table \ref{tab:splitwise_quan_results}, our method achieved the best results across all metrics on both UC+SM and UC+UM. Our method also achieved the highest Precision and Accuracy on all splits, while remaining competitive on the SC splits, where R$^2$ET yielded a slightly lower penetration rate on SC+SM and MotionBuilder achieved higher Recall on SC+UM. These results further support the generalization capability of the proposed method beyond the character identities observed during training, especially in the important setting where both the target character and the motion are unseen.

\begin{figure} [!t]
    \centering
    \begin{minipage}{0.24\columnwidth}
        \centering
        \includegraphics[width=\columnwidth]{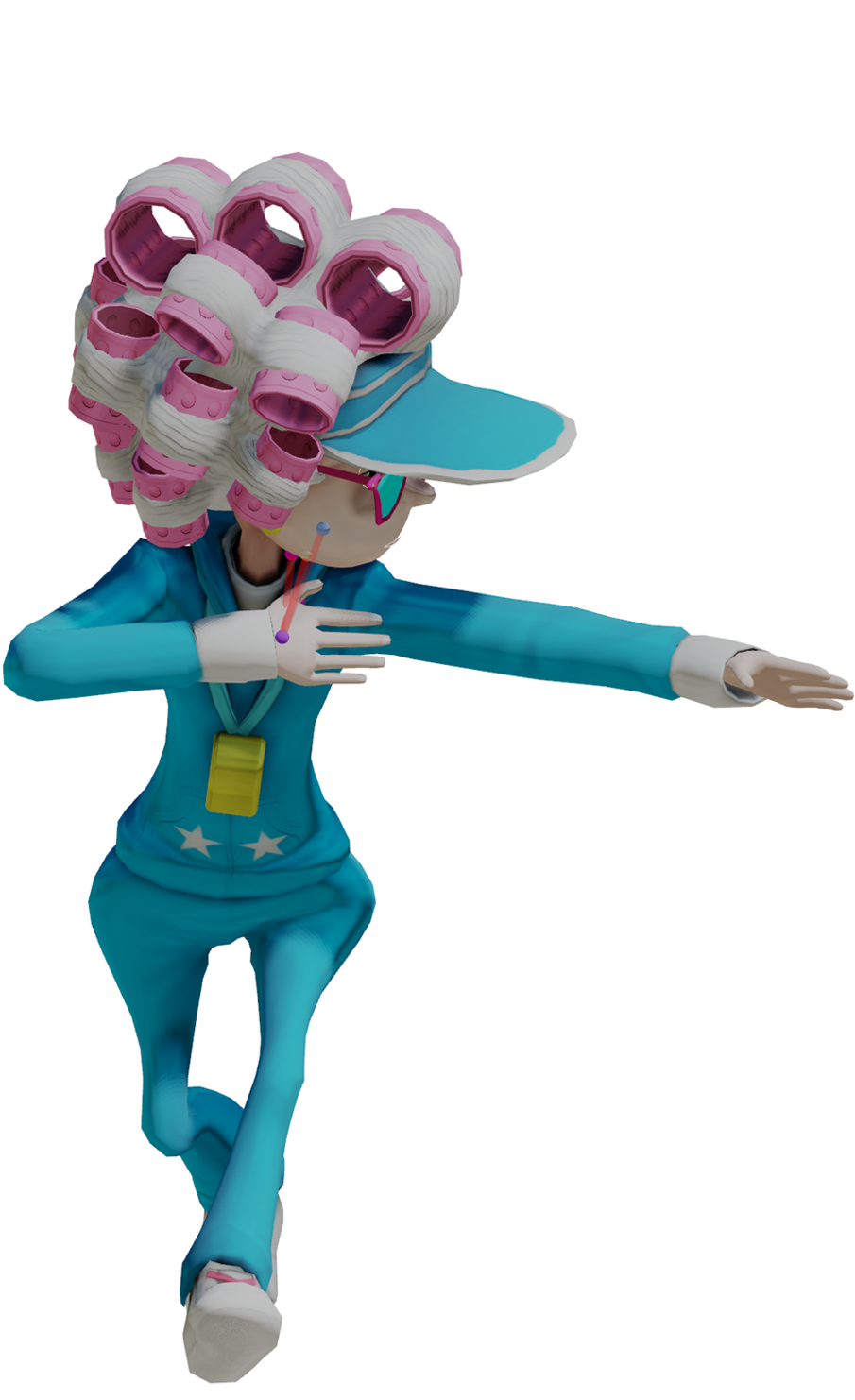}
        \caption*{\centering  \textit{Source Pose} \par}   
        \label{fig:OOD_Source}
    \end{minipage}
    % \hspace{0.5em}
    \begin{minipage}{0.37\columnwidth}
        \centering
        \includegraphics[width=\columnwidth]{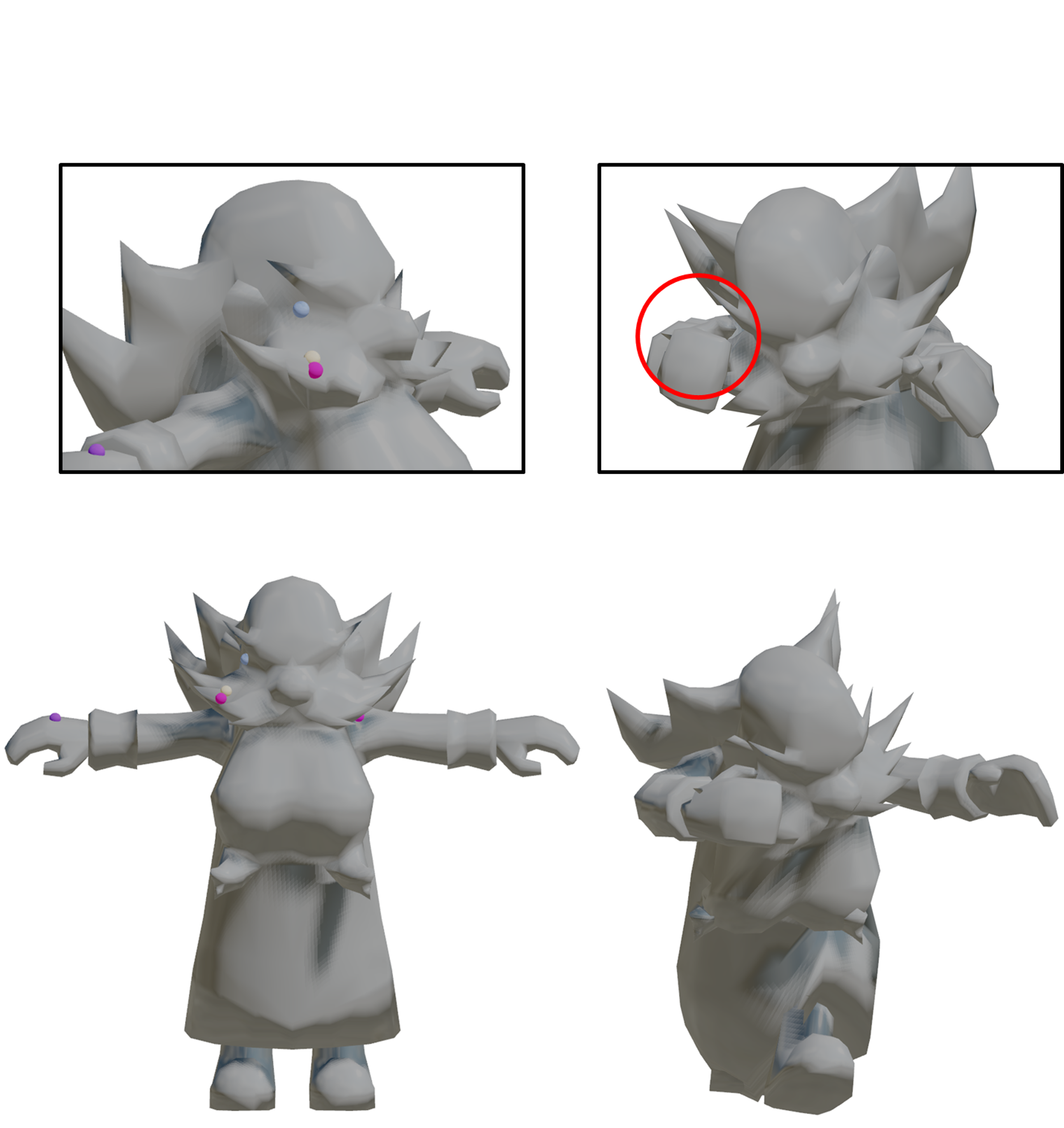}
        \caption*{\centering {(a) Retargeted Pose \\ w/ Initial Anchors}\par}
        \label{fig:OOD_Initial}
    \end{minipage}
    % \hspace{0.5em}
    \begin{minipage}{0.37\columnwidth}
        \centering 
        \includegraphics[width=\columnwidth]{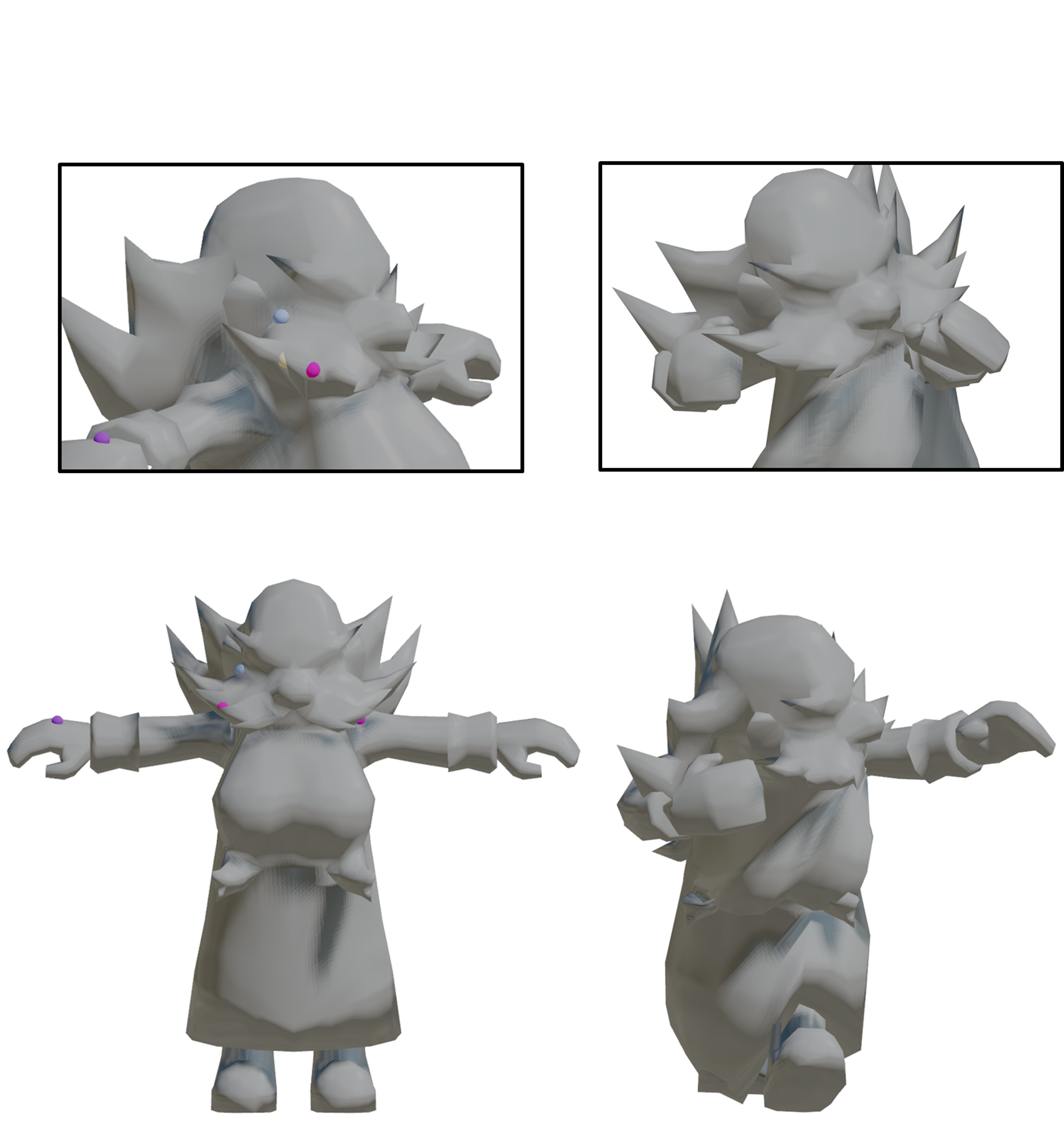}
        \caption*{\centering {(b) Retargeted Pose \\ w/ Adapted Anchors}\par}
        \label{fig:OOD_Adapted}
    \end{minipage}
    \caption{Qualitative results on an out-of-domain character from the RigNet-v1 dataset. Figures enclosed in boxes depict the corresponding results of a closer view whereas the red circle indicates the penetration artifact.}
    \label{fig:OOD_Character}
\end{figure}

\paragraph{Evaluation on Out-of-Domain Character} 
We further conducted an experiment an on out-of-domain character to evaluate the generalization capability of our method beyond the Mixamo \cite{Mixamo} characters used for training. We used an unseen character from the RigNet-v1 dataset \cite{xu2020rignet} as the target character. Because our method assumes a fixed joint topology, where the number of joints and their connectivity are consistent across characters, we extracted the common joints corresponding to the training topology and used them as the target skeleton. Figure \ref{fig:OOD_Character} shows how the initial anchors of the out-of-domain target character, characterized by spiky hair and a beard, are adapted according to the source pose in which the right hand is positioned close to the head. Variants (a) and (b) present the initial and adapted anchors of the target character corresponding to the interacting anchors in the source pose, respectively, along with the retargeted poses obtained using each anchor configuration.

As shown in Figure \ref{fig:OOD_Character}, the source pose contains a query anchor located on the right hand and its interacting counterpart anchors around the head region. The retargeted pose through the initial anchors produced penetration between the right hand and hair, as shown in variant (a). In contrast, the adapted anchors are repositioned toward the right side of the character's head, guiding the target pose to better follow the source pose without such penetration artifacts, while also better preserving the spatial relationships between the hand and head in the source pose, as shown in variant (b). For animation results, please see the supplementary video. 

\paragraph{Training on Larger Dataset}
To examine the effect of scaling the number of training characters, we additionally trained the model with 10 and 11 characters by adding Mixamo characters \cite{Mixamo} with large shape variations that were not included in the original 8-character training set. For a controlled comparison, we kept the training configuration unchanged across all settings. As shown in Table \ref{tab:larger_dataset}, simply increasing the number of characters under the same training configuration did not lead to a consistent improvement. The model trained with 8 characters achieved the best average performance in penetration avoidance, contact precision, and contact accuracy, whereas the model trained with 10 characters showed marginal improvement in contact recall. 

This observation suggests that increasing the number of characters alone is insufficient to improve generalization performance of the proposed method. Under the expanded training distribution, the original balance among objective terms and the optimization hyperparameters may no longer be optimal. In particular, because there is no paired supervision for adapted anchors, the proposed method relies on multiple loss terms to learn anchor adaptation in an unsupervised manner while providing task-relevant guidance for motion retargeting. As the number of training characters increases, retuning the loss weights and other optimization hyperparameters may therefore be required to better accommodate the broader training distribution and to more effectively benefit from the additional training characters. These results suggest that effective scaling may require adjusting the training objectives and optimization settings to account for increased data diversity, rather than simply increasing the number of training characters.
\begin{table}
    \caption{Quantitative comparison between our method and baselines trained with different numbers of characters. The best result for each metric is highlighted in bold.}
    \centering
    \setlength{\tabcolsep}{16pt}
    \resizebox{\columnwidth}{!}{%
    \begin{tabular}{l|c|cll}
    \hline
    \multicolumn{1}{c|}
    {\multirow{2}{*}{Methods}} 
    & \multirow{2}{*}{\begin{tabular}[c]{@{}c@{}}Pen (\%) ↓ \end{tabular}} 
    & \multicolumn{3}{c}{Contact Preservation}\\ \cline{3-5} 

    \multicolumn{1}{c|}{} 
    & & Prec ↑ & Rec ↑ & Acc ↑ \\ \hline 
    8 (Ours)      
    & \textbf{14.730} & \textbf{0.415} & 0.350 & \textbf{0.948} \\
    10			
    & 16.227 & 0.374 & \textbf{0.371} & 0.944 \\
    11     			
    & 15.878 & 0.357 & 0.342 & 0.942 \\
    \hline
    \end{tabular}%
    }
    \label{tab:larger_dataset}
\end{table}

\subsection{Comparison with Optimization-based Baseline}
To evaluate the benefit of the learning-based framework, we additionally compared our method with a direct optimization baseline that minimizes the same objectives at test time without using the learned modules. In this baseline, the Adaptive Anchor Sampling and Proximity-based Retargeting modules were removed, and the optimization was directly performed over per-anchor displacements $\Delta \textbf{A}$, the target pose $\hat{D}^{tgt}$, and the temperature parameter $\tau$. To ensure a fair comparison, we retained the Soft Projection operation, which projects the translated anchors onto the target mesh surface through interpolation over neighboring vertices. The parameters were optimized in an alternating manner for 500 steps, using a learning rate of 0.001 for both the anchor and pose variables. The initial values of the anchor displacements, target pose, and temperature parameter were set to the zero matrix $\mathbf{0} \in \mathbb{R}^{N_A \times 3}$, the ground truth motion $D^{tgt}$, and $1.0$, respectively.

\begin{figure}
    \centering
    \begin{minipage}{0.32\columnwidth}
        \centering
        \includegraphics[width=\columnwidth]{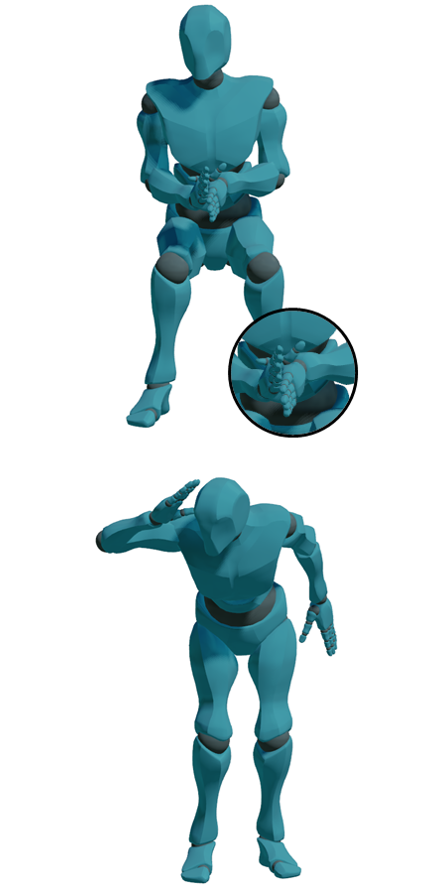}
        \caption*{\centering  \textit{Source Pose} \par}   
        \label{fig:DirectOpt_Comparison_Source}
    \end{minipage}
    \begin{minipage}{0.32\columnwidth}
        \centering
        \includegraphics[width=\columnwidth]{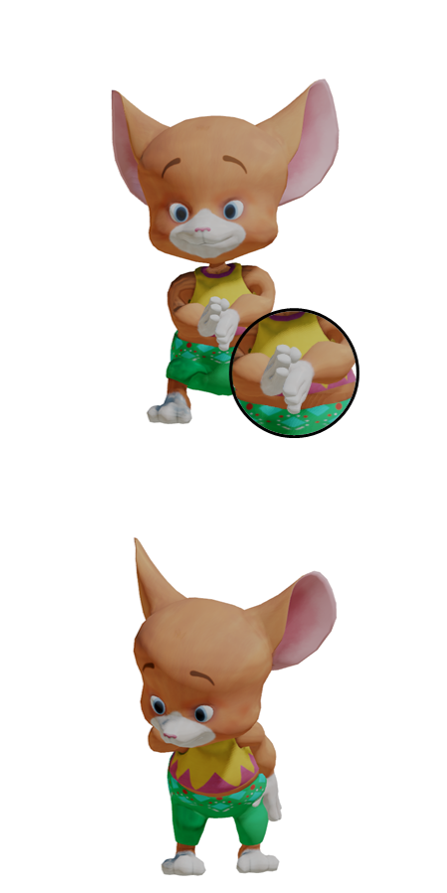}
        \caption*{\centering {(a) Direct Optimization}}
        \label{fig:DirectOpt_Comparison_Opt}
    \end{minipage}
    \begin{minipage}{0.32\columnwidth}
        \centering 
        \includegraphics[width=\columnwidth]{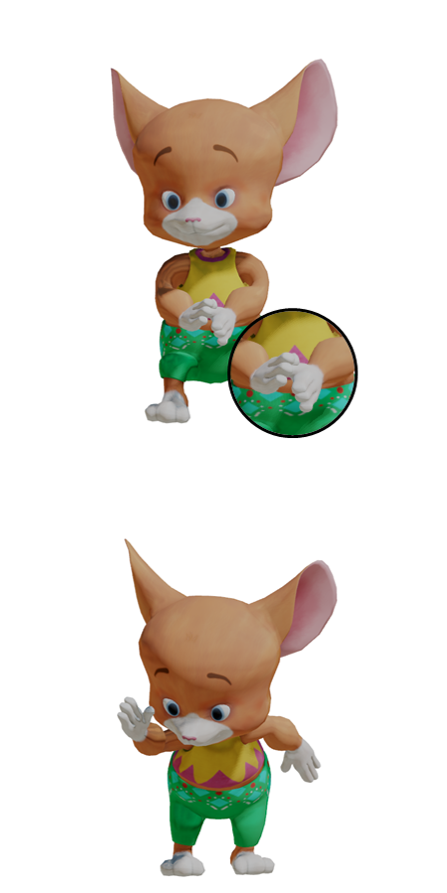}
        \caption*{\centering {(b) Learning-based}}
        \label{fig:DirectOpt_Comparison_Ours}
    \end{minipage}
    \caption{Qualitative comparison between direct optimization and our learning-based model. The leftmost figure shows the source pose, and subfigures (a) and (b) show the retargeted results corresponding to the variants in Table \ref{tab:comparison_with_optimization_based}. The upper row shows a simpler case where successful retargeting can be achieved without anchor adaptation, whereas the bottom row shows a more challenging case where anchor adaptation is required to transfer the source pose.}
    \label{fig:DirectOptimization_Comparison}
\end{figure}

\begin{figure} 
    \centering
    \begin{minipage}{0.09\textwidth}
        \centering
        \includegraphics[width=\textwidth]{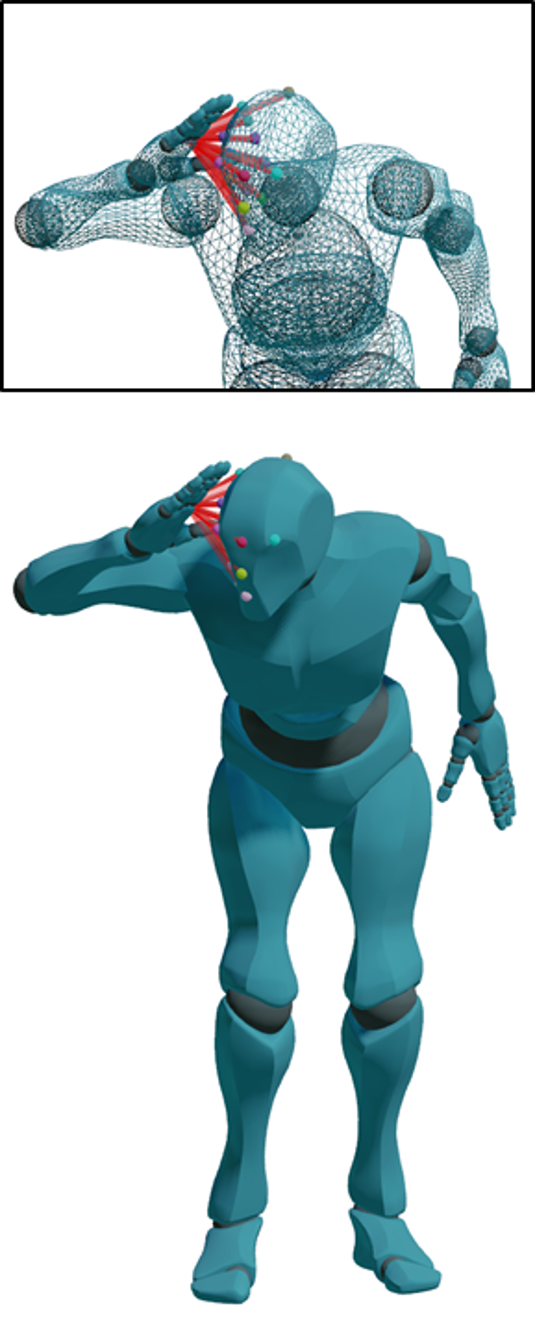}
        \caption*{\textit{Source Pose}} \label{fig:DirectOpt_Ablation_Anclr_Source}
    \end{minipage}
    % \hspace{2em}
    \begin{minipage}{0.09\textwidth}
        \centering
        \includegraphics[width=\textwidth]{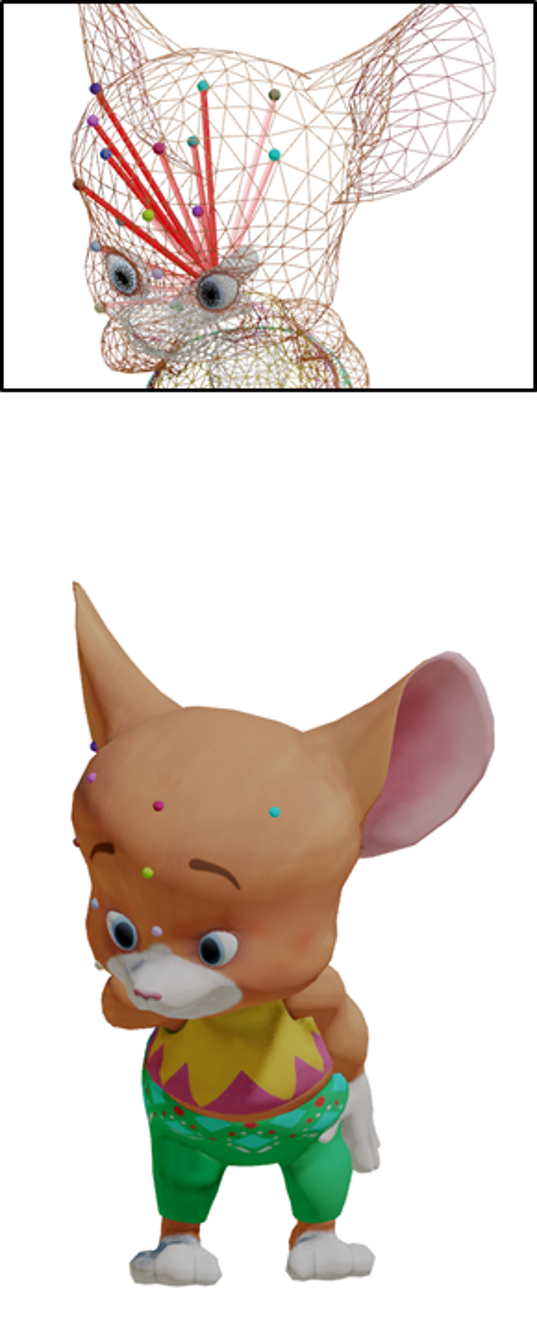}
        \caption*{(a) lr=0.001} \label{fig:DirectOpt_Ablation_Anclr_Opt0.001}
    \end{minipage}
    % \hspace{2em}
    \begin{minipage}{0.09\textwidth}
        \centering 
        \includegraphics[width=\textwidth]{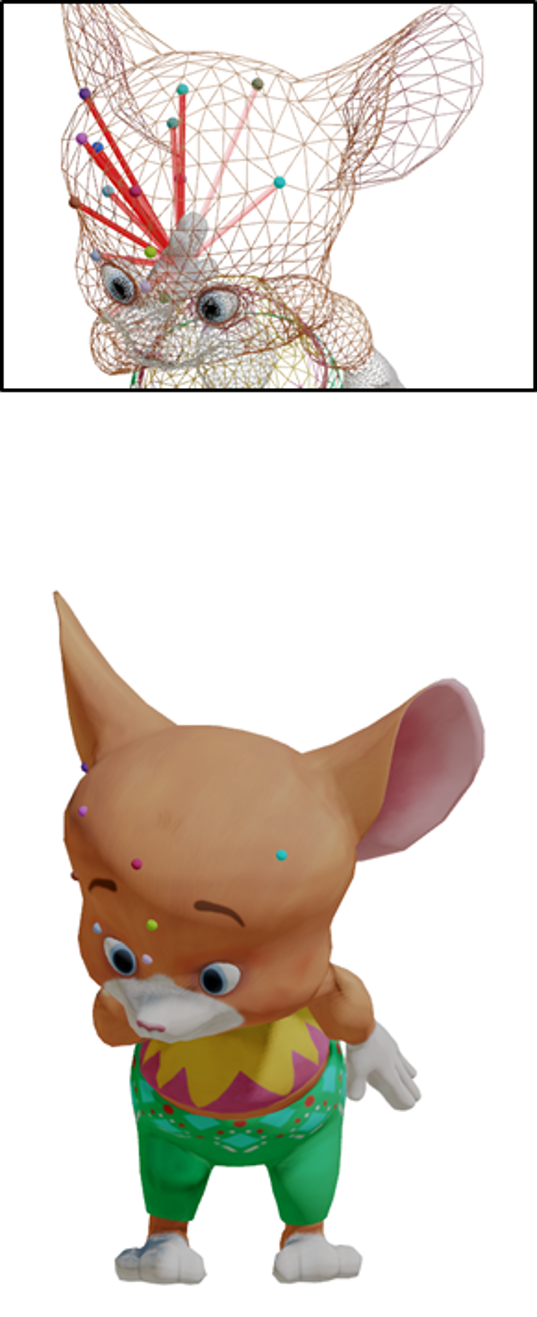}
        \caption*{(b) lr=0.01} \label{fig:DirectOpt_Ablation_Anclr_Opt0.01}
    \end{minipage}
    \begin{minipage}{0.09\textwidth}
        \centering 
        \includegraphics[width=\textwidth]{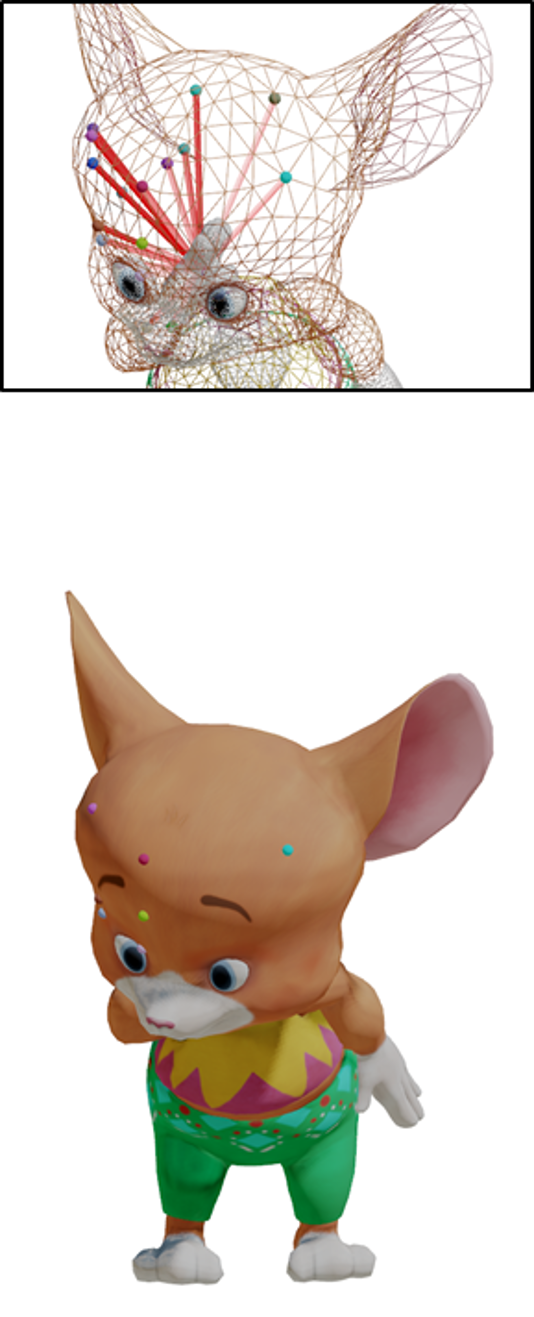}
        \caption*{(c) lr=0.1} \label{fig:DirectOpt_Ablation_Anclr_Opt0.1}
    \end{minipage}
    \begin{minipage}{0.09\textwidth}
        \centering 
        \includegraphics[width=\textwidth]{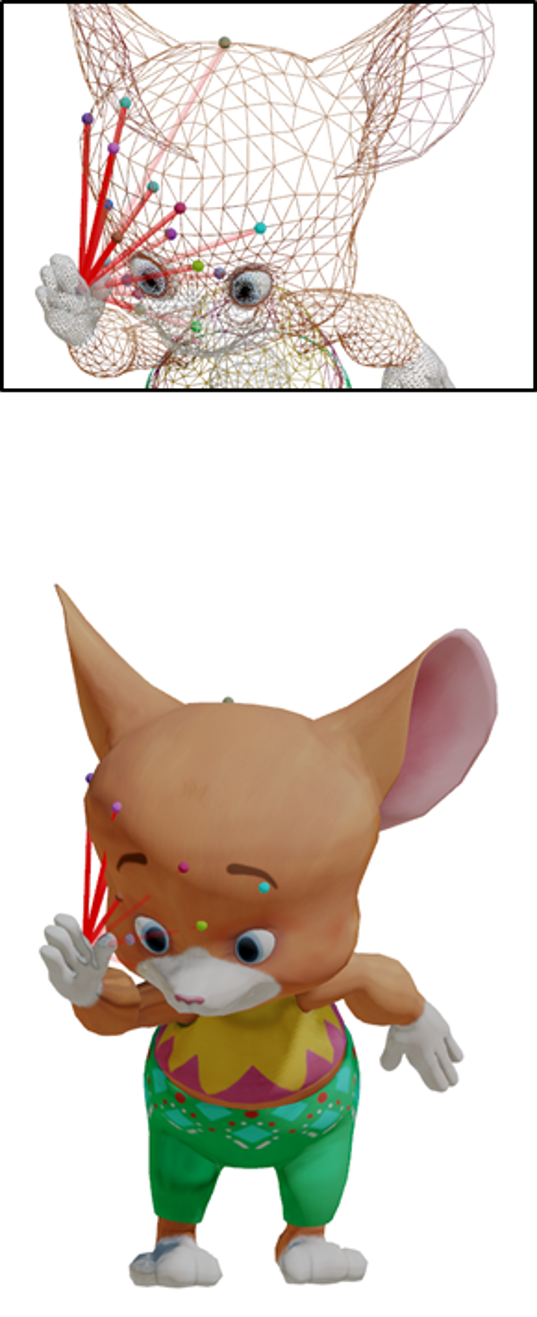}
        \caption*{(d) Ours} \label{fig:DirectOpt_Ablation_Anclr_Ours}
    \end{minipage}
    \caption{Qualitative results of the ablation on the learning rate (lr) for anchor optimization. Figures enclosed in boxes provide inner views of the characters. Subfigures (a) to (c) correspond to retargeted poses produced by the direct optimization with different learning rates, while subfigure (d) corresponds to the result of our learning-based model.}
    \label{fig:DirectOpt_Ablation_Anclr}
\end{figure}

As shown in Table \ref{tab:comparison_with_optimization_based}, the direct optimization baseline achieved higher contact preservation scores than our learning-based model, as it directly minimizes the geometric proximity objectives at test time. However, this improvement comes at the cost of a higher penetration rate, which can be attributed to limited anchor adaptation. Although we used the same alternating update strategy as in our learning-based method, the anchors remained close to their initial positions, and the proximity constraints were mainly satisfied through pose updates. Consequently, direct optimization preserved precise contact in less challenging cases where substantial anchor repositioning was not required, but became less reliable in challenging cases that required anchor adaptation, as shown in Figure \ref{fig:DirectOptimization_Comparison}. We further examined whether the limited anchor adaptation in direct optimization could be alleviated by increasing the number of optimization steps or the learning rate of the anchor. Increasing the number of steps distorted the target pose and introduced temporal instability, while the anchor configuration remained largely unchanged. As shown in Figure \ref{fig:DirectOpt_Ablation_Anclr}, although a higher anchor learning rate produced larger anchor displacements, these changes did not consistently lead to task-relevant anchor adaptation or plausible poses guided by the adapted anchors.

\begin{table} [!t]
    \caption{Quantitative comparison between our method and optimization-based baseline. The best result for each metric is highlighted in bold.}
    \centering
    \setlength{\tabcolsep}{10pt}
    \resizebox{\columnwidth}{!}{%
    \begin{tabular}{l|c|cll}
    \hline
    \multicolumn{1}{c|}
    {\multirow{2}{*}{Methods}} 
    & \multirow{2}{*}{\begin{tabular}[c]{@{}c@{}}Pen (\%) ↓\end{tabular}} 
    & \multicolumn{3}{c}{Contact Preservation}             
    \\ \cline{3-5} 
    \multicolumn{1}{c|}{}    
    &   & Prec ↑   & Rec ↑   & Acc ↑ \\ \hline
    
    (a) \ Direct Optimization          
    & 16.350 & \textbf{0.452} & \textbf{0.528} & \textbf{0.950} \\ 
    (b) \ Learning-based (Ours)                     
    & \textbf{14.730} & 0.415 & 0.350 & 0.948 \\ \hline    
    \end{tabular}%
    } 
    \label{tab:comparison_with_optimization_based}
\end{table}
\begin{table} [!t]
    \caption{Comparison of the processing time required for direct optimization and inference using our learning-based method for the sequence shown in Figure \ref{fig:DirectOpt_Ablation_Anclr}.}
    \centering
    \setlength{\tabcolsep}{10pt}
    \resizebox{0.8\columnwidth}{!}{%
    \begin{tabular}{l|c}
    \hline
    Methods & Avg. Time / Frame (sec) $\downarrow$ \\ \hline
    (a) \ Direct Optimization 
    & 6.237 \\
    (b) \ Learning-based (Ours) 
    & \textbf{0.00545} \\ \hline
    \end{tabular}%
    }
    \label{tab:DirectOptimization_Comparison_Computation_Time}
\end{table}
These results suggest that the limited anchor adaptation observed in direct optimization is not merely due to insufficient optimization steps or small anchor updates. Rather, without a data-driven prior learned from the training distribution, the coupled objectives are mainly satisfied through pose adjustment instead of task-relevant anchor adaptation. In contrast, our learning-based approach learns from diverse character-motion pairs through shared network parameters. This provides implicit data-driven regularization for plausible anchor adaptation and pose retargeting, leading to a more balanced trade-off between contact preservation and penetration avoidance. In addition, our method avoids iterative per-sequence optimization at test time. For the sequence shown in Figure \ref{fig:DirectOpt_Ablation_Anclr}, the average processing time per frame is reduced from $6.237$ seconds for direct optimization to $0.00545$ seconds for our learning-based method, as shown in Table \ref{tab:DirectOptimization_Comparison_Computation_Time}. Please refer to the supplementary video for the animation results.

\end{document}